\documentclass{IISERB}

\usepackage{extraipa}

\graphicspath{{figures/}}

\newcommand{\be}{\begin{equation}}
\newcommand{\ee}{\end{equation}}

\newcommand{\etal}{\textit{et al}.}

\newcommand{\intp}[1]{\int\frac{d^4{#1}}{(2\pi)^4}}
\newcommand{\intx}[1]{\int d^4{#1} }

\newcommand{\thesistitle}{Covariant Quantum Gravitational Corrections to Scalar and Tensor Field Models}
\newcommand{\studentname}{Sandeep Aashish}
\newcommand{\studentrollno}{1610511}
\newcommand{\advisorname}{Dr. Sukanta Panda}
\newcommand{\subject}{Physics}
\newcommand{\thesisdate}{April, 2020}

\begin{document}

\begin{titlepage}
\maketitle
\end{titlepage}
\newpage
\begin{titlepage}
\maketitle
\end{titlepage}

\blankpage
\addcontentsline{toc}{chapter}{Academic Integrity and Copyright Disclaimer}

\begin{center}
{\textbf{\large{ACADEMIC INTEGRITY AND COPYRIGHT DISCLAIMER}}}
 \end{center}
I hereby declare that this thesis is my own work and, to the best of my knowledge, it contains no materials previously published or written by any other person, or substantial proportions of material which have been accepted for the award of any other degree or diploma at IISER Bhopal or any other educational institution, except where due acknowledgement is made in the thesis. \\ \\
I certify that all copyrighted material incorporated into this thesis is in compliance with the Indian Copyright (Amendment) Act, 2012 and that I have received written permission from the copyright owners for my use of their work, which is beyond the scope of the law. I agree to indemnify and save harmless IISER Bhopal from any and all claims that may be asserted or that may arise from any copyright violation.

\vspace{2cm}
\parbox{0.7\textwidth}{ 
	\flushleft{\thesisdate\\IISER, Bhopal}
}
\hfill 
\parbox{0.3\textwidth}{ 
\mbox{\studentname}
}\\ \\ \\ \\ \\ \\
\hphantom{blablablablablablablablablaballabalbalbalabblablablablablablabl} 
\advisorname\\
\hphantom{blablablablablablablablablaballabalbalbalabblablablablablablabl} \ \ \ \ \ (Supervisor)
\blankpage
\addcontentsline{toc}{chapter}{Acknowledgement}

\begin{center}
{\textbf{\large{ACKNOWLEDGEMENTS}}}
 \end{center}
This thesis is the culmination of contributions from a number of people, who have knowingly and unknowingly encouraged me to persist through this incredible journey spanning a plethora of experiences. First and foremost, I am indebted to my advisor Dr. Sukanta Panda for his unwavering support and guidance. Thank you for giving me the freedom to pursue these problems, and for allowing me to fail and learn from my failures. I also express my sincere gratitude to Dr. Asrarul Haque for an early introduction to the way of life that is research, during my undergraduate studies. Those lessons have undeniably helped shape my endeavours ever since. 

I am thankful to Dr. Ambar Jain for his patience and help on numerous occasions throughout my time at IISER Bhopal, starting from teaching me quantum field theory to introducing me to effective field theory and several insightful discussions. I am thankful to Dr. Rajib Saha, Dr. Phani K. Peddibhotla, Prof. Subhash Chaturvedi and Dr. Sebastian W{\"u}ster for the opportunity to assist with lectures and tutorials. I also thank Dr. Snigdha Thakur, Dr. Suvankar Dutta and Dr. Ritam Mallick for the administrative support time and again. I humbly acknowledge the financial and infrastructural support from IISER Bhopal. I am grateful to Prof. Alan Kostelecky and Prof. David J. Toms for their insights and help related to my works. 

I am fortunate to have had the opportunity to start a journal club known as Physics Studio along with my seniors, most notably Dr. Arghya Chattopadhyay, Dr. Sreeraj Nair, Dr. Surajit Sarkar, Prabha Chuphal, Ujjal Purkayastha and Shalabh Anand. It has been an honour sharing this incredible journey with my friends Abhilash, Avani, Pawan, Arun, Devendra, Ritu and Soudamini. Thank you for the great memories! I also thank my dearest friends Pranay, Shivam, Vaishak and Vikrant from before IISER, for brief but joyous moments whenever we got the chance. 

I cannot possibly describe in words how grateful I am to my family. My selfless parents have endured many hardships to ensure that I get to follow my dreams. I am thankful to my lovely sisters Supriya and Shephali for cheering me up by making me cheer them up! And of course, to our pet Chiku, who deserves a lot of credit for making us all happy!

Finally to nature, the greatest teacher. Thank you for inspiring this journey, and for teaching me humility.

\blankpage
\addcontentsline{toc}{chapter}{Dedication}

\begin{dedication}
    Dedicated to the average people of this world,\\
    for their average, unrelenting, unacknowledged hard work\\
    that makes dreams of visionaries come true.
\end{dedication}

\blankpage 
\addcontentsline{toc}{chapter}{Abstract}

\begin{center}\large{\textbf{ABSTRACT}}\end{center}

    Recent and upcoming experimental data as well as the possibility of rich phenomenology have spiked interest in studying the quantum effects in cosmology at low (inflation-era, typically four orders of magnitude lower than Planck scale) energy scales. While Planck scale physics is under development, it is still possible to incorporate quantum gravity effects at relatively low energies using the framework of Quantum Field Theory in Curved Spacetime. It serves as a low-energy limit of Planck scale physics, making it particularly useful for studying physics in the early universe. One of the approaches to find covariant quantum corrections is the DeWitt-Vilkovisky's (DV) covariant effective action formalism that is gauge invariant and background field invariant. 
    
    We use the DeWitt-Vilkovisky method to study formal and cosmological aspects of quantum fields in curved spacetime, and take initial steps towards studying quantum gravitational corrections in cosmological setting. The thesis comprises of mainly two parts. We first study the formal aspects of rank-2 antisymmetric tensor field which appear in the low energy limit of superstring models and are thus relevant in the early universe, in particular the quantization and quantum equivalence properties, for the case with and without spontaneous Lorentz violation. The effective action is generalized for gauge theories whose gauge parameters possess additional symmetries. When used in the case of spontaneously Lorentz violating antisymmetric tensor field model, it is found that classical equivalence with a vector theory breaks down at one-loop level due to the presence of Lorentz violating terms. The final chapter of this thesis is devoted to taking first steps towards exploring applications of DV method in early universe cosmology. We calculate perturbatively the covariant one-loop quantum gravitational effective action for a scalar field model inspired by the recently proposed nonminimal natural inflation model. The effective potential is evaluated taking into account the finite corrections, and an order-of-magnitude estimate of the one-loop corrections reveals that gravitational and non-gravitational corrections have same or comparable magnitudes.  
\blankpage
\addcontentsline{toc}{chapter}{List of Publications}

\begin{center}\large{\textbf{LIST OF PUBLICATIONS}}\end{center}

\label{AppendixE}

Publications included in this thesis:
\begin{enumerate}
\item S. Aashish and S. Panda, 2018, Covariant effective action for an antisymmetric tensor field, \textit{Phys. Rev. D} \textbf{97}, 125005.

\item S. Aashish and S. Panda, 2019, Quantum aspects of antisymmetric tensor field with spontaneous Lorentz violation, \textit{Phys. Rev. D} \textbf{100}, 065010.

\item S. Aashish and S. Panda, 2019, One-Loop Effective Action for Nonminimal Natural Inflation Model, arXiv:1905.12249 (to appear in \textit{Springer Proc. Phys.}).

\item S. Aashish and S. Panda, 2020, On the quantum equivalence of an antisymmetric tensor field with spontaneous Lorentz violation, \textit{Mod. Phys. Lett. A} \textbf{33}, 1 2050087.

\item S. Aashish and S. Panda, 2020, Covariant quantum corrections to a scalar field model inspired by nonminimal natural inflation, \textit{JCAP} \textbf{06}, 009.
\end{enumerate}

Other publications on the cosmological aspects of antisymmetric tenosr fields, from the work carried out at IISERB outside of this thesis:
\begin{enumerate}
\item S. Aashish, A. Padhy, S. Panda and A. Rana, 2018, Inflation with an antisymmetric tensor field, \textit{Eur. Phys. J. C} \textbf{78}: 887.
\item S. Aashish, A. Padhy and S. Panda, 2019, Avoiding instabilities in antisymmetric tensor field driven inflation, \textit{Eur. Phys. J. C} \textbf{79}: 784.
\item S. Aashish, A. Padhy and S. Panda, 2020, Gravitational waves from inflation with antisymmetric tensor field, arXiv:2005.14673 (under revision, JCAP)
\end{enumerate}

\addcontentsline{toc}{chapter}{Table of Contents}
\tableofcontents
\newpage






\addcontentsline{toc}{chapter}{List of Figures}
\listoffigures

\addcontentsline{toc}{chapter}{List of Tables}
\listoftables

\pagestyle{fancy}
\clearpage
\pagenumbering{arabic}

\newpage


\chapter{Introduction}

In this chapter, we introduce the background and general motivations for this thesis. Sec. \ref{thesisintro} contains an introduction to the thesis. In Sec. \ref{thesisreview}, we review the covariant effective action formalism, which forms the basis of this thesis. 

\section{\label{thesisintro}Motivations}
When Newton first published a comprehensive theory of gravity way back in 1687 \cite{newton}, little was known about what other forces exist in nature. More than three centuries later, we now know that gravity is one of the four fundamental forces of nature. In fact, Newtonian gravity is a low length-scale \footnote{Newtonian gravity is valid in the solar system length scale.} approximation of general relativity, developed by Einstein in early twentieth century, which describes gravity as a consequence of the geometry of spacetime and its relation to the energy-momentum tensor of matter. Ironically however, gravity is also the least understood of all, especially since we lack an understanding of gravity at energies close to Planck scale. A major hindrance is our inability to successfully quantize gravity. Unlike the other three fundamental forces, namely the electromagnetic, strong and weak forces which come within the purview of Standard Model \cite{schwartz2013}, the energy scale at which quantum gravity effects might become relevant (Planck scale) is much higher than what is attainable in laboratory, for example with the LHC. Hence, any theoretical formulation of quantum gravity is untestable and consequently there exist several competing theories for quantizing gravity, such as string theory and loop quantum gravity.   

Moreover, due to the smallness of gravitational constant, according to effective field theory treatment it should be possible to obtain perturbative quantum corrections to gravity at low energy scales (compared to Planck scale) which are in principle more accurate than quantum electrodynamics or quantum chromodynamics  (see Ref. \cite{woodard2009} for a comprehensive review). Unfortunately, efforts towards quantizing gravity perturbatively have failed to consistently absorb the divergences, giving rise to non-renormalizability. In the past two decades or so, a more modern view has developed where general relativity is studied as a quantum effective field theory at low energies \cite{donoghue1994}. This treatment allows separation of quantum effects from known low energy physics from those that depend on the ultimate high energy completion of the theory of gravity (see Ref. \cite{donoghue2015} for a review by Donoghue and Holstein). 

One of the well-known methods employed in such studies is to compute the effective action, which is known to be the generator of 1PI diagrams \cite{buchbinder1992,vilkovisky1992}. An advantage of this technique is that one can directly obtain divergence structure at a given loop order, without going through the hassle of summing over individual Feynman diagram contributions. Other applications include the calculation of effective potential \cite{coleman1973}. The computation of effective action is most commonly carried out using the background field method, where small fluctuations about a classical background field are quantized, not the total field. In general, it turns out that the results consequently depend on the choice of background field \cite{falkenberg1998,labus2016,ohta2016}. In case of gravity, which is treated as a gauge theory, it is therefore important to ensure that there are no fictitious dependence of conclusions on the choice of gauge and background. The subject of this thesis is to systematically develop the computation of one-loop effective action for theories relevant in the early universe using DeWitt-Vilkovisky's approach that yields gauge and background independent effective action \cite{dewitt1967b,dewitt1967c,parker2009,dewitt1964,dewitt1967a,vilkovisky1984a,vilkovisky1984b}. Moreover, features such as frame independence can be introduced in this formalism by taking into account the conformal transformations in addition to field reparametrizations\cite{steinwachs2013,steinwachs2015,karamitsos2018,karamitsos2017}, making it an ideal tool to study quantum gravitational effects in the context of spontaneously Lorentz-violating models of antisymmetric fields, and other cosmological models in general.

In what follows, we will briefly review the antisymmetric tensor fields which are the subject of next two chapters, and inflationary cosmology which inspires the final chapter. In the next section, we review the effective action formalism used throughout this work. In chapter 2, we set up the formalism to write the effective action of theories where gauge parameters have additional symmetries, and apply it to the case of rank-2 antisymmetric tensor field. In chapter 3, we compute the one-loop effective action for a spontaneously Lorentz violating model of antisymmetric tensor field, in a nearly flat spacetime while keeping gravitational perturbations classical. In chapter 4, we include quantum gravitational corrections in the computation of effective action and effective potential for a scalar field model inspired by nonminimal natural inflation.

\subsection{Antisymmetric Tensors}
Antisymmetric tensor field appears in most superstring theories in the low-energy limit corresponding to four dimensional spacetime \cite{rohm1986,ghezelbash2009}. They have been studied in the past in several contexts, including strong-weak coupling duality and phase transitions \cite{quevedo1996,olive1995,polchinski1995,siegel1980,hata1981,buchbinder1988,duff1980,bastianelli2005a,bastianelli2005b}. In the recent past, interest has grown towards studying $2-$forms (and, by extension $n-$forms) in the context of early universe physics. There are no observational signatures of antisymmetric fields in the present universe \cite{das2018}, but it has been shown that $2-$forms might play a significant role in the early universe \cite{elizalde2018}. This line of thought along with challenges faced by scalar and vector models of primordial inflation has fuelled exploration of inflation models driven by antisymmetric tensors \cite{markou2019a,elizalde2018,aashish2018c,aashish2019a,koivisto2009a}, and is under active development. 

A majority of this thesis focuses on the formal aspects of antisymmetric tensors, and is inspired by the work carried out by Altschul \etal \cite{altschul2010}, where spontaneous Lorentz violation with various rank-2 antisymmetric field models minimally and non minimally coupled to gravity was investigated. A remarkable feature of that study is the presence of distinctive physical features with phenomenological implications for tests of Lorentz violation, even with relatively simple antisymmetric field models with a gauge invariant kinetic term. More recently, quantization and propagator for such theories have been studied in Refs. \cite{maluf2019,aashish2019b}. Lorentz violation is also a strong candidate signal for quantum gravity, and is part of the Standard Model Extension research program \cite{bonder2015}. Such interesting phenomenological possibilities have been a strong motivation for various works on spontaneous Lorentz violation (SLV) \cite{hernaski2016,azatov2006,kostelecky1989b,kostelecky1998,kostelecky2004,bluhm2005,carroll1990,jackiw1999,coleman1999,bertolami1999}.

Antisymmetric tensors, and $n-$forms in general, also display interesting properties with regard to their equivalence with scalar and vector fields. For instance, in four dimensions theory of a massless 2-form field (with a gauge-invariant kinetic term) is classically equivalent to a massive nonconformal scalar field, while a massless 3-form theory does not have any physical degrees of freedom (see \cite{buchbinder2008} and references therein). Likewise, a massive rank-2 antisymmetric field is classically equivalent to a massive vector field, and a rank-3 antisymmetric field is equivalent to massive scalar field \cite{buchbinder2008}. Such properties are useful in the analysis of degrees of freedom of these theories \cite{altschul2010}. Classical equivalence implies that the actions of two theories are equivalent. However, quantum equivalence is established at the level of effective actions, and it is in general not straightforward to check especially in curved spacetime. Moreover, classical equivalence between two theories does not necessarily carry over to the quantum level, particularly in the case of spontaneously broken Lorentz symmetry \cite{seifert2010a,aashish2019b}, and thus makes for an interesting study. 

Quantum equivalence in the context of massive rank-2 and rank-3 antisymmetric fields in curved spacetime, without SLV, was first studied by Buchbinder \etal \cite{buchbinder2008} and later confirmed in Ref. \cite{shapiro2016}. The proof of quantum equivalence in Ref. \cite{buchbinder2008} was based on the zeta-function representation of functional determinants of $p$-form Laplacians appearing in the 1-loop effective action, and identities satisfied by zeta-functions for massless case \cite{rosenberg1997,elizalde1994,hawking1977}. Quantum equivalence results from these identities generalized to the massive case. In flat spacetime though, the proof is trivial as operators appearing in the effective action reduce to d'Alembertian operators due to vanishing commutators of covariant derivatives and equivalence follows by taking into account the independent components of each field.

In the subsequent chapters, we consider two simple models of rank-2 antisymmetric field minimally coupled to gravity: first with a massive potential term upon which we apply the general quantization procedure developed in Chapter 2; and then with the simplest choice of spontaneously Lorentz violating potential \cite{altschul2010} to investigate its equivalence properties. Its classical equivalence was studied in Ref. \cite{altschul2010} in terms of an equivalent Lagrangian consisting of a vector field $A_{\mu}$ coupled to auxiliary field $B_{\mu\nu}$ in Minkowski spacetime. Our interest is to take first steps to extend the classical analysis in Ref. \cite{altschul2010} to quantum regime. However, checking the quantum equivalence of such classically equivalent theories is not straightforward, in flat as well as curved spacetime. We find that the simple structure of operators breaks down due to the presence of SLV terms. As a result, the difference of their effective actions does not vanish in Minkowski spacetime, contrary to the case without SLV. However, this does not threaten quantum equivalence due to a lack of field dependence in the effective actions, which will therefore cancel after normalization.

In curved spacetime, making a conclusive statement about quantum equivalence is a nontrivial task for the following reasons. First, directly comparing effective actions using known proper time methods as in Ref. \cite{shapiro2016} is a difficult mathematical problem. Unlike the minimal operators (of the form $g_{\mu\nu}\nabla^{\mu}\nabla^{\nu} + Q$, where $Q$ is a functional without any derivative terms) found in \cite{buchbinder2008} for instance, we encounter nonminimal operators in functional determinants of the effective action, for which finding heat kernel coefficients to evaluate the determinants is a highly nontrivial task. Second, the formal arguments made in Ref. \cite{buchbinder2008} do not apply to the present case due to the non-trivial structure of operators appearing in effective actions. Therefore, we adopt a perturbative approach wherein the effective action is computed in a nearly flat spacetime perturbatively in orders of the (classical) metric perturbations.

\subsection{Quantum Gravity and Inflationary Cosmology}
More recently, with the availability of high precision data from experiments probing the early universe, especially inflation era, it has become important to consider quantum gravitational corrections in early universe cosmology \cite{fabris2012,krauss2014,woodard2014}. This has motivated several studies of aspects of quantum gravitational corrections in inflationary universe, see for example Refs. \cite{klemm2004,cognola2005,cognola2012,hebecker2017,herranen2017,bounakis2018,markkanen2018,ruf2018,heisenberg2019}. Here, we touch upon some motivations for the work carried out in fourth chapter. 

The most phenomenologically accessible information about the early universe comes from a nearly uniform background electromagnetic radiation dating back to the epoch of recombination, known as the cosmic microwave background (CMB) \cite{jones1998,baumann2009,fixsen2009}. The surprisingly uniform nature of CMB, among other problems (namely the flatness of early universe), is explained by a paradigm called inflation, first introduced by Guth \cite{guth1981}. This proposal has since led to more than three decades of effort to build models of inflation that fit well with the observed CMB data (see Ref. \cite{martin2016} for a review). The simplest models of inflation consist of one or more scalar fields driving the inflation. With the advent of high-precision observational data (like the recent Planck 2018 results \cite{planck2018x}), majority of scalar field driven inflation models have been ruled out while the ones in agreement are tightly constrained. More recently, new set of theoretical conditions called the Swampland criteria arise from the requirements for any effective field theory to admit string theory UV completion \cite{brennan2018,andriot2018,garg2018,obied2018,kallosh2019}, and further constrain scalar field potentials. For a comprehensive review, see Ref. \cite{martin2014}.
   
Among the theories not involving scalar fields, in particular those with vector fields \cite{ford1989,burd1991,golovnev2008,darabi2014,bertolami2015}, constructing successful models is often marred by ghost and gradient instabilities \cite{peloso2009,emami2017} that lead to unstable vacua. Inflation with non-Abelian gauge fields have been shown to be free from these instabilities \cite{jabbari2011,jabbari2012,jabbari2013a,jabbari2013b}, but are in tension with Planck data and hence ruled out \cite{peloso2013}. In the recent past, inflation models with rank-2 antisymmetric tensor fields have been explored \cite{markou2019a,elizalde2018,aashish2018c,aashish2019a,koivisto2009a} and efforts are on to perform phenomenological studies in the near future \cite{aashish2020b}.

The CMB, and hence the physics of inflation, is one of the very few realistic avenues of detecting quantum gravity signatures \cite{woodard2009}. We are far off from the possibility of probing energy scales of quantum gravity (Planck scale) in laboratory, but it may be possible to detect these in the primordial gravitational waves generated during the phase of inflation in near future \cite{marc1997,woodard2009,woodard2014}, since the energy scales during inflation era ($~10^{16} GeV$) is high enough for perturbative quantum gravity effects to be relevant \footnote{From dimensional analysis in the natural units, it can be seen that perturbative quantum gravitational corrections should be suppressed by a factor of $GE^2$, where $G$ is the gravitational constant and $E$ is the energy scale. To overcome the smallness of $G$, one thus has to increase $E$.}. This is possible because the massless graviton modes (a direct consequence of quantum gravity) produced during inflation were frozen when they crossed the cosmic horizon. These modes re-entering the horizon today, if detected, would confirm the existence of quantum gravity. 

Among the approaches for studying perturbative quantum gravitational corrections are the diagrammatic calculations in the EFT framework pioneered by Donoghue \cite{donoghue1994} (See also Ref. \cite{donoghue2003} for calculation of correction to Newtonian potential) used to study UV corrections, and the deep IR corrections using the approach developed by Woodard and collaborators \cite{woodard2014,prokopec2002,janssen2008}. We use the covariant effective action approach (see Ref. \cite{saltas2017} for a recent example) that yields quantum gravitational corrections including off-shell contributions, unlike non-covariant methods\footnote{The diagrammatic computations of one-loop gravitational corrections in the past have been non-covariant, for instance in Ref. \cite{rodigast2010a}. However, the issue of covariance only arises when off-shell effects are considered, for example in \cite{bounakis2018}.}. As a starting step, we perform the calculations in a Minkowski background.

\section{\label{thesisreview}Covariant Effective Action Formalism}
In this section, we review the formalism used throughout this thesis largely inspired by Parker and Toms \cite{parker2009}. We begin with introducing the \textit{condensed} or geometric notations introduced by DeWitt \cite{dewitt1967b} to write a gauge- and background- independent effective action. 

\subsection{Geometric notations}
Invariance of the effective action under coordinate transformations and field redefinitions is effected by going to the space of fields with field components as the coordinates in field space. In DeWitt's condensed notation \cite{dewitt1964}, field components are denoted by local coordinates $\varphi^{i}$ in field space. The index $i$ in field space corresponds to all gauge indices and coordinate dependence of fields. 

All the field components (variables) in the action are represented by $\varphi^{i}$. For example, if a field variable $A_{\mu}(x)$ is denoted by $\varphi^{i}$ in field space, then $i$ is mapped to both the tensor index and coordinate index i.e. $i\longrightarrow (\mu,x)$. Similarly, a set of multiple fields like $\phi(x),A_{\mu}(x),h_{\mu\nu}(x)$ when represented by $\varphi^{i}$ in the condensed notation, implies that $i$ runs over indices of all fields i.e. $\{x\},\{\mu,x\},\{\mu\nu,x\}$. The Einstein summation convention still follows here, which implies that repeated (or contracted) indices in the condensed notation represent a sum over all the associated gauge or tensor indices and integral over all coordinate indices, i.e.
\begin{equation}
\label{geo2}
g_{ij}v^{i}w^{j} = \int d^{n}x d^{n}x' g_{IJ}(x,x')v^{I}(x)w^{J}(x').
\end{equation}
The Dirac $\delta$-distribution in field space is defined as 
\begin{equation}
\label{geo3}
\delta^{i}_{j} = |g(x')|^{1/2}\delta^{I}_{J}\delta(x,x') \equiv \delta^{I}_{J} \tilde{\delta}(x,x'),
\end{equation}
where, $\tilde{\delta}(x,x')$ transforms as a scalar in first argument, and scalar density in second argument. This definition of $\tilde{\delta}(x,x')$ carries over to the field space Dirac $\delta$-distribution as well. 
The functional derivative is given by
\begin{eqnarray}
\label{geo4}
\varphi^{i}_{,j} \equiv \dfrac{\delta\varphi^{I}(x)}{\delta\varphi^{J}(x')} = |g(x')|^{1/2}\delta^{I}_{J}\delta(x,x') = \delta^{i}_{j}.
\end{eqnarray}
A metric $g_{ij}$ can be defined in the field space with properties analogous to the spacetime metric: $g_{ij}g^{jk} = \delta_{i}^{k}$. Analogous to the coordinate space treatment the structure of field-space metric can be read off of the invariant length element in field space,
\begin{eqnarray}
    ds^2 = g_{ij}d\varphi^{i} d\varphi^{j}.
\end{eqnarray}
As an example, let us consider the case of electromagnetic theory with the action, 
\begin{eqnarray}
    S[A] = -\dfrac{1}{4} \intx{x} F_{\mu\nu}F^{\mu\nu},
\end{eqnarray} 
where $F_{\mu\nu} = \partial_{\mu}A_{\nu} - \partial_{\nu}A_{\mu}$. The field-space length element can be written as,
\begin{eqnarray}
    \label{ds2}
    ds^2 = \intx{x}\intx{x'}G^{\mu\nu}(x,x') dA_{\mu}(x) dA_{\nu}(x').
\end{eqnarray}
The simplest choice for field-space metric $G^{\mu\nu}(x,x')$ is,
\begin{eqnarray}
    \label{ex}
    G^{\mu\nu}(x,x') = |g(x)|^{1/2} g^{\mu\nu}(x) \tilde{\delta}(x,x');
\end{eqnarray}
where $g^{\mu\nu}(x)$ is the spacetime metric. In principle, other choices of field space metric are possible but would necessitate the introduction of extra dimensional parameters to balance the dimensions on both sides\footnote{The dimension of $ds^2$ is conventionally chosen to be length-squared.} of Eq. (\ref{ds2}) \cite{parker2009}. According to Vilkovisky's prescription \cite{vilkovisky1984a,vilkovisky1984b} the field space metric can be read off from the highest derivative terms in the classical action functional; for the electromagnetic theory, this prescription can be seen to lead to Eq. (\ref{ex}). 

For any field-space metric $g_{ij}$, the Christoffel connections are defined as,
\begin{equation}
\label{geo5}
\Gamma^{k}_{ij} = \frac{1}{2}g^{kl}(g_{il,j} + g_{lj,i}-g_{ij,l}).
\end{equation}
Since we will be dealing with covariant quantities in our calculations, we will denote the invariant volume element $\sqrt{-g(x)}d^{n}x$ with $dv_{x}$ henceforth ($g(x)$ is the determinant of spacetime metric).

A gauge transformation is given by 
\begin{eqnarray}
\label{gaugetr}
\delta\varphi^{i} = K^{i}_{\alpha}[\varphi]\delta\epsilon^{\alpha},
\end{eqnarray}
where $\epsilon^{\alpha}$ is the gauge parameter and $K^{i}_{\alpha}$ are generators of gauge transformation. For a covariant field-space calculation, one needs to use covariant intervals $\sigma^{i}[\varphi_{*};\varphi]$, which are a generalization of flat field space intervals $\varphi^{i} - \varphi^{i}_{*}$ (where $\varphi_{*}^{i}$ are fixed points in field space) and are defined as,
\begin{eqnarray}
\sigma^{i}[\varphi_{*};\varphi] = g^{ij}\dfrac{\delta}{\delta\varphi^{j}}\sigma[\varphi_{*};\varphi],
\end{eqnarray}
where $\sigma[\varphi_{*};\varphi]$ is the geodetic interval defined as,
\[
    \label{geoint}
\sigma[\varphi_{*};\varphi] = \dfrac{1}{2}(length \ of \ geodesic \ from \ \varphi^{i}_{*} \ to \ \varphi^{i})^{2}.
\]

\subsection{Effective Action}
The Feynman path integral constitutes the fundamental object in quantum field theory that leads to observable quantum properties of a theory. Also known as transition amplitude in scattering theory, or partition function in the statistical physics nomenclature, the path integral has the following form:
\begin{eqnarray}
    \label{partfn}
    Z[J] = \int D\varphi e^{i ( S[\varphi] + \intx{x} J\varphi)},
\end{eqnarray}
where $J(x)$ is the source for field $\varphi(x)$. In the limit $J\to 0$, $Z[J]$ generates the usual green's function. Similarly the generator of connected green's functions, $W[J]$, is defined as\footnote{See Chap. 2 of Ref. \cite{buchbinder1992} for a nice introduction.} 
\begin{eqnarray}
    \exp\left(iW[J]\right) \equiv Z[J].
\end{eqnarray}
Effective action, $\Gamma[\bar{\varphi}]$, is the generator of one-particle irreducible (1PI) diagrams and is defined by the Legendre transform of $W[J]$ that replaces dependence on $J(x)$ in favour of the mean field $\bar{\varphi}$, 
\begin{eqnarray}
    \Gamma[\varphi] = W[J] - J_{i}\varphi^{i};
\end{eqnarray}
where 
\begin{eqnarray}
    \label{legendre}
    \bar{\varphi}^{i} \equiv \langle \varphi^{i} \rangle [J] = \dfrac{\delta W[J]}{\delta J_{i}}. 
\end{eqnarray}
Notice that due to Eq. (\ref{legendre}), the effective action by definition acquires a dependence on a new field $\bar{\varphi}^{i}$ in addition to the fields in action $S[\varphi]$, which in most calculations is the background field around which the theory is quantized. This is precisely the motivation for a covariant formalism that gets rid of background field dependence. 

The expression of effective action in terms of fields and action is given by,
\begin{equation}
\label{geo1}
\exp\left(i\Gamma [\bar{\phi}]\right)= \mathcal{N}\int D\phi\exp\left\{iS[\phi ]+i(\phi-\bar{\phi})\Gamma_{,\phi}[\bar{\phi}]\right\}
\end{equation}
The above expression is not generally invariant under coordinate transformations and field re-parametrizations, both of which are natural requirements for a physical theory. Moreover, the treatment of gauge theories using the Faddeev-Popov method in (\ref{geo1}) leads to a background and gauge condition dependent effective action\cite{buchbinder1992}. 
A covariant effective action, free of gauge and background dependence was achieved by DeWitt\cite{dewitt1964,dewitt1967a,dewitt1967b,dewitt1967c} and Vilkovisky\cite{vilkovisky1984a,vilkovisky1984b} is derived using the geodetic intervals defined in Eq. (\ref{geoint}) and has the form,
\begin{eqnarray}
    \label{effac}
    \exp\left(i\Gamma[\bar{\varphi};\varphi_{*}]\right) = \int d\mu[\varphi_{*}; \varphi] \exp i\left[S[\varphi] + \dfrac{\delta\Gamma[\bar{\varphi};\varphi_{*}]}{\delta \sigma^{i}[\varphi_{*}; \bar{\varphi}]} (\sigma^{i}[\varphi_{*}; \bar{\varphi}] - \sigma^{i}[\varphi_{*}; \varphi])\right].
\end{eqnarray}
The functional measure $d\mu[\bar{\varphi};\varphi_{*}]$ is given by,
\begin{eqnarray}
    d\mu[\varphi_{*}; \varphi] = \left(\prod_{i}d\varphi^{i}|g(\varphi)|^{1/2} |\Delta[\varphi_{*}; \varphi]|\right);
\end{eqnarray}
where, 
\begin{eqnarray}
    \Delta[\varphi_{*}; \varphi] = |g(\varphi_{*})|^{-1/2} |g(\varphi)|^{-1/2} \det (-\sigma_{;ij'}[\varphi_{*},\varphi]).
\end{eqnarray}
Here, $;ij'$ represents covariant derivative with respect first argument (unprimed index) followed by second argument (primed index). Notice the presence of $\varphi_{*}$, which is an arbitrary background field in the vicinity of $\varphi$ and is key to the background independence of $\Gamma[\bar{\varphi};\varphi_{*}]$. Clearly, $\Gamma[\bar{\varphi};\varphi_{*}]$ cannot be computed exactly due to its presence on both sides of Eq. (\ref{effac}). Therefore, the effective action is evaluated perturbatively, in orders of number of loops, as follows. In Eq. (\ref{effac}), the action $S[\varphi]$ is rewritten as a functional of $\varphi_{*}$ and $\sigma^{i}[\varphi_{*}; \varphi]$, and Taylor expanded about $v^{i}\equiv\sigma^{i}[\varphi_{*}; \bar{\varphi}]$:
\begin{eqnarray}
    \label{add1}
S[\varphi_{*}; \sigma^{i}[\varphi_{*}; \varphi]] &=& S[\varphi_{*};v^i] + \sum_{n=1}^{\infty}\dfrac{1}{n!}\dfrac{\delta^n S[\varphi_{*};v^i]}{\delta v^{i_1}\cdots\delta v^{i_n}}(\sigma^{i_1}-v^{i_1})\cdots (\sigma^{i_n}-v^{i_n}),
\end{eqnarray}
followed by a rescaling $\sigma^{i}\to \hbar^{1/2}\sigma^{i} + v^{i}$ so that order of $\hbar$ matches the loop-order. Similarly, expanding $\Gamma[\bar{\varphi};\varphi_{*}]$ in orders of $\hbar$, 
\begin{eqnarray}
    \label{add2}
    \Gamma[\bar{\varphi};\varphi_{*}] = \sum_{n=0}^{\infty}\hbar^{n}\Gamma^{(n)}[\bar{\varphi};\varphi_{*}],
\end{eqnarray}
which gives rise to a series of $n-$loop effective actions $\Gamma^{(n)}$. Using Eqs. (\ref{add1}) and (\ref{add2}) in Eq. (\ref{effac}), and comparing both sides for $n=1$ yields, after some algebra,
\begin{eqnarray}
    \label{1leffac}
    \Gamma^{(1)}[\bar{\varphi};\varphi_{*}] = \dfrac{i\hbar}{2}\ln\det(l^{2} S^{,i}_{j}),
\end{eqnarray}
where $S^{,i}_{j} = g^{ik}[\varphi_{*}]S_{,kj}$, and $l^2$ is an arbitrary factor for fixing dimensionality. 
The DeWitt one-loop effective action can be obtained by taking the limit $\varphi_{*}\to \bar{\varphi}$ in Eq. (\ref{1leffac}). The integral representation of one-loop effective action can also be obtained straightforwardly from Eq. (\ref{1leffac}),
\begin{eqnarray}
    \label{1lint}
    \Gamma^{(1)}[\bar{\varphi}] = -\ln\int[d\zeta]\exp\left[-\dfrac{1}{2}\zeta^{i}\zeta^{j}\Big(S_{,ij}[\bar{\varphi}] - \Gamma^{k}_{ij}S_{,k}[\bar{\varphi}]\Big)\right] .
\end{eqnarray}
where the term inside the exponential is the field-space covariant derivative of functional $S[\varphi]$ with respect to connections $\Gamma^{k}_{ij}$.

\begin{figure}
    \centering
    \includegraphics[scale=0.5]{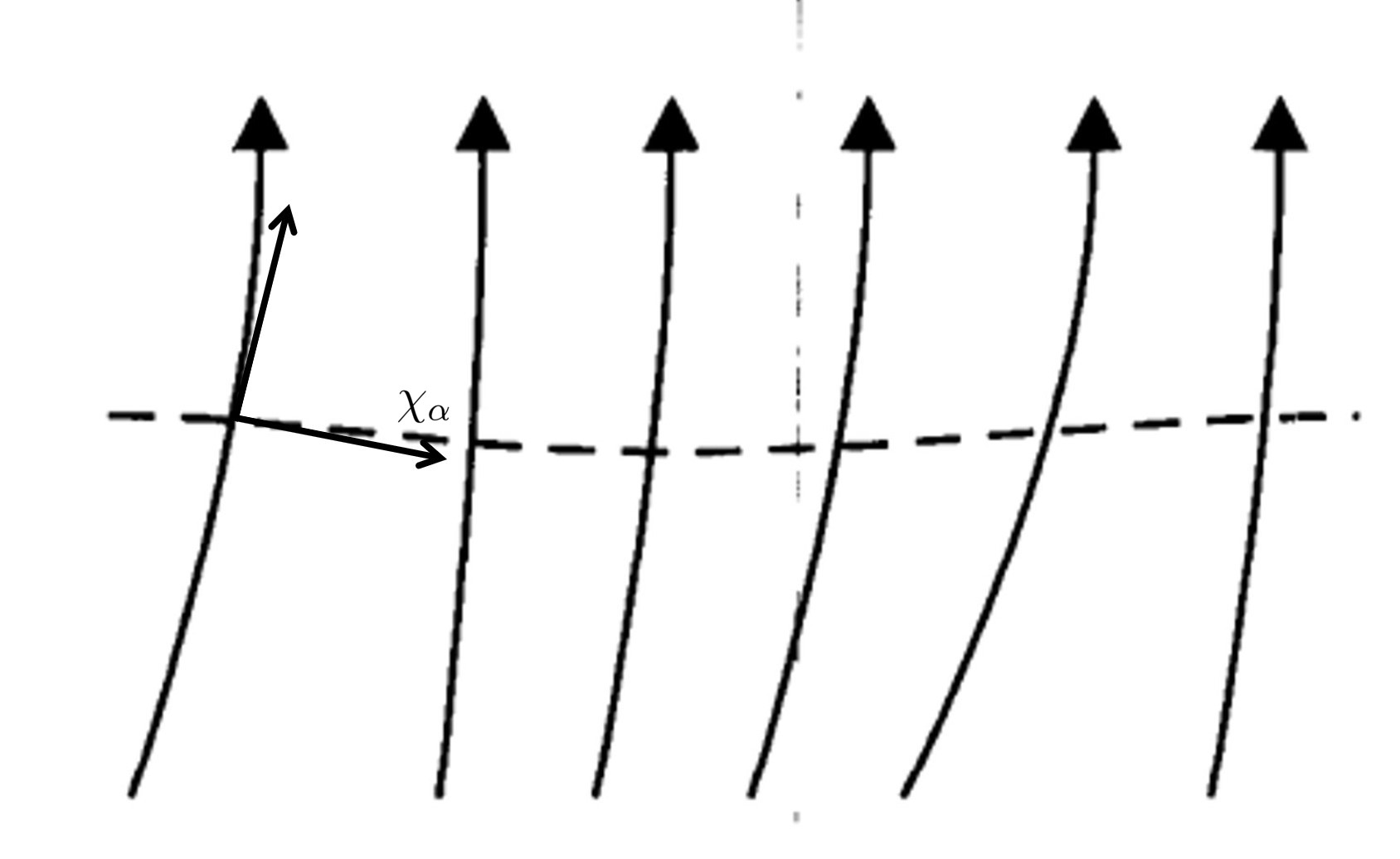}
    \caption{\label{gaugefig} An illustration of gauge orbits in field space (solid vertical lines). The dashed line represents the gauge-fixing procedure, intersecting each orbit at a unique point. Fixing a value of the functional $\chi_{\alpha}$ amounts to fixing a gauge.}
\end{figure}
    
For gauge theories, there exist orbits in field space in which all points are connected by transformations of the form Eq. (\ref{gaugetr}). Fixing a gauge then refers to selecting a unique point from each gauge orbit (see Fig. \ref{gaugefig}). The infinitesimal gauge transformations are represented by Eq. (\ref{gaugetr}), wherein $K^{i}_{\alpha}$ is identified as the generator of gauge transformations, while $\delta\epsilon^{\alpha}$ are the gauge parameters. The gauge fixing condition is given by fixing a functional $\chi_{\alpha}[\bar{\varphi}]$ so that it intersects each gauge orbit in field space only once. Consequently, the functional measure in Eq. (\ref{effac}) receives a modification so as to exclude the contributions from rest of the field coordinates in the gauge orbits. Including the gauge-fixing condition(s) and corresponding ghost determinant(s), the covariant one-loop effective action is given by \cite{huggins1987,toms2007}
\begin{eqnarray}
\label{aeaf2}
\Gamma = -\ln\int[d\zeta]\exp\left[\dfrac{1}{2}\left(-\zeta^{i}\zeta^{j}\Big(S_{,ij}[\bar{\varphi}] - \Gamma^{k}_{ij}S_{,k}[\bar{\varphi}]\Big) - \frac{1}{2\alpha}f_{\alpha\beta}\chi^{\alpha}\chi^{\beta}\right)\right]-\ln\det Q_{\alpha\beta}[\bar{\varphi}],
\end{eqnarray}
as $\alpha\longrightarrow 0$ (Landau gauge). Here, $[d\zeta]\equiv \prod_{i} d\zeta$. A few comments on Eq. (\ref{aeaf2}) are in order. The first term inside the exponential is the covariant derivative of the action functional with respect to $\zeta^{i}$ in field space. $\Gamma^{k}_{ij}$ are the field-space connections defined with respect to the field-space metric $G_{ij}$, and are responsible for general covariance of Eq. (\ref{aeaf2}). In general, the field-space connections have complicated, non-local structure especially in presence of a gauge symmetry. However, they reduce to the standard Christoffel connections, in terms of $G_{ij}$, when $\chi_{\alpha}$ is chosen to be the Landau-DeWitt gauge i.e. $\chi_{\alpha} = K_{\alpha i}[\bar{\varphi}]\zeta^{i} = 0$, along with $\alpha\to 0$ \cite{mackay2010,bounakis2018}. $f_{\alpha\beta}$ is any symmetric, positive definite operator and makes no non-trivial contribution to effective action \cite{parker2009}. Note also that the contributions from connection terms, and hence the question of covariance, is relevant for off-shell analyses, since $S_{,i} = 0$ on-shell. $\det Q_{\alpha\beta}$ is the ghost determinant term that appears during quantization. This term is absorbed into the exponential by introducing Faddeev-Popov ghosts, $c^{\alpha}$ and $\bar{c}^{\alpha}$, so that \cite{parker2009},
\begin{eqnarray}
    \label{aeaf3}
    \ln\det Q_{\alpha\beta} = \ln\int [d \bar{c}^{\alpha}] [d c^{\beta}] \exp \left(- \bar{c}^{\alpha} Q_{\alpha\beta} c^{\beta} \right).
\end{eqnarray}

In the subsequent chapters, we will build on the foundations presented here to explore formal and cosmological applications. We have refrained from reviewing the derivation of integral measure here because this is the subject of next chapter, although in a more general context. Similarly, the computation of effective action is not touched upon here because it is part of chapters three and four. Although the formalism presented here is in general valid for any dimensionality, throughout this thesis we work in four spacetime dimensions, hence it is understood that in what follows, the corresponding integrals and expressions have dimensionality $d=4$. 


\chapter{Covariant Effective Action for an Antisymmetric Tensor Field}

Unlike 1-forms (vector fields), quantization of rank-2 or higher antisymmetric fields is non-trivial especially because of the additional symmetries of gauge parameters of the theory\cite{buchbinder2007}. A simple, ad hoc resolution applicable to the case of antisymmetric fields was discussed by Buchbinder and Kuzenko\cite{buchbinder1988}. More general but complex resolutions to this problem have been discussed before in literature\cite{schwarz1978,schwarz1979,batalin1983}. Moreover, quantization of massive antisymmetric models has an additional challenge, as these models suffer from redundant degrees of freedom even though they are not gauge-invariant, and require the use of St{\"u}ckelberg procedure\cite{stuckelberg1957} to restore \textit{softly} broken gauge freedom before quantizing the theory. 

In this chapter, we will use DeWitt-Vilkovisky's geometrical understanding of field space and gauge fixing to generalize the covariant effective action formalism for quantizing massive rank-2 antisymmetric fields. In doing so, we present an intuitive, geometric prescription for dealing with symmetries of gauge parameters while quantizing the theory. We then write the covariant effective action for a massive rank-2 antisymmetric field. In an attempt to be pedagogical, major steps leading to the effective action have also been presented.

In the next section, we describe the set-up of our problem, including the geometric notations used and the action for the antisymmetric field. The subsequent sections are devoted to generalizing quantization of theories with gauge parameters having additional symmetries followed by an application to the case of antisymmetric tensor field, where the calculation of the covariant one-loop effective action is carried out. The contents of this chapter originally appeared in \cite{aashish2018a}, though some additional calculations and paragraphs have been presented here for the reader's convenience.

\section{\label{sec2} Action for the free antisymmetric rank-2 tensor field}
In our calculations we follow the general procedure of the book by Toms and Parker\cite{parker2009}. This section contains a brief review of the action of the rank-2 antisymmetric field to be quantized. 

Although motivated by superstring models, our interest here differs from the usual Kalb-Rammond fields (massless rank-2 antisymmetric tensor field) that appear in the low energy limit of superstring theories. We consider the action for a minimal model of massive antisymmetric tensor field discussed by Altschul \etal \cite{altschul2010} in the context of spontaneous Lorentz violation,
\begin{equation}
\label{act1}
S[B] = \int d v_{x} \left\{-\dfrac{1}{12}F^{\mu\nu\lambda}[B]F_{\mu\nu\lambda}[B] - \dfrac{1}{4}m^{2}B^{\mu\nu}B_{\mu\nu}\right\},
\end{equation}
where, 
\begin{equation}
\label{act2}
F_{\mu\nu\lambda}[B] \equiv \nabla_{\mu}B_{\nu\lambda} + \nabla_{\lambda}B_{\mu\nu} + \nabla_{\nu}B_{\lambda\mu}.
\end{equation}
The motivation for studying such theories stems from the fact that they possess significant phenomenological consequences, as pointed out in Ref. \cite{altschul2010}. Recent progress in this direction, especially in the context of cosmology and gravitation (see, for example Refs. \cite{koivisto2009a,elizalde2018,aashish2018c,aashish2019a,almeida2019,seifert2019,petrov2019,markou2019a,markou2019b}), validate this motivation. The action (\ref{act1}) belongs to a class of theories having gauge-invariant kinetic term (first term in Eq. \ref{act1}) with a gauge-breaking potential (second term in Eq. \ref{act1}) \cite{buchbinder2007}. The kinetic term is invariant under the transformation,
\begin{equation}
\label{act3}
B_{\mu\nu}\longrightarrow B^{\xi}_{\mu\nu} = B_{\mu\nu} + \nabla_{\mu}\xi_{\nu}-\nabla_{\nu}\xi_{\mu},
\end{equation}
while the mass term $m^{2}B^{\mu\nu}B_{\mu\nu}/4$ is not. It has been shown that, such theories contain the redundant degrees of freedom but cannot be dealt with using traditional Faddeev-Popov method\cite{buchbinder2007,barvinsky1985}. A convenient way to deal with this problem is to restore the softly broken symmetry\cite{buchbinder2007} of the theory using St\"{u}ckelberg procedure\cite{stuckelberg1957}. One introduces a new field $C_{\mu}$ such that, 
\begin{eqnarray}
\label{act4}
S[B,C] = \int d v_{x} \left\{-\dfrac{1}{12}F^{\mu\nu\lambda}[B]F_{\mu\nu\lambda}[B] - \dfrac{1}{4}m^{2}(B^{\mu\nu} + \frac{1}{m}H^{\mu\nu}[C])^{2}\right\},
\end{eqnarray}
where, $H_{\mu\nu}[C]\equiv \nabla_{\mu}C_{\nu}-\nabla_{\nu}C_{\mu}$. The new action (\ref{act4}) has the following symmetries:
\begin{eqnarray}
\label{act5}
B_{\mu\nu}&\longrightarrow & B^{\xi}_{\mu\nu} = B_{\mu\nu} + \nabla_{\mu}\xi_{\nu}-\nabla_{\nu}\xi_{\mu}, \nonumber \\
C_{\mu}&\longrightarrow & C^{\xi}_{\mu} = C_{\mu} -m\xi_{\mu},
\end{eqnarray}
and 
\begin{eqnarray}
\label{act6}
C_{\mu}&\longrightarrow & C^{\Lambda}_{\mu} = C_{\mu} + \nabla_{\mu}\Lambda, \nonumber \\
B_{\mu\nu}&\longrightarrow & B^{\Lambda}_{\mu\nu} = B_{\mu\nu},
\end{eqnarray}
and reduces to the original theory (\ref{act1}) in the gauge $C_{\mu}=0$. Since our approach is gauge invariant, we can work with the full theory (\ref{act4}) instead of (\ref{act1}) and choose any suitable gauge condition. 
As is encountered later, particular choices of gauge condition lead to further softly broken symmetry in the St\"{u}ckelberg field, and successive application of St\"{u}ckelberg procedure is the key to resolving such cases.\\
The theory (\ref{act4}) is, however, still not free from degeneracies because of the extra symmetry of the gauge parameter $\xi_{\mu}$, 
\begin{equation}
\label{act7}
\xi_{\mu} \longrightarrow \xi^{\psi}_{\mu} = \xi_{\mu} + \nabla_{\mu}\psi, \quad
\Lambda \longrightarrow \Lambda + m\psi .
\end{equation} 
leaving fields $B_{\mu\nu}$, $C_{\mu}$ invariant. We give a geometric prescription for dealing with this issue and generalize the quantization of such theory in the next section.

\section{Dealing with the symmetries of gauge parameters}
In the field space, set of points $\{\varphi^{i}_{\epsilon}\}$ connected by the gauge parameter $\epsilon^{\alpha}$ form an orbit called the \textit{gauge orbit}. So, fixing a gauge is equivalent to selecting one point from each gauge orbit. This is achieved by setting up a coordinate system $(\xi^{A},\theta^{\alpha})$ such that coordinates $\theta^{\alpha}$ are along the orbit (longitudinal) while coordinates $\xi^{A}$ are transverse to the orbit. Fixing $\theta^{\alpha}$ is then equivalent to choosing one point from the orbit. Gauge invariant quantities are defined as having no $\theta^{\alpha}$ dependence. As is standard practice, we assign $\theta^{\alpha} = \chi^{\alpha}[\varphi]$, where $\chi^{\alpha}[\varphi]$ is the gauge-fixing condition for fields $\varphi^{i}$.

In the case of (\ref{act4}), however, the gauge parameters $\xi^{\mu}$ too have symmetry given by (\ref{act7}). This means, for every choice of $\theta^{\alpha}$ there exists an equivalence class (a set of points $\{\xi^{\psi}_{\mu}\}$) in the space of gauge parameter $\xi_{\mu}$. To deal with this issue, we follow the familiar procedure of `fixing the gauge' in parameter space. What this means in the geometric picture is as follows. We will work in condensed notation for this purpose. 

\begin{figure}
\centering
\includegraphics[scale=0.5]{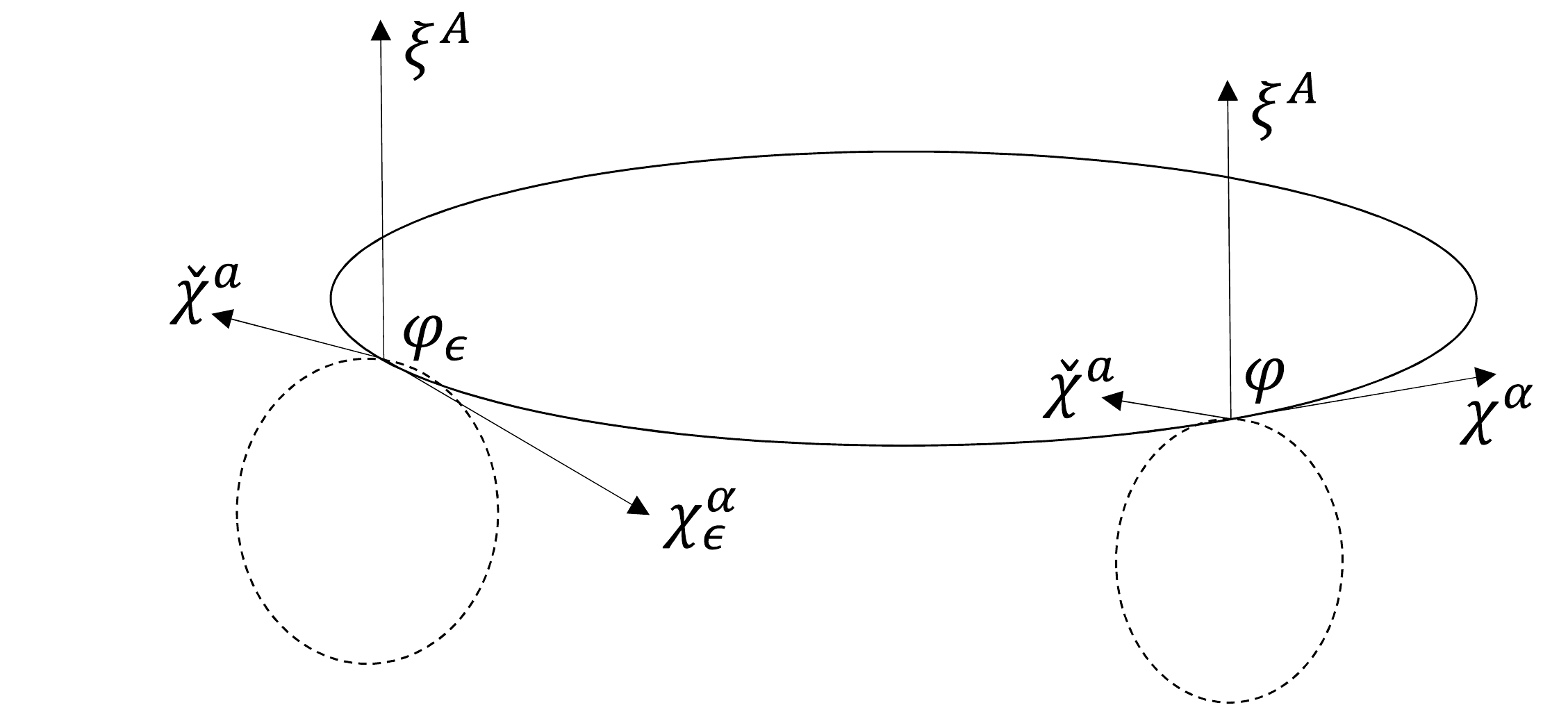}
\caption{\label{fig1} An illustration of gauge orbit in field space (solid line) and parameter space orbit (dashed line). At different points on gauge orbit ($\varphi$ and $\varphi_{\epsilon}$ respectively), a point $\check{\chi}^{a}$ on parameter-space orbit remains fixed.}
\end{figure}

Let us denote gauge parameters by $\epsilon^{\alpha}$ where $\alpha$ is a condensed index mapped to $(\mu,x)$. We are interested in the case where $\epsilon^{\alpha}$ has a gauge freedom that leaves $\varphi^{i}$ unchanged, having a general form
\begin{eqnarray}
\label{dgp1}
\delta\epsilon^{\alpha} = \check{K}^{\alpha}_{a}[\epsilon]\delta\lambda^{a},
\end{eqnarray}
where $\lambda^{a}$ parametrizes the transformations of $\epsilon^{\alpha}$. It is assumed that $\lambda^{a}$ are free of any such symmetry. 
The usual Faddeev-Popov method of gauge fixing involves introducing a factor, 
\begin{eqnarray}
\label{dgpn1}
1 = \int \left(\prod_{\alpha}d\chi^{\alpha}\right)\tilde{\delta}[\chi^{\alpha}[\varphi_{\epsilon}];0] ,
\end{eqnarray}
in the path integral, to calculate an appropriate gauge-fixed measure. However, in the present case, a technical difficulty with (\ref{dgpn1}) is that the measure spans all $\epsilon^{\alpha}$ including points on parameter-space orbit ($\epsilon^{\alpha}_{\lambda}$).
In order to deal with this issue, we start by revisiting the condition for gauge fixing in field space, that is, the requirement for $\chi^{\alpha}[\varphi]$ to be a gauge-fixing condition. $\chi^{\alpha}[\varphi]$ is required to be unique at each point $\varphi_{\epsilon}$ on a gauge orbit. This translates to requiring that the equation
\begin{equation}
\label{dgp2}
\chi^{\alpha}[\varphi_{\epsilon}] = \chi^{\alpha}[\varphi],
\end{equation} 
have a unique solution $\delta\epsilon^{\alpha} = 0$. Expanding left hand side about $\varphi$ yields, 
\begin{equation}
\label{dgpr1}
Q^{\alpha}_{\beta}[\varphi]\delta\epsilon^{\beta} = 0,
\end{equation}
where, $Q^{\alpha}_{\beta}[\varphi] = \chi^{\alpha}_{,i}[\varphi]K^{i}_{\beta}[\varphi]$. The condition for unique solution is $\det{Q^{\alpha}_{\beta}}\neq 0$. But, it turns out that for theories with degeneracy in the gauge parameter (of the form eq. (\ref{dgp1})), determinant of $Q^{\alpha}_{\beta}$ vanishes, making condition (\ref{dgp2}) insufficient for gauge fixing in this case\cite{buchbinder1992}. Indeed, for theory (\ref{act4}), it can be explicitly checked that (\ref{dgp1}) is a solution to eq. (\ref{dgpr1})\cite{buchbinder1988,buchbinder1992}. It is clear that the source of this problem is the symmetry of $\epsilon^{\alpha}$. Geometrically, it can be understood as $\epsilon^{\alpha}$ taking all possible values on the orbit spanned by $\check{\chi}^{a}$ in parameter space (dashed orbit), as illustrated in FIG. \ref{fig1}.
The key to resolving this issue is to simultaneously fix a point in the parameter-space orbit while demanding condition (\ref{dgp2}).\\
Let, $\check{\chi}^{a}[\epsilon]$ be the coordinates on parameter-space orbit. We demand that the equation,
\begin{eqnarray}
\label{dgpr2}
\chi^{\alpha}[\varphi_{\epsilon}]|_{\check{\chi}^{a}[\epsilon]=\check{\chi}^{a}} = \chi^{\alpha}[\varphi]|_{\check{\chi}^{a}[\epsilon]=\check{\chi}^{a}},
\end{eqnarray}
have unique solution $\delta\epsilon^{\alpha} = 0$. Condition (\ref{dgpr2}) ensures that the change from $\varphi$ to $\varphi_{\epsilon}$ in the field-space orbit happens with respect to a fixed point $\check{\chi}^{a}$ in parameter-space orbit. Hence, eq. (\ref{dgpr2}) is the correct requirement for gauge fixing in the case where $\delta\epsilon^{\alpha}$ has additional symmetry. \\
A more useful form of (\ref{dgpr2}) can be obtained by expressing $\chi^{\alpha}[\varphi]$ as a functional of $\varphi^{i}$, $\epsilon^{\alpha}$ and $\check{\chi}^{a}$ so that,
\begin{eqnarray}
\label{dgpr3}
\chi^{\alpha}[\varphi_{\epsilon}] = \chi^{\alpha}[\varphi,\epsilon,\check{\chi}].
\end{eqnarray}
Substituting in (\ref{dgp2}) and expanding about $\epsilon = 0$ keeping $\varphi$ and $\check{\chi}$ constant gives,
\begin{eqnarray}
\label{dgpr4}
\left(\dfrac{\delta}{\delta\epsilon^{\beta}}\chi^{\alpha}[\varphi,\epsilon ,\check{\chi}]\right)_{\epsilon = 0}\delta\epsilon^{\beta} \equiv Q'^{\alpha}_{\beta}\delta\epsilon^{\beta} = 0 .
\end{eqnarray}
Moreover, expanding both sides of eq. (\ref{dgpr3}) about $\varphi$ results in, 
\begin{eqnarray}
\label{dgpr5}
Q^{\alpha}_{\beta}\delta\epsilon^{\beta} = Q'^{\alpha}_{\beta}\delta\epsilon^{\beta} + \dfrac{\delta\chi^{\alpha}}{\delta\check{\chi}^{a}}\delta\check{\chi}^{a}.
\end{eqnarray}
Using $\delta\check{\chi}^{a} = \check{\chi}^{a}_{,\beta}\delta\epsilon^{\beta}$ in eq. (\ref{dgpr5}), we get a relation between $Q^{\alpha}_{\beta}$ and $Q'^{\alpha}_{\beta}$, 
\begin{eqnarray}
\label{dgpr6}
Q'^{\alpha}_{\beta} = Q^{\alpha}_{\beta} - \dfrac{\delta\chi^{\alpha}}{\delta\check{\chi}^{a}}\check{\chi}^{a}_{,\beta}.
\end{eqnarray}
We would like to make a couple of comments about $Q'^{\alpha}_{\beta}$. Firstly, $Q'^{\alpha}_{~\beta}$ defines the functional derivative $\chi^{\alpha}_{~,\beta}$, and it can be explicitly checked for theory (\ref{act4}) that $\det Q'^{\alpha}_{~\beta} \neq 0$. Secondly, eq. (\ref{dgpr4}) gives a general expression for calculating ghost determinant. Traditional methods for calculating such determinants involve working out integrals of Faddeev-Popov factor, and are thus specific to a particular theory as well as gauge conditions\cite{buchbinder1992, shapiro2016}. In contrast, a remarkable feature of the result (\ref{dgpr4}) is that, in addition to being independent of gauge conditions and any particular theory, it gives geometric meaning to the resolution of degeneracy in such determinants.

Next step towards writing the effective action is to find the appropriate path integral measure, including the Faddeev-Poppov factor (\ref{dgpn1}). We follow the standard procedure of ref. \cite{parker2009}. Full field space volume element can be written in terms of ($\xi^{A}, \epsilon^{\alpha}$) using the length element,
\begin{eqnarray}
\label{dgpr7}
ds^2 = g_{ij}d\varphi^{i}d\varphi^{j} = h_{AB}d\xi^{A}d\xi^{B} + \gamma_{\alpha\beta}d\epsilon^{\alpha}d\epsilon^{\beta}.
\end{eqnarray}
Also, in the parameter space,
\begin{eqnarray}
\label{dgpr8}
\gamma_{\alpha\beta}d\epsilon^{\alpha}d\epsilon^{\beta} = \gamma^{\perp}_{\alpha\beta}d\epsilon_{\perp}^{\alpha}d\epsilon_{\perp}^{\beta} + \check{\gamma}_{ab}d\lambda^{a}d\lambda^{b}.
\end{eqnarray}
Here, $\xi^{A}$ and $\epsilon^{\alpha}$ orthogonal; and so are $d\epsilon_{\perp}^{\alpha}$ and $d\lambda^{a}$ by definition. $h_{AB}$, $\gamma_{\alpha\beta}$, $\gamma^{\perp}_{\alpha\beta}$ and $\check{\gamma}_{ab}$ are the corresponding induced metric on respective coordinates, whereas $g_{ij}$ is the metric on the full field space. The field-space volume element is given by,
\begin{equation}
\label{dgpr9}
\prod_{i}d\varphi^{i}(\det g_{ij})^{1/2} = \prod_{A}d\xi^{A}\prod_{\alpha}d\epsilon^{\alpha} (\det h_{AB}\det \gamma_{\alpha\beta})^{1/2}.
\end{equation}
The technical difficulty we mentioned earlier is due to the non-trivial structure of parameter space, as shown by (\ref{dgpr8}). For a trivial parameter space, where there are no symmetries, one can show that none of the factors in (\ref{dgpr9}) depend on $\epsilon^{\alpha}$\cite{parker2009} and hence $\prod_{\alpha} d\epsilon^{\alpha}$ integrates out. But, for (\ref{dgpr8}) the gauge group volume element is not trivial. So, to determine which factor integrates out of (\ref{dgpr9}), we must calculate the gauge group volume element first. 
We start with parameter space volume element, 
\begin{equation}
\label{dgprr1}
\prod_{\alpha}d\epsilon^{\alpha}(\det \gamma_{\alpha\beta})^{1/2} = \prod_{\alpha}d\epsilon_{\perp}^{\alpha}\prod_{a}d\lambda^{a}(\det \gamma^{\perp}_{\alpha\beta}\det \check{\gamma}_{ab})^{1/2}.
\end{equation}
Following the arguments of \cite{parker2009}, it can be shown that determinants appearing in the right hand side of eq. (\ref{dgprr1}) are independent of $\lambda^{a}$, thereby making the relevant parameter space measure,
\begin{eqnarray}
\label{dgpr10}
\prod_{\alpha}d\epsilon_{\perp}^{\alpha}(\det \gamma^{\perp}_{\alpha\beta})^{1/2}(\det \check{\gamma}_{ab})^{1/2}.
\end{eqnarray}
Now, we introduce the Faddeev-Popov factor,
\begin{eqnarray}
\label{dgp7}
1 = \int \left(\prod_{a}d\check{\chi}^{a}\right)\tilde{\delta}[\check{\chi}^{a}[\epsilon_{\lambda}];0] ,
\end{eqnarray}
so that, the parameter space measure becomes, 
\begin{equation}
\label{dgprr2}
[d\epsilon] = \prod_{\alpha}d\epsilon_{\perp}^{\alpha}\prod_{a}d\check{\chi}^{a}(\det \gamma^{\perp}_{\alpha\beta})^{1/2}(\det \check{\gamma}_{ab})^{1/2}\tilde{\delta}[\check{\chi}^{a}[\epsilon_{\lambda}];0].
\end{equation}
To calculate the Jacobian for transformation from ($\epsilon_{\perp}^{\alpha}, \check{\chi}^{a}$) to $\epsilon^{\alpha}$, we use,
\begin{eqnarray}
\label{dgprr3}
d\check{\chi}^{a} = \check{\chi}^{a}_{,\alpha}d\epsilon^{\alpha} = \check{\chi}^{a}_{,\alpha}d\epsilon_{\perp}^{\alpha} + \check{Q}^{a}_{b}d\lambda^{b},
\end{eqnarray}
where, $\check{Q}^{a}_{b} = \check{\chi}^{a}_{,\alpha}\check{K}^{\alpha}_{b}$. Solving for $d\lambda^{a}$ gives,
\begin{eqnarray}
\label{dgprr4}
d\lambda^{a} = \left(\check{Q}^{-1}\right)^{a}_{b}(d\check{\chi}^{b} - \check{\chi}^{a}_{,\alpha}d\epsilon_{\perp}^{\alpha}).
\end{eqnarray}
Substituting (\ref{dgprr4}) in (\ref{dgpr8}), length element in parameter space is obtained as,
\begin{eqnarray}
\label{dgprr5}
\gamma_{\alpha\beta}d\epsilon^{\alpha}d\epsilon^{\beta} = \gamma^{\perp}_{\alpha\beta}d\epsilon_{\perp}^{\alpha}d\epsilon_{\perp}^{\beta} + \check{\gamma}_{ab}\left(\check{Q}^{-1}\right)^{a}_{c}\left(\check{Q}^{-1}\right)^{b}_{d}  (d\check{\chi}^{c} - \check{\chi}^{c}_{,\alpha}d\epsilon_{\perp}^{\alpha})(d\check{\chi}^{d} - \check{\chi}^{d}_{,\alpha}d\epsilon_{\perp}^{\alpha}).
\end{eqnarray}
The metric in (\ref{dgprr5}) has the form of that in Kaluza-Klein theory\cite{parker2009}, so it is straightforward to read off the relation between volume elements,
\begin{eqnarray}
\label{dgprr6}
(\det\gamma_{\alpha\beta})^{1/2}\prod_{\alpha}d\epsilon^{\alpha} = \left(\prod_{\alpha}d\epsilon_{\perp}^{\alpha}\prod_{a}d\check{\chi}^{a}\right) (\det \gamma^{\perp}_{\alpha\beta})^{1/2}(\det \check{\gamma}_{ab})^{1/2}(\det\check{Q}^{a}_{b})^{-1}.
\end{eqnarray}
Hence, the gauge group volume is found to be,
\begin{eqnarray}
\label{dgprr7}
[d\epsilon] = \prod_{\alpha}d\epsilon^{\alpha} (\det\gamma_{\alpha\beta})^{1/2} (\det\check{Q}^{a}_{b}[\epsilon])\tilde{\delta}[\check{\chi}^{a}[\epsilon_{\lambda}];0].
\end{eqnarray}
It is safe to say now, that the factor $\prod_{\alpha}d\epsilon^{\alpha}(\det\check{Q}^{a}_{b}[\epsilon])\tilde{\delta}[\check{\chi}^{a}[\epsilon_{\lambda}];0]$ will integrate out of the field space measure. Therefore, we express the field-space measure as,
\begin{eqnarray}
\label{dgprr8}
\prod_{A}d\xi^{A} (\det h_{AB})^{1/2}(\det \gamma_{\alpha\beta})^{1/2} (\det\check{Q}^{a}_{b})^{-1}.
\end{eqnarray}
At this point, it is standard to introduce the Faddeev-Popov factor given by (\ref{dgpn1}), at $\check{\chi}^{a}=0$, and calculate path integral measure by working out Jacobian of coordinate transformations to full field space coordinates $\varphi^{i}$\cite{parker2009}. We need to calculate the field-space measure,
\begin{eqnarray}
\label{dgp9}
[d\varphi] = \int \left(\prod_{A}d\xi^{A}\prod_{\alpha}d\chi^{\alpha}\right)\tilde{\delta}[\chi^{\alpha}[\varphi_{\epsilon}]|_{\check{\chi}^{a}=0};0] (\det h_{AB} \det \gamma_{\alpha\beta})^{1/2}(\det\check{Q}^{a}_{b})^{-1}.
\end{eqnarray}
Calculation of Jacobian proceeds in a similar way as in eqs. (\ref{dgprr2}) to (\ref{dgprr6}). Using the result (\ref{dgpr6}), 
\begin{eqnarray}
\label{dgpn2}
d\chi^{\alpha} = \chi^{\alpha}_{,A}d\xi^{A} + Q'^{\alpha}_{\beta}d\epsilon^{\beta},
\end{eqnarray}
and solving for $d\epsilon^{\alpha}$ followed by substituting in (\ref{dgpr7}), one obtains the relation, 
\begin{eqnarray}
\label{dgpn3}
\prod_{i}d\varphi^{i}(\det g_{ij})^{1/2}(\det Q'^{\alpha}_{\beta}) = \prod_{A}d\xi^{A}\prod_{\alpha}d\chi^{\alpha} \times (\det h_{AB})^{1/2}(\det \gamma_{\alpha\beta})^{1/2}.
\end{eqnarray}
Finally, substituting (\ref{dgpn3}) in (\ref{dgp9}), we obtain the path integral measure for fields $\varphi^{i}$,
\begin{eqnarray}
\label{dgpn5}
[d\varphi] = \int \left(\prod_{i}d\varphi^{i}\right)(\det g_{ij})^{1/2}(\det Q'^{\alpha}_{\beta}) (\det\check{Q}^{a}_{b})^{-1}\tilde{\delta}[\chi^{\alpha}[\varphi_{\epsilon}]|_{\check{\chi}^{a}=0};0].
\end{eqnarray}

\section{Effective Action for rank-2 antisymmetric field}
Due to the properties of covariant effective action approach, the above result (\ref{dgpn5}) is useful to study quantum properties of theories with degeneracy in gauge parameters in curved spacetime, including studies of quantum gravitational corrections as will be seen in later chapters. For the purpose of present chapter, we now use this result to derive the one-loop effective action for a massive antisymmetric field in curved spacetime as described by Eq. (\ref{act1}) (and equivalently, Eq. (\ref{act4})), previously obtained in Refs. \cite{buchbinder2008,shapiro2016}. 

The first step is to choose the field space metric. We write the length element in the field space, 
\begin{eqnarray}
\label{eac1}
ds^{2} = \int dv_{x} \left\{g^{\mu\rho}(x)g^{\sigma\nu}(x)dB_{\mu\nu}(x)dB_{\rho\sigma}(x) + g^{\mu\nu}(x)dC_{\mu}(x)dC_{\nu}(x)\right\}.
\end{eqnarray}
From (\ref{eac1}), we read off the field space metric components $G_{I(x)J(x^{\prime})}$, taking into account the antisymmetrization of metric component for $B_{\mu\nu}$ fields, 
\begin{eqnarray}
\label{eac2}
G_{B_{\mu\nu}(x)B_{\rho\sigma}(x')} &=& \sqrt{-g(x)} g^{\mu[\rho}(x)g^{\sigma]\nu}(x')\tilde{\delta}(x,x'), \nonumber \\
G_{C_{\mu}(x)C_{\nu}(x')} &=& \sqrt{-g(x)} g^{\mu\nu}(x)\tilde{\delta}(x,x'),
\end{eqnarray}
where $g^{\mu\nu}$ is the spacetime metric and $\tilde{\delta}(x,x')$ is the invariant delta function.
We can calculate the inverse field space metric using the identity 
\begin{equation}
\label{eac3}
\int d^{4}x' G^{I(x)J(x^{\prime})}G_{J(x^{\prime})K(y)} = \delta^{I}_{K}\delta(x,y),
\end{equation}
which gives,
\begin{eqnarray}
\label{eac4}
G^{B_{\mu\nu}(x)B_{\rho\sigma}(x')} &=& \dfrac{1}{-g(x)}g_{\mu[\rho}(x)g_{\sigma]\nu}(x')\delta(x,x'), \nonumber \\
G^{C_{\mu}(x)C_{\nu}(x')} &=& \dfrac{1}{-g(x)} g_{\mu\nu}(x)\delta(x,x').
\end{eqnarray}
Since, in (\ref{eac2}) and (\ref{eac4}), there is no dependence on the fields, the field space Christoffel connections will vanish, 
\begin{equation}
\label{eac5}
\Gamma^{i}_{jk} = 0 \quad \forall \quad  i,j,k \in (B_{\mu\nu}, C_{\mu}).
\end{equation}
Note that the Christoffel connections will be nonzero if we choose to quantize gravity as well. For the present case, however, we treat gravity as a classical field so that field space only has components of  $B$ and $C$ fields and the connections vanish. Another case where Christoffel connections would be nonzero is if we choose a different parametrization for the fields. This corresponds to choosing another coordinate system in field space. \\
Next, we find the gauge generators $K^{i}_{\alpha}$ from the relation
\begin{eqnarray}
\label{eac6}
\delta\varphi^{I} = \int d^{4}x' K^{\varphi^{I}(x)}_{\bar{\alpha}(x')}\delta\epsilon^{\bar{\alpha}(x')},
\end{eqnarray}
where $\varphi^{I(x)} = \{B_{\mu\nu}(x), C_{\mu}(x)\}$ and $\delta\epsilon^{\bar{\alpha}(x')} = \{\xi_{\mu}(x), \Lambda(x)\}$. The generators can be read off from eqs. (\ref{act5}) and (\ref{act6}), 
\begin{eqnarray}
\label{eac7}
K^{B_{\mu\nu}(x)}_{\xi_{\rho}(x')} &=& (\nabla_{\mu}\delta^{\rho}_{\nu} - \nabla_{\nu}\delta^{\rho}_{\mu})\tilde{\delta}(x,x'), \nonumber \\
K^{C_{\mu}(x)}_{\xi_{\rho}(x')} &=& -m\delta^{\rho}_{\mu}\tilde{\delta}(x,x'), \\
K^{C_{\mu}(x)}_{\Lambda(x')} &=& \nabla_{\mu}\tilde{\delta}(x,x'). \nonumber
\end{eqnarray}
For future use, we also calculate $K_{\alpha i}$ below, using the condensed notation identity $K_{\alpha i} = g_{ij}K^{j}_{\alpha} = \int d^{4}y G_{I(x)J(y)}K^{J(y)}_{\bar{\alpha}(x')}$:
\begin{eqnarray}
\label{eac7b}
K_{\xi_{\mu}(x')B_{\nu\rho}(x)} &=& \sqrt{-g(x)}(\nabla^{\nu}\delta^{\rho\mu}-\nabla^{\rho}\delta^{\nu\mu})\tilde{\delta}(x,x'), \nonumber \\
K_{\Lambda(x')C_{\mu}(x)} &=& \sqrt{-g(x)} \nabla^{\mu} \tilde{\delta}(x,x'), \\
K_{\xi_{\rho}(x')C_{\mu}(x)} &=& -m \sqrt{-g(x)} \delta^{\mu\rho} \tilde{\delta}(x,x'). \nonumber
\end{eqnarray}
Keeping in mind the geometrical interpretation of gauge fixing, we understand that for effective action only a sub-space of field space constrained by the gauge condition will be integrated over, and hence the metric with which covariant derivatives and connections are calculated in field space is not the full field space metric $G_{I(x)J(x')}$, but that which describes the space of transverse fields $\xi^{A}$. 
The induced metric is found by first calculating the metric along the gauge orbit (corresponding to points $\theta^{\alpha}$), given by $\gamma_{\alpha\beta} = K^{i}_{\alpha}g_{ij}K^{j}_{\beta}$ in the condensed notation\cite{parker2009}. 
Rewriting in terms of spacetime indices, 
\begin{eqnarray}
\label{eac8}
\gamma_{\bar{\alpha}(x)\bar{\beta}(x')} = \int d^{4}y d^{4}y' K^{I(y)}_{\bar{\alpha}(x)} G_{I(y)J(y')}K^{J(y')}_{\bar{\beta}(x')} .
\end{eqnarray}
Substituting (\ref{eac2}) and (\ref{eac7}) in (\ref{eac8}), we get
\begin{eqnarray}
\label{eac9}
\gamma_{\xi_{\mu}(x)\xi_{\nu}(x')} &=& \sqrt{-g(x)}\Box^{\mu\nu}_{\xi}\tilde{\delta}(x,x'), \nonumber \\
\gamma_{\Lambda(x)\Lambda(x')} &=& \sqrt{-g(x)} \Box_{x}\tilde{\delta}(x,x'), \\
\gamma_{\xi_{\mu}(x)\Lambda(x')} &=& -m\sqrt{-g(x)}\nabla^{\mu}\tilde{\delta}(x,x') =   \gamma_{\Lambda(x)\xi_{\mu}(x')}, \nonumber
\end{eqnarray}
where $\Box_x$ is the de'Alembertian operator with respect to coordinates $x$, and, 
\begin{equation}
\label{eac12}
\Box^{\mu\nu}_{\xi} \equiv (\nabla_{\rho}\delta^{\mu}_{\sigma} - \nabla_{\sigma}\delta^{\mu}_{\rho})(\nabla^{\rho}\delta^{\nu\sigma} - \nabla^{\sigma}\delta^{\nu\rho}) + m^{2}g^{\mu\nu}.
\end{equation} 
One can find the inverse of $\gamma_{\alpha\beta}$ using the property
\begin{eqnarray}
\label{eac10}
\int d^{4}y \gamma_{\bar{\alpha}(x)\bar{\beta}(y)}\gamma^{\bar{\beta}(y)\bar{\gamma}(x')} = \delta_{\bar{\alpha}}^{\bar{\gamma}}\tilde{\delta}(x,x').
\end{eqnarray}
Now it is possible to find the connection $\tilde{\Gamma}^{m}_{ij}$ on the restricted space over which we quantize the fields. Since $\Gamma^{m}_{ij} = 0$, it is given by\cite{parker2009}
\begin{eqnarray}
\label{eac13}
\tilde{\Gamma}^{m}_{ij} = \dfrac{1}{2}\gamma^{\alpha\varepsilon}\gamma^{\beta\sigma}K_{\alpha i}K_{\beta j} (K^{n}_{\varepsilon}K^{m}_{\sigma ;n} + K^{n}_{\sigma}K^{m}_{\varepsilon ;n}) - \gamma^{\alpha\beta}(K_{\alpha i}K^{m}_{\beta ;j} + K_{\alpha j}K^{m}_{\beta ;i}).
\end{eqnarray}
Here, covariant derivative is with respect to the full field space metric, $g_{ij}$ and thus contains connection $\Gamma^{m}_{ij}$. Since, $\Gamma^{m}_{ij}=0$ covariant derivatives in (\ref{eac13}) can be replaced with ordinary derivatives.  
However, for the present example, since $K^{i}_{\alpha}$ does not have any dependence on the fields $B_{\mu\nu}$ and $C_{\mu}$, it turns out the induced connection vanishes for all combinations of $B_{\mu\nu}$ and $C_{\mu}$:
\begin{equation}
\label{eac14}
\tilde{\Gamma}^{m}_{ij} = 0.
\end{equation}
This simplifies the present problem a lot. In order to find the 1-loop corrections, covariant derivatives of the action become merely simple functional derivatives.\\
The covariant field-space intervals also reduce to flat geodetic intervals due to vanishing connections, 
\begin{equation}
\label{eac15}
\sigma^{i} = -\eta^{i} = \varphi^{i}_{*} - \varphi^{i}.
\end{equation}

Next, we use the result (\ref{dgpn5}) to write the gauge-fixed measure. For convenience, we choose for $B_{\mu\nu}$ the gauge condition,
\begin{eqnarray}
\label{eac20}
\chi_{\xi_{\nu}} = \nabla^{\mu}B_{\mu\nu} + m C_{\nu},
\end{eqnarray}
and for gauge parameters,
\begin{eqnarray}
\label{eacf5}
\check{\chi}_{\psi} = \nabla^{\mu}\xi_{\mu} - m\Lambda
\end{eqnarray}
This results in the action (\ref{act4}) being appended by a gauge fixing term,
\begin{eqnarray}
\label{eacf1}
-\frac{1}{2}(\chi_{\xi_{\nu}})^{2} = -\frac{1}{2}\nabla^{\mu}B_{\mu\nu}\nabla_{\rho}B^{\rho\nu}  - \frac{1}{2}m^{2}C_{\nu}C^{\nu} - \frac{1}{2}m C_{\nu}\nabla_{\mu}B^{\mu\nu}.
\end{eqnarray}
The second term on the right hand side of eq. (\ref{eacf1}) induces soft breaking of gauge symmetry in field $C_{\nu}$. As pointed out earlier, one has to apply the St\"{u}ckelberg procedure of sec. \ref{sec2} to restore gauge symmetry. As a result, a second St\"{u}ckelberg field $\Phi$ is introduced so that,
\begin{eqnarray}
\label{eacf2}
- \frac{1}{2}m^{2}C_{\nu}C^{\nu} \ \longrightarrow \ - \frac{1}{2}m^{2}\left(C^{\nu} + \frac{1}{m}\nabla^{\nu}\Phi\right)^{2}.
\end{eqnarray}
Hence, an appropriate gauge condition for $C_{\mu}$ is,
\begin{eqnarray}
\label{eacf3}
\chi_{\Lambda} = \nabla^{\mu}C_{\mu} + m\Phi .
\end{eqnarray}
In order to calculate the components of $Q'^{\alpha}_{\beta}$, we use eq. (\ref{dgpr3}) to write, 
\begin{eqnarray}
\label{eacf4}
\chi_{\xi_{\nu}} = \nabla^{\mu}B_{\mu\nu} + m C_{\nu} + \Box_{x}\xi_{\nu} - [\nabla^{\mu},\nabla_{\nu}]\xi_{\mu} - m^{2}\xi_{\nu} - \nabla_{\nu}\check{\chi}_{\psi}
\end{eqnarray}
From the definition (\ref{dgpr4}), we find,
\begin{eqnarray}
\label{eacf6}
Q'^{\xi_{\mu}(x)}_{\xi_{\nu}(x')} &=& (\Box_{x}\delta_{\nu\mu} - [\nabla^{\alpha},\nabla_{\nu}]\delta_{\alpha\mu} - m^{2}\delta_{\nu\mu})\tilde{\delta}(x,x') \nonumber \\ 
&\equiv & (\Box_{1} - m^{2}\delta_{\nu\mu})\tilde{\delta}(x,x').
\end{eqnarray}
Calculation of $Q'^{\Lambda(x)}_{\Lambda(x')}$ is straightforward, and yields,
\begin{eqnarray}
\label{eac23}
Q'^{\Lambda(x)}_{\Lambda(x')} &=& (\Box_{x}-m^2)\tilde{\delta}(x,x').  
\end{eqnarray}
A similar calculation results in $\check{Q}^{\psi(x)}_{\psi(x')} = (\Box_{x} - m^{2})\tilde{\delta}(x,x')$.
With all the ingredients in place, we obtain the effective action by substituting eqs. (\ref{eac20}), (\ref{eacf3}), and (\ref{eac23}) in (\ref{dgpn5}) and working in spacetime (uncondensed) coordinates, 
\begin{eqnarray}
\label{eac25}
\exp(i\Gamma[\bar{B},\bar{C}]) = \int\prod_{\mu}dC_{\mu}\prod_{\rho\sigma}dB_{\rho\sigma}\prod_{x}d\Phi
\delta[\chi_{\Lambda(x)};0]\delta[\chi_{\xi_{\mu}(x)};0]\det(\Box_{1} - m^{2})\times \nonumber \\  
 \exp\left\{i \int d v_{x} \left(-\dfrac{1}{12}F^{\mu\nu\lambda}[B]F_{\mu\nu\lambda}[B] 
-\dfrac{1}{4}m^{2}(B^{\mu\nu} + \frac{1}{m}H^{\mu\nu}[C])^{2}\right) \right. \nonumber \\ \left. + (\bar{B}_{\mu\nu}-B_{\mu\nu})\dfrac{\delta}{\delta \bar{B}_{\mu\nu}}\Gamma[\bar{B},\bar{C}] + (\bar{C}_{\mu}-C_{\mu})\dfrac{\delta}{\delta \bar{C}_{\mu}}\Gamma[\bar{B},\bar{C}] \right\}.
\end{eqnarray}
We have ignored the field-space metric determinant here, because it does not affect the result apart from raising and lowering indices inside determinants. As usual, the Dirac $\delta$-distributions in (\ref{eac25}) give rise to the gauge-fixed action, 
\begin{eqnarray}
S_{GF} =  \int d v_{x} \Bigg(-\dfrac{1}{12}F^{\mu\nu\lambda}[B]F_{\mu\nu\lambda}[B] - \dfrac{1}{4}m^{2}(B^{\mu\nu} + \frac{1}{m}H^{\mu\nu}[C])^{2} \nonumber \\ -\frac{1}{2}(\chi_{\xi_{\mu}(x)})^{2} - \frac{1}{2}(\chi_{\Lambda(x)})^{2}\Bigg),
\end{eqnarray}
which can be cast into the form,
\begin{eqnarray}
\label{eacr1}
S_{GF} =  \int d v_{x} \Bigg(\frac{1}{4}B^{\mu\nu}\Box_{2}B_{\mu\nu} - \frac{1}{4}m^{2}B^{\mu\nu}B_{\mu\nu} + \frac{1}{2}C^{\mu}\Box_{1}C_{\mu} - \frac{1}{2}m^{2}C^{\mu}C_{\mu} \nonumber \\ + \frac{1}{2}\Phi(\Box_{x}-m^2)\Phi\Bigg),
\end{eqnarray}
where, 
\begin{eqnarray}
\Box_{2}B_{\mu\nu} = \Box_{x}B_{\mu\nu} - [\nabla^{\rho},\nabla_{\nu}]B_{\mu\rho} - [\nabla^{\rho},\nabla_{\mu}]B_{\rho\nu} .
\end{eqnarray}
Substituting eq. (\ref{eacr1}) in (\ref{eac25}), one obtains the effective action as,
\begin{eqnarray}
\label{eac28}
\Gamma = S[\bar{B},\bar{C}] + \hbar\dfrac{i}{2}\left(\ln\det(\Box_{2} - m^{2}) - \ln\det(\Box_{1} - m^{2}) + \ln\det(\Box_{x}-m^2)\right).
\end{eqnarray}
For the present case of free theory (\ref{act4}), there are no contributions at higher loop orders.\\ The result (\ref{eac28}) is in line with that obtained earlier in \cite{buchbinder2008}, and later confirmed in \cite{shapiro2016}.

\section{Summary}
We quantized a massive rank-2 antisymmetric field by finding a general path integral measure for theories with degeneracy in gauge parameter, using DeWitt-Vilkovisky's covariant effective action approach. In the process, we arrived at a simple resolution to the problem of dealing with additional symmetries of gauge parameter, through a geometric understanding of gauge-fixing. 
In particular, the ghost determinant calculation receives a simple geometric meaning, and generalizes traditional methods\cite{buchbinder1992,shapiro2016} which are specific to a particular theory and gauge condition.

For the simple case of free theory (\ref{act4}), where gravity is classical, we find that the covariant effective action is identical to that obtained in earlier works\cite{buchbinder2008,shapiro2016}, up to a difference in sign of $m^{2}$ due to corresponding sign of potential. More applications of this formalism lie in the study of gravitational corrections to models of antisymmetric tensor fields, and can in principle be extended to n-forms and other fields with similar characteristics. For instance, a rather simple generalization of the result (\ref{eac28}) is for the nonminimal model considered in \cite{altschul2010} with a coupling term $\zeta R B_{\mu\nu}B^{\mu\nu}$, which can be absorbed in the mass term with $m\longrightarrow m - \zeta R$. Similarly, the next chapter deals with one of the applications of these results to the case of antisymmetric tensor with spontaneous Lorentz violation. 


\chapter{Quantum Aspects of Antisymmetric Tensor Field with Spontaneous Lorentz Violation}

Here, we study the quantum aspects of a simple model of antisymmetric tensor field with spontaneous Lorentz violation in curved spacetime. We begin with some background and motivations in the next section. In Sec. \ref{sec3a}, we briefly review spontaneous Lorentz violation in antisymmetric tensor and introduce the classical action considered in this work. The notations used here are largely inspired by Ref. \cite{altschul2010}. We discuss the covariant effective action technique and its application to derive 1-loop corrections in Sec. \ref{sec3b}. We also calculate the various propagators required to solve the 1-loop integrals. In Sec. \ref{sec3c}, we consider the classically equivalent vector theory and calculate 1-loop corrections to compare with the results of Sec. \ref{sec3b}, to check the quantum equivalence. The contents of this chapter and appendix \ref{AppendixB} originally appeared in Refs. \cite{aashish2018b,aashish2019b}.

\section{\label{sec3intro}Introduction}

The quest for quantizing gravity is ultimately related to understanding physics at the Planck scale, candidates for which include string theory and loop quantum gravity. A difficulty that the development of such theories faces, is our inability to probe high energy scales, owing to the limitations of current particle physics experiments \cite{zimmermann2018}. This has led to significant efforts towards finding low energy signatures using effective field theory tools that could be relevant in current and near future experiments in both particle physics and early universe cosmology. Phenomenologically, this amounts to detecting Planck supressed variations to standard model and general relativity while maintaining observer independence, termed as standard model extension (SME) \cite{kostelecky1995,colladay1997,colladay1998,kostelecky2004,bluhm2006}. 

There is substantial evidence of SME effects from string theory and quantum gravity, according to which certain mechanisms could lead to violation of Lorentz symmetry \cite{kostelecky1989b,kostelecky1991a,kostelecky1991b,kostelecky1996,kostelecky2001,gambini1999,alfaro2002, sudarsky2002,sudarsky2003,myers2003}, which is a fundamental symmetry in general relativity that relates all physical local Lorentz frames. In principle, Lorentz violation can be introduced in a theory either explicitly, in which case the Lagrange density is not Lorentz invariant, or spontaneously, so that the Lagrange density is Lorentz invariant but the physics can still display Lorentz violation \cite{kostelecky1991a,kostelecky1989b}. However, theories with explicit Lorentz violation have been found to be problematic due to their incompatibility with Bianchi identities in Riemann geometry \cite{kostelecky2004}, and are therefore not favourable for studies involving gravity.

Another consequence of string theory, at low energies, is the appearance of antisymmetric tensor field along with a symmetric tensor (metric) and a dilaton (scalar field) as a result of compactification of higher dimensions \cite{rohm1986,ghezelbash2009}. Until recently, antisymmetric tensor had not received serious consideration in studies of early universe cosmology, in particular inflation, due to some generic instability issues \cite{koivisto2009a,aashish2018c,prokopec2006}, but some recent studies have shown that presence of antisymmetric tensor field is likely to play a role during inflation era \cite{aashish2019a,elizalde2018}. Hence, as a natural extension, an interesting exercise is to consider Lorentz violation in conjunction with antisymmetric tensor (see, for example Ref. \cite{petrov2019}).

Altschul \etal \ in Ref. \cite{altschul2010} explored in detail spontaneous Lorentz violation with antisymmetric tensor fields, and found the presence of distinctive physical features with phenomenological implications for tests of Lorentz violation, even with relatively simple antisymmetric field models with a gauge invariant kinetic term. 

Our interest in this chapter is to take first steps to extend the classical analysis in Ref. \cite{altschul2010} to quantum regime. We focus on the formal aspects of quantization of antisymmetric tensor field with spontaneous Lorentz violation, and primarily restrict ourselves to dealing with two issues. First, we set up the framework to evaluate the one-loop effective action using covariant effective action approach \cite{dewitt1964,dewitt1967a,dewitt1967b,dewitt1967c,vilkovisky1984a,vilkovisky1984b,parker2009}. For simplicity, we consider an action with only quadratic order terms, but in a \textit{nearly flat} spacetime (Minkowski metric $\eta_{\mu\nu}$ plus a classical perturbation $\kappa h_{\mu\nu}$). This yields one-loop corrections at $O(\kappa \hbar)$, involving terms up to first order in $h_{\mu\nu}$. Second, we check the quantum equivalence of the quadratic action considered in the first part with a classically equivalent vector theory, at 1-loop level. The issue of quantum equivalence in curved spacetime is interesting because a free massive antisymmetric tensor theory (no Lorentz violation) is known to be equivalent to a massive vector theory at classical and quantum level due to topological properties of zeta functions \cite{buchbinder2008} but, such properties do not hold when Lorentz symmetry is spontaneously broken \cite{aashish2018b}. In fact, it was demonstrated by Seifert in Ref. \cite{seifert2010a} that interaction of vector and tensor theories with gravity are different when topologically nontrivial monopole-like solutions of the spontaneous symmetry breaking equations exist. The method presented here is quite general in terms of its applicability to models with higher order terms in fields.

\section{\label{sec3a}Spontaneous Lorentz Violation and classical action}
Spontaneous symmetry breaking occurs when the equations of motion obey a symmetry but the solutions do not, and is effected via fixing a preferred value of vacuum (ground state) solutions. In general relativity, physically equivalent coordinate (or observer) frames are related via general coordinate transformations and local Lorentz transformations. Hence, the condition for a physical theory is that observer Lorentz symmetry must hold true. However, by breaking the Lorentz symmetry \textit{spontaneously}, we choose to break the particle Lorentz symmetry whilst keeping the observer Lorentz symmetry intact. That is, in a given observer frame, we fix the vacuum expectation value (vev) of a tensor or vector field leading to spontaneous breaking of Lorentz symmetry, since all couplings with vev have preferred directions in spacetime \cite{bluhm2005,bluhm2006}. 

Spontaneous Lorentz violation in tensor field Lagrangians can be introduced through a potential term that drives a nonzero vacuum value of tensor field. For an antisymmetric 2-tensor field $B_{\mu\nu}$, we assume,
\begin{eqnarray}
\label{slva0}
\langle B_{\mu\nu}\rangle = b_{\mu\nu}.
\end{eqnarray}
It is possible to attain a special observer frame in a local Lorentz frame in Riemann spacetime or everywhere in Minkowski spacetime, in which $b_{\mu\nu}$ takes a simple block-diagonal form \cite{altschul2010},
\begin{eqnarray}
\label{slva1}
b_{\mu\nu} = 
\left(\begin{matrix}
0 & -a & 0 & 0\\
a & 0 & 0 & 0\\
0 & 0 & 0 & b\\
0 & 0 & -b & 0
\end{matrix}\right),
\end{eqnarray}
provided at least one of the quantities $X_{1}=-2(a^2-b^2)$ and $X_{2}=4ab$ is nonzero, where $a$ and $b$ are real numbers. Moreover, the analysis of monopole solutions of antisymmetric tensor in Ref. \cite{seifert2010b} showed that for a spherically symmetric nontrivial solution of the equation of motion of $B_{\mu\nu}$ that asymptotically approaches vev, the potential of the form considered below (Eq. (\ref{slva2})) requires putting $a=0$. As will be seen later on, this choice of $a$ also ensures positivity of certain determinants appearing in loop integral calculations. We thus assume $a=0$ in the present analysis, although most of the calculations presented here are independent of the structure of $b_{\mu\nu}$. For later convenience, we also choose $b_{\mu\nu}b^{\mu\nu}=1$. 

We consider a simple model of a rank-2 antisymmetric tensor field, $B_{\mu\nu}$, with a spontaneous Lorentz violation inducing potential \cite{altschul2010},  
\begin{eqnarray}
\label{slva2}
V(B) = \frac{1}{16}\alpha^2 \Big(B_{\mu\nu}B^{\mu\nu} - b_{\mu\nu}b^{\mu\nu}\Big)^{2}.
\end{eqnarray}
Again, for the purpose of present analysis, we would like to consider only quadratic order terms in $B_{\mu\nu}$. To this end, we consider small fluctuations of $B_{\mu\nu}$ about a background value $b_{\mu\nu}$ \cite{altschul2010}, 
\begin{eqnarray}
\label{slva3}
B_{\mu\nu} = b_{\mu\nu} + \tilde{B}_{\mu\nu}.
\end{eqnarray} 
and neglect quartic and cubic terms in fluctuations $\tilde{B}_{\mu\nu}$ assuming $|| \tilde{B}_{\mu\nu}|| \ll ||b_{\mu\nu}||$. The resulting potential is,
\begin{eqnarray}
\label{slva4}
V(B) \approx \dfrac{1}{4}\alpha^2\Big(b_{\mu\nu}\tilde{B}^{\mu\nu}\Big)^2.
\end{eqnarray}
Although it may seem at this point that a quadratic Lagrangian might not lead to any significant physical result upon quantization, and it is actually true in case of a flat spacetime, nontrivial physical contributions appear in the 1-loop effective action in curved spacetime as demonstrated in the next section. For notational convenience, we do not explicitly write the \textit{tilde} symbol for field fluctuations, and assume its use throughout. We thus work with the Lagrangian,
\begin{eqnarray}
\label{slva5}
\mathcal{L} = -\frac{1}{12}H_{\mu\nu\lambda}H^{\mu\nu\lambda} - \dfrac{1}{4}\alpha^{2} \Big(b_{\mu\nu}{B}^{\mu\nu}\Big)^{2}.
\end{eqnarray}
The first term in Eq. (\ref{slva5}) is the gauge invariant kinetic term, 
\begin{equation}
\label{slva6}
H_{\mu\nu\lambda} \equiv \nabla_{\mu}B_{\nu\lambda} + \nabla_{\lambda}B_{\mu\nu} + \nabla_{\nu}B_{\lambda\mu},
\end{equation}
obeying the symmetry: $B^{\mu\nu} \longrightarrow B^{\mu\nu} + \nabla_{\mu}\xi_{\nu} - \nabla_{\nu}\xi_{\mu}$ for a gauge parameter $\xi_{\mu}$. The gauge invariance of kinetic term in an otherwise non-gauge invariant Lagrangian (\ref{slva5}) gives rise to redundancy problems in the energy spectrum \cite{buchbinder2007}, and cannot be removed via usual quantization method. A consistent method to treat this redundancy is given by the St{\"u}ckelberg procedure \cite{stuckelberg1957}. According to this procedure, a strongly coupled field called the St{\"u}ckelberg field is introduced in the symmetry breaking potential term such that the gauge symmetry is restored in a given Lagrangian. The original theory is still recovered in a special gauge (where St{\"u}ckelberg field is put to zero), however, the advantage is that the redundant degrees of freedom are now encompassed in the St{\"u}ckelberg field, and can be dealt with using well known quantization frameworks like the Faddeev-Popov method. For a detailed account of this procedure applied to massless and massive antisymmetric tensors, interested reader is referred to Refs. \cite{buchbinder1992,buchbinder2008} respectively, and to Ref. \cite{aashish2018a} for a more recent analysis in the context of covariant effective action.  

The above procedure is applied to (\ref{slva5}) via the introduction of a strongly coupled vector field $C_{\mu}$:
\begin{eqnarray}
\label{slva7}
\mathcal{L} = -\frac{1}{12}H_{\mu\nu\lambda}H^{\mu\nu\lambda} - \frac{1}{4}\alpha^{2}\Big[b_{\mu\nu}\Big(B^{\mu\nu} + \frac{1}{\alpha}F^{\mu\nu}[C]\Big)\Big]^{2},
\end{eqnarray}
so that the Lagrangian (\ref{slva7}) becomes gauge invariant (here, $F_{\mu\nu}\equiv\partial_{\mu}C_{\nu}-\partial_{\nu}C_{\mu}$), and reduces to original Lagrangian (\ref{slva5}) in the gauge $C_{\mu}=0$. The new Lagrangian is invariant under two sets of transformations: $(i)$ gauge transformation of $B_{\mu\nu}$ and shift of field $C_{\mu}$, 
\begin{eqnarray}
\label{slvb0}
B^{\mu\nu} &\longrightarrow & B^{\mu\nu} + \nabla_{\mu}\xi_{\nu} - \nabla_{\nu}\xi_{\mu}, \nonumber \\
C_{\mu} &\longrightarrow & C_{\mu} - \alpha\xi_{\mu},
\end{eqnarray}
and, $(ii)$ under the gauge transformation of St{\"u}ckelberg field $C_{\mu}$
\begin{eqnarray}
\label{slvb1}
C_{\mu} &\longrightarrow & C_{\mu} + \nabla_{\mu}\Lambda , \nonumber \\
B^{\mu\nu} &\longrightarrow & B^{\mu\nu} ,
\end{eqnarray}
where, $\xi_{\mu}$ and $\Lambda$ are the corresponding gauge parameters. 
In addition to the above symmetries of fields, there exists a set of transformation of gauge parameters $\Lambda$ and $\xi_{\mu}$ that leaves the fields $B_{\mu\nu}$ and $C_{\mu}$ invariant,
\begin{eqnarray}
\label{slvb2}
\xi_{\mu} &\longrightarrow & \xi_{\mu} + \nabla_{\mu}\psi , \nonumber \\
\Lambda &\longrightarrow & \Lambda + \alpha\psi ,
\end{eqnarray}
which means that the gauge generators are linearly dependent \cite{buchbinder2008}. The gauge fixing procedure requires that a gauge condition be chosen for each of the fields $B_{\mu\nu}$ and $C_{\mu}$ as well as for the parameter $\xi_{\mu}$, so that the redundant degrees of freedom due to symmetries (\ref{slvb0}), (\ref{slvb1}) and (\ref{slvb2}) are taken care of. An important consideration while choosing a gauge condition is to ensure that all cross terms of fields in the Lagrangian cancel out or lead to a total derivative term, so that path integral can be computed with ease. Keeping this in mind, we choose the gauge condition for $B_{\mu\nu}$ to be (a similar choice for gauge condition in the context of Bumblebee model was considered in Ref. \cite{escobar2017})
\begin{eqnarray}
\label{aeaa6}
\chi_{\xi_{\nu}} = n_{\mu\nu}n_{\rho\sigma}\nabla^{\mu} B^{\rho\sigma} + \alpha C_{\nu}.
\end{eqnarray}
It turns out that the gauge fixing action term corresponding to Eq. (\ref{aeaa6}) introduces yet another soft symmetry breaking in $C_{\mu}$ \cite{aashish2018a}, so one has to introduce another St{\"u}ckelberg field $\Phi$ so that, 
\begin{eqnarray}
\label{aeaa7}
C_{\mu} \longrightarrow C_{\mu} + \dfrac{1}{\alpha}\nabla_{\mu}\Phi.
\end{eqnarray}
This modifies the symmetry in Eq. (\ref{slvb1}) by an additional shift transformation, 
\begin{eqnarray}
\label{aeaa8}
\Phi \longrightarrow \Phi - \alpha\Lambda.
\end{eqnarray}
From Eqs. (\ref{slvb1}) and (\ref{aeaa8}), the gauge condition for $C_{\mu}$ can be chosen to be,
\begin{eqnarray}
\label{aeaa9}
\chi_{\Lambda} = \nabla^{\mu}C_{\mu} + \alpha\Phi.
\end{eqnarray}
Similarly, for the symmetry of parameters, Eq. (\ref{slvb2}), we choose 
\begin{eqnarray}
\label{beaa0}
\check{\chi}_{\psi} = \nabla^{\mu}\xi_{\mu} - \alpha\Lambda.
\end{eqnarray}
The gauge conditions chosen above are incorporated in the action through ``gauge-fixing Lagrangian" terms of the form $-\frac{1}{2}\chi_{(\cdot)}^{2}$ and $-\frac{1}{2}\check{\chi}_{(\cdot)}^{2}$ for each of the conditions (\ref{aeaa6}), (\ref{aeaa9}) and (\ref{beaa0}). The final result for the total gauge fixed Lagrangian is given by,
\begin{eqnarray}
\label{slvb3}
\mathcal{L}_{GF} = -\dfrac{1}{12}H_{\mu\nu\lambda}H^{\mu\nu\lambda} - \frac{1}{4}\alpha^{2}\Big(b_{\mu\nu}B^{\mu\nu}\Big)^{2} - \dfrac{1}{4}\Big(b_{\mu\nu}F^{\mu\nu}\Big)^{2} - \frac{1}{2}\Big(b_{\mu\nu}b_{\rho\sigma}\nabla^{\mu}B^{\rho\sigma}\Big)^{2} \nonumber \\ - \frac{1}{2}\alpha^{2}C_{\nu}C^{\nu} - \frac{1}{2}(\nabla_{\mu}\Phi)^{2} - \frac{1}{2}(\nabla^{\mu}C_{\mu})^{2} - \frac{1}{2}\alpha^{2}\Phi^{2}.
\end{eqnarray}
The presence of a new scalar field $\Phi$ is a direct consequence of gauge-fixing of St{\"u}ckelberg field, and explicitly displays a scalar degree of freedom that remains hidden in the original Lagrangian (\ref{slva5}) with broken gauge symmetry.

\section{\label{sec3b}1-Loop Effective Action}
The quantization of theories such as (\ref{slva7}) is tricky, because of the symmetries in gauge parameters, as in Eq. (\ref{slvb2}). Such symmetries lead to a degeneracy in the ghost determinant appearing in the Faddeev-Popov procedure \cite{buchbinder1992}, and require special treatment for quantization \cite{buchbinder1988,buchbinder1992,aashish2018a}. We follow a general quantization procedure developed in Ref. \cite{aashish2018a} based on DeWitt-Vilkovisky's approach \cite{dewitt1967a,dewitt1967b,dewitt1967c,vilkovisky1984a} that yields covariant and background independent results, to deal with the additional symmetries of gauge parameters and derive the 1-loop effective action. 

For a quadratic action not involving quantization of metric the expression for 1-loop effective action in the DeWitt-Vilkovisky's field space notation, about a set of background fields $\bar{\varphi}^{i}$, is given by \cite{aashish2018a},
\begin{eqnarray}
\label{aea0}
\Gamma_{1}[\bar{\varphi}] = -\ln\det Q_{\alpha\beta}[\bar{\varphi}] + \ln\det \check{Q}_{a b} + \dfrac{1}{2}\ln\det\left(S^{GF}_{,ij}[\bar{\varphi}]]\right).
\end{eqnarray}
where $S^{GF}$ is the gauge-fixed action. Let us briefly explain the various (field-space) notations in Eq. (\ref{aea0}) (see \cite{parker2009} for a detailed introduction). The index $i$ in field space corresponds to all the tensor indices and spacetime dependence of fields in the coordinate space. For example, fields ($B_{\mu\nu}(x), C_{\mu}(x),\Phi(x)$) are denoted by components of $\varphi^{i}$ ($i = 1,2,3$) in field space, where $\varphi^{1}\leftrightarrow B_{\mu\nu}(x)$, $\varphi^{2}\leftrightarrow C_{\mu}(x)$, and $\varphi^{3}\leftrightarrow \Phi(x)$. The rest of the constructions in field space (tensors, scalar products, connections, field space metric, etc.) are similar to that in a coordinate space. The background fields in this notation, $\bar{\varphi}^{i}$, too carry all the indices of their respective counterparts including coordinate dependence. The object $S_{,ij}$ represents a derivative in field space, define by,
\begin{eqnarray}
\label{aea1}
S_{,ij}[\bar{\varphi}] = \left(\dfrac{\delta^2}{\delta \varphi^{j}\delta \varphi^{i}} S[\varphi]\right)_{\varphi = \bar{\varphi}}.
\end{eqnarray}
Let, $\check{\epsilon}^{a}$ parametrize the symmetry of gauge parameters, as in Eq. (\ref{slvb2}), and $\check{\chi}^{a}$ be the corresponding fixing condition for gauge parameters $\epsilon^{\alpha}$ (can be read off of Eqs. (\ref{slvb0}) and (\ref{slvb1}) ), then \cite{aashish2018a}
\begin{eqnarray}
\label{aea3}
Q^{\alpha}_{\beta} = \left(\dfrac{\delta}{\delta\epsilon^{\beta}}\chi^{\alpha}[\varphi,\epsilon,\check{\chi}]\right)_{\epsilon = 0},
\end{eqnarray}
where $\chi^{\alpha}$ is the gauge fixing condition for fields $\varphi^{i}$. $\det Q^{\alpha}_{\beta}$ is the ghost determinant factor. 

In the present case, corresponding to the symmetries (\ref{slvb0}), (\ref{slvb1}) and (\ref{slvb2}), there are two gauge conditions $\chi_{\xi_{\nu}} $ and $ \chi_{\Lambda}$, along with a condition $\check{\chi}_{\psi}$ on the parameters, that lead to three operators $Q^{\xi_{\mu}}_{\xi_{\nu}}$, $Q^{\Lambda}_{\Lambda}$ and $\check{Q}_{\psi}^{\psi}$ respectively. The results are displayed in Table \ref{tab1}.
\begin{table}[h!]
  \begin{center}
    \begin{tabular}{|c|c|} 
    \hline
      $\chi_{\xi_{\nu}} = b_{\mu\nu}b_{\rho\sigma}\nabla^{\mu} B^{\rho\sigma} + \alpha C_{\nu}$ & \ $Q^{\xi_{\mu}}_{\xi_{\nu}} = 2b_{\alpha\mu}b_{\beta\nu}\nabla^{\alpha}\nabla^{\beta} + \nabla_{\mu}\nabla_{\nu} - \alpha^{2}\delta_{\mu\nu}$ \ \\
      \hline
      $\chi_{\Lambda}= \nabla^{\mu}C_{\mu} + \alpha\Phi$ & $Q^{\Lambda}_{\Lambda} = \Box - \alpha^{2}$ \\
      \hline
      $\check{\chi}_{\psi}=\nabla^{\mu}\xi_{\mu} - \alpha\Lambda$ & $\check{Q}_{\psi}^{\psi}=\Box - \alpha^{2}$ \\
      \hline
    \end{tabular}
    \caption{Results for $Q$ operators corresponding to choices of fixing conditions $\chi$. }
    \label{tab1}
  \end{center}
\end{table}\\
Using these results in Eq. (\ref{aea0}), we get
\begin{eqnarray}
\label{aea4}
\Gamma_{1} = -\ln\det Q^{\xi_{\mu}}_{\xi_{\nu}} + \dfrac{1}{2}\ln\det\left(S^{GF}_{,ij}[\bar{\varphi}]]\right).
\end{eqnarray}
$S^{GF}$ is of course quadratic in fields, and the value of $\Gamma_{1}$ in operator form turns out to be, 
\begin{eqnarray}
\label{aea5}
\Gamma_{1} = \frac{i\hbar}{2}\Big[\ln\det(\Box_{2}{}^{\mu\nu}_{\ \ \rho\sigma} - \alpha^{2}b^{\mu\nu}b_{\rho\sigma}) - \ln\det(\Box_{1}{}^{\mu}_{\ \nu} -\alpha^{2}\delta^{\mu}_{\nu}) + \ln\det(\Box - \alpha^{2})\Big],
\end{eqnarray}
where,
\begin{eqnarray}
\label{aea6}
\Box_{2}{}^{\mu\nu}_{\ \ \rho\sigma}B^{\rho\sigma} &\equiv & \nabla_{\alpha}\nabla^{\alpha}B^{\mu\nu} + \nabla_{\alpha}\nabla^{\mu}B^{\nu\alpha} + \nabla_{\alpha}\nabla^{\nu}B^{\alpha\mu} + 2b^{\mu\nu}b_{\rho\sigma}b^{\alpha\sigma}b_{\beta\gamma}\nabla^{\rho}\nabla_{\alpha}B^{\beta\gamma}, \nonumber \\
\Box_{1}{}^{\mu}_{\ \nu}C^{\nu} &\equiv & 2b^{\nu\mu}b_{\rho\sigma}\nabla_{\nu}\nabla^{\rho}C^{\sigma} + \nabla^{\mu}\nabla_{\nu}C^{\nu};
\end{eqnarray}
and $\Box$ is the de'Alembertian operator. In flat spacetime, no physically interesting inferences can be extracted from the above expression. However, in curved spacetime, the operators in Eq. (\ref{aea5}) are coupled to the metric $g_{\mu\nu}$. So, addressing certain issues, like that of quantum equivalence, then becomes nontrivial \footnote{see Appendix \ref{AppendixB} for a detailed account of this issue}. Unfortunately, effective action cannot be calculated exactly in such cases \cite{aashish2018b}, and the best way forward is to perform a perturbative study. Therefore, we will consider a \textit{nearly} flat spacetime instead of a general curved one, so that,
\begin{eqnarray}
\label{aea7}
g_{\mu\nu}(x) = \eta_{\mu\nu} + \kappa h_{\mu\nu}(x).
\end{eqnarray}
$\eta_{\mu\nu}$ is the Minkowski metric and $h_{\mu\nu}$ is a perturbation, while $\kappa = 1/M_{p}$($M_{p}$ is Planck mass) parametrizes the scale of perturbation.

We can rewrite $\Gamma_{1}$ in integral form by introducing ghost fields $c_{\mu}$ and $\bar{c}_{\mu}$, 
\begin{eqnarray}
\label{aea8}
\Gamma_{1} = -\ln\int [d\eta][dc_{\mu}] [d\bar{c}_{\mu}] e^{-S_{GH}},
\end{eqnarray}
where, 
\begin{eqnarray}
\label{aea9}
S_{GH} = \eta^{i}S^{GF}_{,ij}\eta^{j} + \bar{c}^{\mu}Q^{\xi_{\mu}}_{\xi_{\nu}}c^{\nu},
\end{eqnarray}
and $\eta^{i}$ are the quantum fluctuations ($\delta B_{\mu\nu}(x), \delta C_{\mu}(x),\delta\Phi(x)$). Now, we use Eq. (\ref{aea7}) in Eq. (\ref{aea9}) and rearrange terms in orders of $h_{\mu\nu}$:
\begin{eqnarray}
\label{bea0}
S_{GH} = S_{0} + S_1 + O(h_{\mu\nu}h_{\alpha\beta}),
\end{eqnarray}
where the subscripts denote the power of $h_{\mu\nu}$. Substituting Eq. (\ref{bea0}) in Eq. (\ref{aea8}), and treating $S_{1}$ as a perturbation, the integrand can be Taylor expanded to write,
\begin{eqnarray}
\label{bea1}
\Gamma_{1} = -\ln\left(1 + \langle S_1\rangle + O(h_{\mu\nu}h_{\alpha\beta})\right),
\end{eqnarray}
where we have used the normalization for path integral of $S_0$. The logarithm can be further expanded to yield, up to first order in $h_{\mu\nu}$,
\begin{eqnarray}
\label{bea2}
\Gamma_1 = -\langle S_1\rangle .
\end{eqnarray}
The calculation of $\Gamma_1$ thus amounts to evaluating $\langle S_1\rangle$, which is a collection of two-point correlation functions of fields. These correlations are just the flat spacetime propagators of fields and can be derived from $S_0$ using projection operator method. We obtained the expansions of $S_{GH}$ using xAct packages \cite{xpert,xact} for Mathematica, results of which are presented below:
\begin{eqnarray}
\label{bea3}
S_0 &=& \int d^{4}x \Big(- \tfrac{1}{2} \alpha^2 \delta C_{\mu } \delta C^{\mu } -  \tfrac{1}{2} \alpha^2 \
\delta \phi^2 -  \tfrac{1}{4} \alpha^2 (\delta B^{\mu \nu } b_{\mu \nu })^2 \nonumber \\ 
&& -  \tfrac{1}{2} (b_{\alpha }{}^{\nu } b_{\beta \gamma } b_{\mu \nu \
} b_{\rho \sigma } \delta B^{\beta \gamma }{}^{,\alpha } \delta B^{\rho \sigma \
}{}^{,\mu }) -  \tfrac{1}{2} (\delta C_{\mu }{}^{,\mu })^2 \nonumber \\ 
&& -  \tfrac{1}{4} \bigl(b_{\mu \nu } (\delta C^{\nu }{}^{,\mu } -  \delta C^{\mu \
}{}^{,\nu })\bigr)^2 -  \tfrac{1}{2} \delta \phi {}_{,\mu } \delta \phi {}^{,\mu } \
\nonumber \\ 
&& + \delta \bar{c}^{\mu } (- \alpha^2 \delta c_{\mu } + \delta c^{\nu }{}_{,\nu }{}_{,\mu \
} + 2 b_{\nu \mu } b_{\rho \sigma } \delta c^{\sigma }{}^{,\rho }{}^{,\nu }) \
\nonumber \\ 
&& + \tfrac{1}{12} (- \delta B_{\nu \rho }{}_{,\mu } -  \delta B_{\rho \mu }{}_{,\nu } -  \delta B_{\mu \nu }{}_{,\rho }) (\delta B^{\nu \rho }{}^{,\mu } + \delta B^{\rho \
\mu }{}^{,\nu } + \delta B^{\mu \nu }{}^{,\rho }) \Big)
\\
S_1 &=& \int d^{4}x \Big( \tfrac{1}{2} \alpha^2 \delta C^{\mu } \delta C^{\nu } h{}_{\mu \nu } -  \alpha^2 \delta c^{\mu } \delta \bar{c}^{\nu } h{}_{\mu \nu } -  \tfrac{1}{4} \alpha^2 \delta \phi^2 h^{\mu }{}_{\mu } \nonumber \\ 
&& -  \tfrac{1}{4} \alpha^2 \delta C{}_{\mu } \delta C^{\mu } h^{\nu }{}_{\nu } -  \tfrac{1}{2} \alpha^2 \delta c^{\mu } \delta \bar{c}{}_{\mu } h^{\nu }{}_{\nu } + \tfrac{1}{2} \alpha^2 \delta B^{\mu \nu } \delta B^{\rho \sigma } h{}_{\rho a} b^{a}{}_{\sigma } n_{\mu \nu } \nonumber \\ 
&& + \tfrac{1}{2} \alpha^2 \delta B^{\mu \nu } \delta B^{\rho \sigma } h{}_{\sigma a} n_{\mu \nu } n_{\rho }{}^{a} -  \tfrac{1}{8} \alpha^2 \delta B^{\mu \nu } \delta B^{\rho \sigma } h^{a}{}_{a} n_{\mu \nu } n_{\rho \sigma } + \delta \bar{c}^{\mu } b^{\nu }{}_{\mu } b^{\rho \sigma } h{}_{\nu \rho }{}_{,a} \delta c{}_{\sigma }{}^{,a} \nonumber \\ 
&& - 2 \delta \bar{c}^{\mu } h{}_{\nu a} b^{\nu }{}_{\mu } b^{\rho \sigma } \delta c{}_{\sigma }{}_{,\rho }{}^{,a} + h{}_{\rho d} b^{bc} b^{\mu \nu } b^{\rho }{}_{\nu } b^{\sigma a} \delta B{}_{bc}{}^{,d} \delta B{}_{\sigma a}{}_{,\mu } + \tfrac{1}{2} h{}_{\rho a} b^{\mu \nu } b^{\rho \sigma } \delta C{}_{\sigma }{}^{,a} \delta C{}_{\nu }{}_{,\mu } \nonumber \\ 
&& + \tfrac{1}{2} \delta \bar{c}^{\mu } h^{\rho }{}_{\rho } \delta c^{\nu }{}_{,\nu }{}_{,\mu } -  \tfrac{1}{4} h^{\nu }{}_{\nu } \delta \phi{}_{,\mu } \delta \phi{}^{,\mu } + \tfrac{1}{2} h{}_{\sigma a} b^{\mu \nu } b^{\rho \sigma } \delta C{}_{\rho }{}^{,a} \delta C{}_{\mu }{}_{,\nu } \nonumber \\ 
&& -  \tfrac{1}{2} h{}_{\rho a} b^{\mu \nu } b^{\rho \sigma } \delta C{}_{\sigma }{}^{,a} \delta C{}_{\mu }{}_{,\nu } -  \tfrac{1}{4} h^{\rho }{}_{\rho } \delta C^{\mu }{}_{,\mu } \delta C^{\nu }{}_{,\nu } -  \tfrac{1}{2} \delta C^{\mu } h^{\rho }{}_{\rho }{}_{,\mu } \delta C^{\nu }{}_{,\nu } \nonumber \\ 
&& -  \tfrac{1}{2} \delta C^{\mu } b^{\nu \rho } b^{\sigma a} h{}_{\mu \sigma }{}_{,a} \delta C{}_{\rho }{}_{,\nu } + \delta \bar{c}^{\mu } b^{\nu }{}_{\mu } b^{\rho \sigma } h{}_{\rho \sigma }{}_{,a} \delta c^{a}{}_{,\nu } -  \delta \bar{c}^{\mu } b^{\nu }{}_{\mu } b^{\rho \sigma } \delta c{}_{\sigma }{}^{,a} h{}_{\rho a}{}_{,\nu } \nonumber \\ 
&& + \tfrac{1}{2} \delta \bar{c}^{\mu } \delta c^{\nu }{}_{,\mu } h^{\rho }{}_{\rho }{}_{,\nu } - 2 \delta \bar{c}^{\mu } h{}_{\rho a} b^{\nu }{}_{\mu } b^{\rho \sigma } \delta c{}_{\sigma }{}^{,a}{}_{,\nu } + \tfrac{1}{2} \delta c^{\mu } \delta \bar{c}^{\nu } h^{\rho }{}_{\rho }{}_{,\mu }{}_{,\nu } \nonumber \\ 
&& + \delta \bar{c}^{\mu } h^{a}{}_{a} b^{\nu }{}_{\mu } b^{\rho \sigma } \delta c{}_{\sigma }{}_{,\rho }{}_{,\nu } + \tfrac{1}{2} h{}_{\mu \nu } \delta \phi{}^{,\mu } \delta \phi{}^{,\nu } -  \tfrac{1}{4} h^{d}{}_{d} b^{bc} b^{\mu \nu } b^{\rho }{}_{\nu } b^{\sigma a} \delta B{}_{\sigma a}{}_{,\mu } \delta B{}_{bc}{}_{,\rho } \nonumber \\ 
&& + h{}_{cd} b^{bc} b^{\mu \nu } b^{\rho }{}_{\nu } b^{\sigma a} \delta B{}_{\sigma a}{}_{,\mu } \delta B{}_{b}{}^{d}{}_{,\rho } + h{}_{bd} b^{bc} b^{\mu \nu } b^{\rho }{}_{\nu } b^{\sigma a} \delta B{}_{\sigma a}{}_{,\mu } \delta B^{d}{}_{c}{}_{,\rho } \nonumber \\ 
&& + \tfrac{1}{2} h{}_{\sigma a} b^{\mu \nu } b^{\rho \sigma } \delta C{}_{\nu }{}_{,\mu } \delta C^{a}{}_{,\rho } + \tfrac{1}{2} \delta C^{\mu } b^{\nu \rho } b^{\sigma a} h{}_{\mu \sigma }{}_{,a} \delta C{}_{\nu }{}_{,\rho } -  \tfrac{1}{2} h{}_{\nu a} b^{\mu \nu } b^{\rho \sigma } \delta C{}_{\mu }{}^{,a} \delta C{}_{\sigma }{}_{,\rho } \nonumber \\ 
&& -  \tfrac{1}{8} h^{a}{}_{a} b^{\mu \nu } b^{\rho \sigma } \delta C{}_{\nu }{}_{,\mu } \delta C{}_{\sigma }{}_{,\rho } + \tfrac{1}{4} h^{a}{}_{a} b^{\mu \nu } b^{\rho \sigma } \delta C{}_{\mu }{}_{,\nu } \delta C{}_{\sigma }{}_{,\rho } + \delta \bar{c}^{\mu } b^{\nu }{}_{\mu } b^{\rho \sigma } h{}_{\nu \sigma }{}_{,a} \delta c^{a}{}_{,\rho } \nonumber \\ 
&& + \delta \bar{c}^{\mu } b^{\nu }{}_{\mu } b^{\rho \sigma } h{}_{\sigma a}{}_{,\nu } \delta c^{a}{}_{,\rho } + \delta C^{\mu } \delta C^{\nu }{}_{,\nu } h{}_{\mu }{}^{\rho }{}_{,\rho } + \tfrac{1}{4} \delta B^{\mu \nu } \delta B^{\rho \sigma }{}_{,\nu } h{}_{\mu \sigma }{}_{,\rho } \nonumber \\ 
&& -  \delta \bar{c}^{\mu } b^{\nu }{}_{\mu } b^{\rho \sigma } \delta c{}_{\sigma }{}^{,a} h{}_{\nu a}{}_{,\rho } + \delta \bar{c}^{\mu } b^{\nu }{}_{\mu } b^{\rho \sigma } \delta c^{a}{}_{,\nu } h{}_{\sigma a}{}_{,\rho } -  \delta c^{\mu } \delta \bar{c}^{\nu } b^{\rho }{}_{\nu } b^{\sigma a} h{}_{\mu \sigma }{}_{,a}{}_{,\rho } \nonumber \\ 
&& + \delta c^{\mu } \delta \bar{c}^{\nu } b^{\rho }{}_{\nu } b^{\sigma a} h{}_{\sigma a}{}_{,\mu }{}_{,\rho } + \delta c^{\mu } \delta \bar{c}^{\nu } b^{\rho }{}_{\nu } b^{\sigma a} h{}_{\mu a}{}_{,\sigma }{}_{,\rho } -  \tfrac{1}{8} g_{\mu \sigma } g_{\nu a} g_{\rho b} h^{c}{}_{c} \delta B^{\sigma a}{}^{,b} \delta B^{\mu \nu }{}^{,\rho } \nonumber \\ 
&& -  \tfrac{1}{4} g_{\mu a} g_{\nu b} g_{\rho \sigma } h^{c}{}_{c} \delta B^{\sigma a}{}^{,b} \delta B^{\mu \nu }{}^{,\rho } + \tfrac{1}{4} g_{\nu a} g_{\rho b} h{}_{\mu \sigma } \delta B^{\sigma a}{}^{,b} \delta B^{\mu \nu }{}^{,\rho } + \tfrac{1}{4} g_{\mu \sigma } g_{\rho b} h{}_{\nu a} \delta B^{\sigma a}{}^{,b} \delta B^{\mu \nu }{}^{,\rho } \nonumber \\ 
&& + \tfrac{1}{2} g_{\mu a} g_{\rho \sigma } h{}_{\nu b} \delta B^{\sigma a}{}^{,b} \delta B^{\mu \nu }{}^{,\rho } + \tfrac{1}{2} g_{\mu b} g_{\rho a} h{}_{\nu \sigma } \delta B^{\sigma a}{}^{,b} \delta B^{\mu \nu }{}^{,\rho } + \tfrac{1}{4} g_{\mu \sigma } g_{\nu a} h{}_{\rho b} \delta B^{\sigma a}{}^{,b} \delta B^{\mu \nu }{}^{,\rho } \nonumber \\ 
&& + \tfrac{1}{2} g_{\mu a} g_{\nu b} h{}_{\rho \sigma } \delta B^{\sigma a}{}^{,b} \delta B^{\mu \nu }{}^{,\rho } + h{}_{\nu \rho } \delta C^{\mu }{}_{,\mu } \delta C^{\nu }{}^{,\rho } + \tfrac{1}{2} \delta B^{\mu \nu } b^{b}{}_{a} b^{cd} b^{\rho }{}_{\nu } b^{\sigma a} h{}_{\mu \rho }{}_{,b} \delta B{}_{cd}{}_{,\sigma } \nonumber \\
&& + \tfrac{1}{2} \delta B^{\mu \nu } b^{b}{}_{a} b^{cd} n_{\mu }{}^{\rho } b^{\sigma a} h{}_{\nu \rho }{}_{,b} \delta B{}_{cd}{}_{,\sigma } -  \tfrac{1}{2} \delta B^{\mu \nu } b^{b}{}_{a} b^{cd} b^{\rho }{}_{\nu } b^{\sigma a} h{}_{\rho b}{}_{,\mu } \delta B{}_{cd}{}_{,\sigma } \nonumber \\ 
&& + \tfrac{1}{2} \delta B^{\mu \nu } b^{b}{}_{a} b^{cd} b^{\rho }{}_{\nu } b^{\sigma a} h{}_{\mu b}{}_{,\rho } \delta B{}_{cd}{}_{,\sigma } + \tfrac{1}{2} \delta B^{\mu \nu } b^{b}{}_{a} b^{cd} n_{\mu }{}^{\rho } b^{\sigma a} h{}_{\nu b}{}_{,\rho } \delta B{}_{cd}{}_{,\sigma } -  \tfrac{1}{2} h{}_{\rho a} b^{\mu \nu } b^{\rho \sigma } \delta C{}_{\nu }{}_{,\mu } \delta C^{a}{}_{,\sigma } \nonumber \\
&& + \tfrac{1}{2} h{}_{\sigma d} b^{ab} b^{cd} b^{\mu \nu } b^{\rho \sigma } \delta B{}_{ab}{}_{,c} \delta B{}_{\mu \nu }{}_{,\rho } - \tfrac{1}{2} \delta B^{\mu \nu } b^{b}{}_{a} b^{cd} n_{\mu }{}^{\rho } b^{\sigma a} h{}_{\rho b}{}_{,\nu } \delta B{}_{cd}{}_{,\sigma } \nonumber 
\end{eqnarray}
\begin{eqnarray}
  \label{bea4}
&& -  \tfrac{1}{2} h{}_{\nu a} b^{\mu \nu } b^{\rho \sigma } \delta C^{a}{}_{,\mu } \delta C{}_{\rho }{}_{,\sigma } + \tfrac{1}{2} h{}_{\mu a} b^{\mu \nu } b^{\rho \sigma } \delta C^{a}{}_{,\nu } \delta C{}_{\rho }{}_{,\sigma } -  \tfrac{1}{8} h^{a}{}_{a} b^{\mu \nu } b^{\rho \sigma } \delta C{}_{\mu }{}_{,\nu } \delta C{}_{\rho }{}_{,\sigma } \nonumber \\ 
&& + \tfrac{1}{2} \delta C^{\mu } b^{\nu \rho } b^{\sigma a} \delta C{}_{\rho }{}_{,\nu } h{}_{\mu a}{}_{,\sigma } -  \tfrac{1}{2} \delta C^{\mu } b^{\nu \rho } b^{\sigma a} \delta C{}_{\nu }{}_{,\rho } h{}_{\mu a}{}_{,\sigma } -  \delta \bar{c}^{\mu } b^{\nu }{}_{\mu } b^{\rho \sigma } \delta c^{a}{}_{,\rho } h{}_{\nu a}{}_{,\sigma } \nonumber \\ 
&& + \tfrac{1}{4} \delta B^{\mu \nu } \delta B^{\rho \sigma }{}_{,\mu } h{}_{\nu \rho }{}_{,\sigma } -  \delta \bar{c}^{\mu } b^{\nu }{}_{\mu } b^{\rho \sigma } \delta c^{a}{}_{,\nu } h{}_{\rho a}{}_{,\sigma } + \tfrac{1}{4} \delta B^{\mu \nu } h{}_{\nu \rho }{}_{,\sigma } \delta B{}_{\mu }{}^{\rho }{}^{,\sigma } \nonumber \\ 
&& + \tfrac{1}{4} \delta B^{\mu \nu } h{}_{\mu \sigma }{}_{,\rho } \delta B{}_{\nu }{}^{\rho }{}^{,\sigma } + \tfrac{1}{4} \delta B^{\mu \nu } h{}_{\nu \sigma }{}_{,\rho } \delta B^{\rho }{}_{\mu }{}^{,\sigma } + \tfrac{1}{4} \delta B^{\mu \nu } h{}_{\mu \rho }{}_{,\sigma } \delta B^{\rho }{}_{\nu }{}^{,\sigma } \Big)
\end{eqnarray}

\subsection{Propagators}
We use the projection operator method \cite{barnes1965} to invert the operators in $S_0$ and derive the Green's functions or propagators. In the operator form, $S_0$ can be recast as
\begin{eqnarray}
\label{apr0}
S_0 =\int d^{4}x \Big(  \dfrac{1}{4} B^{\mu\nu}\mathcal{O}^{B}_{\mu\nu,\alpha\beta}B^{\alpha\beta} + \dfrac{1}{2} C^{\mu}\mathcal{O}^{C}_{\mu\nu}C^{\nu} + \frac{1}{2}\Phi\mathcal{O}^{\Phi}\Phi\Big)
\end{eqnarray}
where,
\begin{eqnarray}
\label{apr1}
\mathcal{O}^{B}_{\mu\nu,\alpha\beta} &=& \frac{\Box}{2}(\eta_{\mu\alpha}\eta_{\nu\beta}-\eta_{\mu\beta}\eta_{\nu\alpha})  + \frac{1}{2}(\partial_{\mu}\partial_{\beta}\eta_{\nu\alpha} - \partial_{\nu}\partial_{\beta}\eta_{\mu\alpha} - \partial_{\mu}\partial_{\alpha}\eta_{\nu\beta} + \partial_{\nu}\partial_{\alpha}\eta_{\mu\beta})\nonumber\\ && -\,\big(\alpha^2 + 2 (b_{\rho\sigma}\partial^{\rho})^2\big) b_{\mu\nu}b_{\alpha\beta},\\
\label{apr2}
\mathcal{O}^{C}_{\mu\nu} &=& 2 b_{\sigma\mu}b_{\rho\nu}\partial^{\sigma}\partial^{\rho} + \partial_{\mu}\partial_{\nu} - \alpha^{2}\eta_{\mu\nu},\\
\label{apr3}
\mathcal{O}^{\Phi} &=& \Box - \alpha^{2}.
\end{eqnarray}
At this point, we would like to point out that a calculation for the propagator of $B_{\mu\nu}$ using projector method was first performed in Ref. \cite{maluf2019} recently. However, their calculation did not account for the St\"uckelberg field and as a result our operator $(\mathcal{O}^{B})_{\mu\nu,\alpha\beta}$ is different from the one in Ref. \cite{maluf2019}, which misses the contribution from gauge-fixing term  $2 (b_{\rho\sigma}\partial^{\rho})^2 b_{\mu\nu}b_{\alpha\beta}$. Fortunately, this term is merely an addition to mass, $\alpha^{2}$, and ends up not contributing to the propagator, $(\mathcal{O}^{B})^{-1}_{\mu\nu,\alpha\beta}$. So, we end up getting an identical result for the propagator, barring complex infinity terms that can be ignored (see Appendix \ref{AppendixA} for details of projection operators $P^{(1)},...,P^{(6)}$),
\begin{eqnarray}
\label{apr4}
(\mathcal{O}^{B})^{-1}_{\mu\nu,\alpha\beta} (x,x')=\int \dfrac{d^{4}p}{(2\pi^{n})}e^{-ip\cdot(x-x')}\left(\frac{1}{p^{2}} P^{(1)}_{\mu\nu, \alpha\beta} + \frac{b^{2}}{(b_{\rho\sigma}p^{\sigma})^{2}}( P^{(4)}_{\mu\nu, \alpha\beta}+ P^{(5)}_{\mu\nu, \alpha\beta})\right),
\end{eqnarray}
There are no massive propagating modes in Eq. (\ref{apr4}) and only one massless mode propagates, as concluded in Ref. \cite{altschul2010,maluf2019}. The second pole describes a massless pole propagating in an anisotropic medium, which for our choice of $b_{\mu\nu}$ gives,
\begin{eqnarray}
\label{bpr0}
b^{2}\big((p^{2})^{2} + (p^{3})^{2}\big) = 0.
\end{eqnarray}
Contrary to the claim in Ref. \cite{maluf2019} where these modes were described as non-physical due to a negative sign appearing in energy-momentum relations as a result of a different choice of $b_{\mu\nu}$, we note that for our choice of $b_{\mu\nu}$ which corresponds to monopole solutions of antisymmetric tensor, energy terms ($p^0$) disappear altogether. 

For the St\"uckelberg field $C_{\mu}$, spontaneous Lorentz violating term appears in the kinetic part (first term in Eq. \ref{apr2}), which makes inverting $\mathcal{O}^{C}_{\mu\nu}$ a little tricky. New projector operators have to be defined apart from the longitudinal and transverse momentum operators, that also have a closed algebra, so that any operator $D_{\mu\nu}$ can be then expanded in terms of these projectors. We define,
\begin{eqnarray}
\label{apr5}
\mathcal{P}^{(1)}_{\mu\nu} = \dfrac{p_{\mu}p_{\nu}}{p^2}; \quad \mathcal{P}^{(2)}_{\mu\nu} = \eta_{\mu\nu} - \dfrac{p_{\mu}p_{\nu}}{p^2}; \quad 
\mathcal{P}^{(3)}_{\mu\nu} = \dfrac{1}{(b_{\rho\sigma}p^{\sigma})^{2}} b_{\sigma\mu}b_{\rho\nu}p^{\sigma}p^{\rho} .
\end{eqnarray}
These operators satisfy a closed algebra, as shown in Table \ref{tab2}.
\begin{table}[h!]
  \begin{center}
    \begin{tabular}{|c|c|c|c|} 
    \hline
      & $\mathcal{P}^{(1)}$ & $\mathcal{P}^{(2)}$ & $\mathcal{P}^{(3)}$ \\
      \hline
    $\mathcal{P}^{(1)}$  & $\mathcal{P}^{(1)}$ & 0 & 0 \\
      \hline
    $ \mathcal{P}^{(2)}$ & 0 & $ \mathcal{P}^{(2)}$ & $\mathcal{P}^{(3)}$ \\
      \hline
     $\mathcal{P}^{(3)}$ & 0 & $\mathcal{P}^{(3)}$ & $\mathcal{P}^{(3)}$ \\
      \hline
    \end{tabular}
    \caption{Algebra of projection operators for the St\"uckelberg field $C_{\mu}$. Tensor indices have not been explicitly written. }
    \label{tab2}
  \end{center}
\end{table}
Using these operators, $\mathcal{O}^{C}_{\mu\nu}$ in momentum space can be written as,
\begin{eqnarray}
\label{apr6}
\mathcal{O}^{C}_{\mu\nu} = - 2(b_{\rho\sigma}p^{\sigma})^2 \mathcal{P}^{(3)}_{\mu\nu} - \left(p^2 + \alpha^2 \right) \mathcal{P}^{(1)}_{\mu\nu} - \alpha^2 \mathcal{P}^{(2)}_{\mu\nu}.
\end{eqnarray}
Assuming that $(\mathcal{O}^{C})^{-1}_{\mu\nu}$ in momentum space has the form,
\begin{eqnarray}
\label{apr7}
(\mathcal{O}^{C})^{-1}_{\mu\nu} = m_{1}\mathcal{P}^{(1)}_{\mu\nu} + m_{2}\mathcal{P}^{(2)}_{\mu\nu} + m_{3} \mathcal{P}^{(3)}_{\mu\nu},
\end{eqnarray}
we use the identity $\mathcal{O}\mathcal{O}^{-1} = \mathcal{I}$ to obtain,
\begin{eqnarray}
\label{apr8}
(\mathcal{O}^{C})^{-1}_{\mu\nu}(x,x') = \int \dfrac{d^{4}p}{(2\pi^{n})}e^{-ip\cdot(x-x')}\left(-\dfrac{1}{p^2 + \alpha^2}\mathcal{P}^{(1)}_{\mu\nu} - \dfrac{1}{\alpha^{2}}\mathcal{P}^{(2)}_{\mu\nu} + \dfrac{1}{\alpha^2}\dfrac{(b_{\rho\sigma}p^{\sigma})^{2}}{(b_{\rho\sigma}p^{\sigma})^{2} + \alpha^2/2} \mathcal{P}^{(3)}_{\mu\nu}\right).
\end{eqnarray}
Here, a massive scalar mode with pole at $\alpha$ propagates while another anisotropic mode propagates with mass $\alpha/\sqrt{2}$. Terms in $\mathcal{P}^{(2)}$ (from Eq. (\ref{apr5})) contain a massless pole and an additive pole-less term which does not contribute to correlations and can be ignored. For $\Phi$, the scalar propagator is given by,
\begin{eqnarray}
\label{bpr1}
(\mathcal{O}^{\Phi})^{-1} (x,x') = \int \dfrac{d^{4}p}{(2\pi^{n})}e^{-ip\cdot(x-x')}\dfrac{1}{p^{2} + \alpha^2}.
\end{eqnarray}

\subsection{Quantum corrections}
Since all terms in $\langle S_{1}\rangle$ are local, they correspond to tadpole diagrams. We solve these integrals in two steps: first, the derivatives of field fluctuations are transformed to momentum space by substituting Eqs. (\ref{apr4}), (\ref{apr8}), and (\ref{bpr1}). We also perform by-parts integrals to get rid of derivatives of $h_{\mu\nu}$, so that in all expressions below, a coefficient $h_{\mu\nu}$ is understood to be present but not explicitly written. The Fourier transformed $\langle S_{1}\rangle$ then has terms of the form,
\begin{eqnarray}
\label{aqc0}
\int d^{4}x A(x) \langle\partial^{m}\delta \ \partial^{n}\delta\rangle \longrightarrow \int d^{4}x \dfrac{d^{4}p}{(2\pi^{n})} A(x) (-ip)^{m} (ip)^{n} \langle \delta_{p}\delta_{p}\rangle,
\end{eqnarray}
where, tensor indices of $A(x)$ and $\delta$ have been omitted for convenience. $\delta$ is the quantum field fluctuation, and $\langle \delta_{p}\delta_{p}\rangle$ represents the propagator(s) in momentum space. 

The second step is to replace $\langle \delta_{p}\delta_{p}\rangle$ with values of Green's function and evaluate the integrals. We primarily use the results in Ref. \cite{bardin1999} to evaluate the divergent terms of most of the integrals, except those involving anisotropic term $(b_{\rho\sigma}p^{\sigma})^{2}$. There are two types of pole-less integrals coming from Eq. (\ref{apr4}):
\begin{eqnarray}
\label{aqc1}
\int d^{4}x \dfrac{d^{4}p}{(2\pi^{n})} A(x)\dfrac{p^{\mu}...p^{\beta}}{p^{2}} ; \quad \int d^{4}x \dfrac{d^{4}p}{(2\pi^{n})} A(x)\dfrac{p^{\mu}...p^{\beta}}{(b_{\rho\sigma}p^{\sigma})^{2}},
\end{eqnarray}
with up to four $p^{\mu}$'s in the numerator. The first integral vanishes due to the lack of a physical scale \cite{bardin1999}. To solve the second integral, we use the approach developed in \cite{george1975,george1987,capper1982a,capper1982b}, and find that it also does not have any physical contribution. 

Next, there are broadly three types of integrals with non-zero poles arising from the rest of propagators:
\begin{eqnarray}
\label{aqc2}
&\int d^{4}x \dfrac{d^{4}p}{(2\pi^{n})} A(x)\dfrac{p^{\mu}...p^{\beta}}{p^{2} + \alpha^2} ; \quad \int d^{4}x \dfrac{d^{4}p}{(2\pi^{n})} A(x)\dfrac{p^{\mu}...p^{\beta}}{p^2(p^{2} + \alpha^2)};& \nonumber \\
 &\int d^{4}x \dfrac{d^{4}p}{(2\pi^{n})} A(x)\dfrac{p^{\mu}...p^{\beta}}{(b_{\rho\sigma}p^{\sigma})^{2} + \alpha^2/2};&
\end{eqnarray}
Again, the solutions to first two types of integrals are available in Ref. \cite{bardin1999}. We solve the third type of integral as follows. Following \cite{george1975}, we write
\begin{eqnarray}
\label{aqc3}
\int d^{4}p \dfrac{1}{(b_{\rho\sigma}p^{\sigma})^{2} + \alpha^2/2} = \int d^{4}p \int_{0}^{\infty} d\theta \exp[-\theta((b_{\rho\sigma}p^{\sigma})^{2} + \alpha^2/2)].
\end{eqnarray}
Integrating over $d^{4}p$, followed by writing the integral over $\theta$ in terms of $\Gamma$ function leads to familiar expressions encountered in dimensional regularization, which finally yields the divergent part as ($\epsilon = n-4$),
\begin{eqnarray}
\label{aqc4}
divp\left(\int d^{4}p \dfrac{1}{(b_{\rho\sigma}p^{\sigma})^{2} + \alpha^2/2}\right) = -\dfrac{\pi^{2}\alpha^{2}}{2\sqrt{\det(b_{\mu\rho}b^{\rho\nu})}}\dfrac{2}{\epsilon},
\end{eqnarray} 
which is identical to that of a scalar propagator integral except for the $\sqrt{\det(b_{\mu\rho}b^{\rho\nu})}$ in the denominator. For our choice of $b_{\mu\nu}$, Eq. (\ref{slva1}) with $a=0$ and $b=1/\sqrt{2}$, this term becomes a diagonal matrix,
\begin{eqnarray}
\label{aqc5}
b_{\mu\rho}b^{\rho\nu} = diag(0 \ \ 0 \ \ 1/2 \ \ 1/2),
\end{eqnarray}
implying that the determinant is zero. It turns out however, that this determinant appears as a factor in the denominator of the divergent part of effective action, and hence we use a regularization factor $\epsilon'$ to write,
 \begin{eqnarray}
\label{aqc6}
b_{\mu\rho}b^{\rho\nu} = \lim_{\epsilon'\to 0} diag(\epsilon' \ \ \epsilon' \ \ 1/2 \ \ 1/2).
\end{eqnarray}

With these inputs in xAct\cite{xact}, the final result for the divergent part of 1-loop effective after some further manipulations, is obtained as,
\begin{eqnarray}
\label{1lea}
divp(\Gamma_{1}) &=& \dfrac{1}{16\pi^{2}\epsilon}\Big( \alpha^4 \kappa h^{a }{}_{a }  + \dfrac{1}{\sqrt{\det(b_{\mu\rho}b^{\rho\nu})}} \Big( -  \tfrac{1}{16} \alpha^4 \kappa h{}_{\mu b} b_{a}{}^{b} b^{\mu a} -  \tfrac{3}{32} \alpha^4 \kappa h^{b}{}_{b} b_{a\mu } b^{\mu a} \nonumber \\ 
&& -  \tfrac{1}{12} \alpha^4 \kappa h{}_{\mu c} b_{a}{}^{b} b_{b}{}^{\nu } b^{c}{}_{\nu } b^{\mu a} -  \tfrac{5}{192} \alpha^4 \kappa h{}_{bc} b_{a\mu } b^{b\nu } b^{c}{}_{\nu } b^{\mu a} + \tfrac{1}{192} \alpha^4 \kappa h{}_{\nu c} b_{b}{}^{c} b^{b\nu } b_{\mu a} b^{\mu a} \nonumber \\ 
&& + \tfrac{1}{96} \alpha^4 \kappa h{}_{bc} b^{b\nu } b^{c}{}_{\nu } b_{\mu a} b^{\mu a} -  \tfrac{1}{384} \alpha^4 \kappa h^{c}{}_{c} b_{a}{}^{\nu } b_{b\nu } b_{\mu }{}^{b} b^{\mu a} -  \tfrac{1}{48} \alpha^4 \kappa h{}_{\nu c} b_{a}{}^{\nu } b^{c}{}_{b} b_{\mu }{}^{b} b^{\mu a} \nonumber \\ 
&& + \tfrac{5}{192} \alpha^4 \kappa h{}_{bc} b_{a}{}^{\nu } b^{c}{}_{\nu } b_{\mu }{}^{b} b^{\mu a} -  \tfrac{1}{384} \alpha^4 \kappa h^{c}{}_{c} b_{a\mu } b^{b\nu } b^{\mu a} b_{\nu b} + \tfrac{5}{384} \alpha^4 \kappa h^{c}{}_{c} b^{b\nu } b_{\mu a} b^{\mu a} b_{\nu b} \nonumber \\ 
&& + \tfrac{5}{192} \alpha^4 \kappa h^{c}{}_{c} b_{a}{}^{\nu } b_{\mu }{}^{b} b^{\mu a} b_{\nu b} + \tfrac{1}{192} \alpha^4 \kappa h{}_{\mu c} b_{a}{}^{b} b_{b}{}^{\nu } b^{\mu a} b_{\nu }{}^{c} + \tfrac{1}{192} \alpha^4 \kappa h{}_{bc} b_{a\mu } b^{b\nu } b^{\mu a} b_{\nu }{}^{c} \nonumber \\ 
&& -  \tfrac{7}{192} \alpha^4 \kappa h{}_{bc} b^{b\nu } b_{\mu a} b^{\mu a} b_{\nu }{}^{c} -  \tfrac{1}{96} \alpha^4 \kappa h{}_{bc} b_{a}{}^{\nu } b_{\mu }{}^{b} b^{\mu a} b_{\nu }{}^{c} -  \tfrac{1}{384} \alpha^4 \kappa h^{c}{}_{c} b_{a}{}^{b} b_{b}{}^{\nu } b^{\mu a} b_{\nu \mu } \nonumber \\ 
&& -  \tfrac{1}{96} \alpha^4 \kappa h{}_{bc} b^{c}{}_{\nu } b_{\mu }{}^{b} b^{\mu a} b^{\nu }{}_{a} + \tfrac{1}{96} \alpha^4 \kappa h{}_{\mu c} b_{a}{}^{b} b^{c}{}_{\nu } b^{\mu a} b^{\nu }{}_{b}\Big)\Big)
\end{eqnarray}
where $\epsilon = n-4$ (as $n\to 4$) is the divergence parameter from dimensional regularization. Eq. (\ref{1lea}) presents the divergent piece of one loop corrections of antisymmetric tensor field theory with spontaneous Lorentz violation at leading order in field fluctuations in a nearly flat spacetime, and is valid for a vacuum value that supports monopole solutions. The one-loop divergence structures in principle lead to corrections to parameters (or couplings) in the classical action through counterterms (for example, in Ref. \cite{mackay2010}). Studying such corrections is interesting at higher orders in background fields, but lie beyond the scope of present work. Also, it is not easy to compare theories with and without spontaneous Lorentz violation in the present context, because the simplest (spontaneously) Lorentz violating potential contains up to quartic order terms in fields; while without Lorentz violation, the potential(s) that have been studied in the past \cite{buchbinder2008} are quadratic in field components.

\section{\label{sec3c}Quantum Equivalence}
The classical Lagrangian (\ref{slva5}) can be written in an equivalent form where the field $B_{\mu\nu}$ can be eliminated through the introduction of a vector field, so that the resulting Lagrangian describes a classically equivalent vector theory with spontaneous Lorentz violation. In this section, we will check their quantum equivalence at one-loop level. 

Checking classical equivalence of two theories is an interesting theoretical exercise, because it provides insight into the degrees of freedom and dynamical properties of theories that may be described by very different fields, like in 2-form, 1-form or a scalar field theories, and thus may lead to several simplifications in a given theory. This problem naturally extends to the quantum regime, and it is certainly not trivial to prove quantum equivalence of two classically equivalent theories especially in curved spacetime. For instance, it can be shown that a massive 2-form field is quantum equivalent to a massive vector field because of some special topological properties of zeta functions \cite{buchbinder2008}. However, it is extremely difficult to perform similar analyses when, for example, the Lorentz symmetry is spontaneously broken \cite{aashish2018b}. In flat spacetime, establishing quantum equivalence is indeed trivial, because there is no field dependence in $\Gamma_{1}$ (Eq. (\ref{aea5})) and hence effective actions of two theories do not possess any physical distinction.

On the contrary, in curved spacetime, the presence of metric makes things interesting. Only problem is, the effective action cannot be calculated exactly. So, our best bet, in this case, is to do a perturbative study like the one in the previous section. 

Classical equivalence of Eq. (\ref{slva5}) was explored in Ref. \cite{altschul2010}, it was found to be equivalent to,
\begin{eqnarray}
\label{eq0}
\mathcal{L} = \dfrac{1}{2}B_{\mu\nu}\mathcal{F}^{\mu\nu} - \frac{1}{2}C^{\mu}C_{\mu} - \dfrac{1}{4}\alpha^{2} \Big(b_{\mu\nu}B^{\mu\nu}\Big)^{2},
\end{eqnarray}
where,
\begin{eqnarray}
\label{eq1}
\mathcal{F}_{\mu\nu} = \dfrac{1}{2}\epsilon_{\mu\nu\rho\sigma}F^{\mu\nu}.
\end{eqnarray}
$C_{\mu}$ is a vector field and $F_{\mu\nu}$ is as defined before. We choose to continue with the same symbol for vector and St\"uckelberg field to avoid unnecessary complications. Eq. (\ref{eq0}) can be written exclusively in terms of $C_{\mu}$ through the use of projection operators,
\begin{eqnarray}
\label{eq2}
T_{||\mu\nu} = b_{\rho\sigma}T^{\rho\sigma} b_{\mu\nu}, \nonumber \\
T_{\perp\mu\nu} = T_{\mu\nu} - T_{||\mu\nu},
\end{eqnarray}
for any two-rank tensor $T_{\mu\nu}$, and subsequently using the equations of motion for $B_{||\mu\nu}$ and $B_{\perp\mu\nu}$, to obtain,
\begin{eqnarray}
\label{eq3}
\alpha^{2}\mathcal{L} = \dfrac{1}{4}\left(\tilde{b}_{\mu\nu}F^{\mu\nu}\right)^{2} - \frac{1}{2}\alpha^{2} C^{\mu}C_{\mu},
\end{eqnarray}
where we have defined $\tilde{b}_{\mu\nu} = \dfrac{1}{2}\epsilon_{\mu\nu\rho\sigma}b^{\rho\sigma}$. Note that Lorentz violation enters Eq. (\ref{eq3}) through the kinetic term, although it is still gauge-symmetric. A similar exercise of applying St\"uckelberg procedure leads to the gauge fixed action in flat spacetime,
\begin{eqnarray}
\label{eq4}
\tilde{S}_{0} = \int d^{4}x \Big( \dfrac{1}{2} C^{\mu}\mathcal{O}^{C'}_{\mu\nu}C^{\nu} + \frac{1}{2}\Phi\mathcal{O}^{\Phi}\Phi \Big)
\end{eqnarray}
where, 
\begin{eqnarray}
\label{eq5}
\mathcal{O}^{C'}_{\mu\nu} = -\dfrac{1}{2}\big(\tilde{b}_{\sigma\mu}\tilde{b}_{\rho\nu}\partial^{\sigma}\partial^{\rho} + \tilde{b}_{\sigma\nu}\tilde{b}_{\rho\mu}\partial^{\sigma}\partial^{\rho}\big) + \dfrac{1}{2}\partial_{\mu}\partial_{\nu} - \dfrac{1}{2}\alpha^{2}\eta_{\mu\nu}.
\end{eqnarray}
A similar calculation of the propagator yields,
\begin{eqnarray}
\label{eq6}
(\mathcal{O}^{C'})^{-1}_{\mu\nu}(x,x') = \int \dfrac{d^{4}p}{(2\pi^{4})}e^{-ip\cdot(x-x')}\left(-\dfrac{1}{p^2 + \alpha^2}\mathcal{P}^{(1)}_{\mu\nu} - \dfrac{1}{\alpha^{2}}\mathcal{P}^{(2)}_{\mu\nu} - \dfrac{1}{\alpha^2}\dfrac{(\tilde{b}_{\rho\sigma}p^{\sigma})^{2}}{-(\tilde{b}_{\rho\sigma}p^{\sigma})^{2} + \alpha^2/2} \tilde{\mathcal{P}}^{(3)}_{\mu\nu}\right),
\end{eqnarray}
where $\tilde{\mathcal{P}}^{(3)}_{\mu\nu}$ has the same form as $\mathcal{P}^{3}_{\mu\nu}$ but with $\tilde{b}_{\mu\nu}$ instead of $b_{\mu\nu}$. Finally, the one-loop effective action is found to be,
\begin{eqnarray}
\label{eq7}
divp(\tilde{\Gamma}_{1}) &=&  \dfrac{1}{16\pi^{2}\epsilon}\Big( \alpha^4 \kappa h^{a }{}_{a }  + \dfrac{1}{\sqrt{-\det(\tilde{b}_{\mu\rho}\tilde{b}^{\rho\nu})}} \Big(\tfrac{1}{96} \alpha^4 \kappa h{}_{ec} \tilde{b}_{ad} \tilde{b}^{ad} \tilde{b}_{b}{}^{c} \tilde{b}^{be} + \tfrac{1}{32} \alpha^4 \kappa h^{b}{}_{b} \tilde{b}^{ac} \tilde{b}_{ca} \nonumber \\ 
&& -  \tfrac{1}{16} \alpha^4 \kappa h{}_{ab} \tilde{b}^{ac} \tilde{b}_{c}{}^{b} + \tfrac{1}{192} \alpha^4 \kappa h{}_{bc} \tilde{b}_{ad} \tilde{b}^{ad} \tilde{b}^{be} \tilde{b}^{c}{}_{e} -  \tfrac{1}{96} \alpha^4 \kappa h{}_{ac} \tilde{b}^{ad} \tilde{b}_{b}{}^{e} \tilde{b}^{c}{}_{e} \tilde{b}_{d}{}^{b} \nonumber \\ 
&& + \tfrac{1}{96} \alpha^4 \kappa h{}_{ec} \tilde{b}_{a}{}^{b} \tilde{b}^{ad} \tilde{b}_{b}{}^{c} \tilde{b}_{d}{}^{e} -  \tfrac{1}{384} \alpha^4 \kappa h^{c}{}_{c} \tilde{b}_{a}{}^{b} \tilde{b}^{ad} \tilde{b}_{be} \tilde{b}_{d}{}^{e} -  \tfrac{1}{96} \alpha^4 \kappa h{}_{ec} \tilde{b}_{a}{}^{b} \tilde{b}^{ad} \tilde{b}^{c}{}_{b} \tilde{b}_{d}{}^{e} \nonumber \\ 
&& + \tfrac{5}{192} \alpha^4 \kappa h{}_{bc} \tilde{b}_{a}{}^{b} \tilde{b}^{ad} \tilde{b}^{c}{}_{e} \tilde{b}_{d}{}^{e} -  \tfrac{1}{384} \alpha^4 \kappa h^{c}{}_{c} \tilde{b}^{ad} \tilde{b}_{b}{}^{e} \tilde{b}_{d}{}^{b} \tilde{b}_{ea} + \tfrac{1}{384} \alpha^4 \kappa h^{c}{}_{c} \tilde{b}_{ad} \tilde{b}^{ad} \tilde{b}^{be} \tilde{b}_{eb} \nonumber \\ 
&& -  \tfrac{1}{384} \alpha^4 \kappa h^{c}{}_{c} \tilde{b}^{ad} \tilde{b}^{be} \tilde{b}_{da} \tilde{b}_{eb} + \tfrac{1}{192} \alpha^4 \kappa h^{c}{}_{c} \tilde{b}_{a}{}^{b} \tilde{b}^{ad} \tilde{b}_{d}{}^{e} \tilde{b}_{eb} -  \tfrac{1}{64} \alpha^4 \kappa h{}_{bc} \tilde{b}_{ad} \tilde{b}^{ad} \tilde{b}^{be} \tilde{b}_{e}{}^{c} \nonumber \\ 
&& + \tfrac{1}{192} \alpha^4 \kappa h{}_{bc} \tilde{b}^{ad} \tilde{b}^{be} \tilde{b}_{da} \tilde{b}_{e}{}^{c} + \tfrac{1}{192} \alpha^4 \kappa h{}_{ac} \tilde{b}^{ad} \tilde{b}_{b}{}^{e} \tilde{b}_{d}{}^{b} \tilde{b}_{e}{}^{c} -  \tfrac{1}{48} \alpha^4 \kappa h{}_{bc} \tilde{b}_{a}{}^{b} \tilde{b}^{ad} \tilde{b}_{d}{}^{e} \tilde{b}_{e}{}^{c} \nonumber \\ 
&& -  \tfrac{1}{192} \alpha^4 \kappa h{}_{db} \tilde{b}_{a}{}^{b} \tilde{b}^{ad} \tilde{b}_{ce} \tilde{b}^{ec}\Big)\Big)
\end{eqnarray}
Upon comparing Eqs. (\ref{1lea}) and (\ref{eq7}), we can immediately notice that the first term is identical, while the rest of terms appearing with $b_{\mu\nu}$ and $\tilde{b}_{\mu\nu}$ do not match. The first term arises from the propagator of non-Lorentz violating modes, while all the other terms correspond to contributions from propagator of (spontaneously) Lorentz violating modes. Hence, the quantum equivalence holds along non-Lorentz violating modes but not along Lorentz violating modes involving $b_{\mu\nu}$. This conclusion is validated by the results of Ref. \cite{seifert2010a}, where it was shown that when there are topologically nontrivial monopole-like solutions of the spontaneous symmetry breaking equations, the interaction with gravity of the vector and tensor theories are different.

\section{Summary}
Study of spontaneous Lorentz violation with rank-2 antisymmetric tensor is interesting because of the possibility of rich phenomenological signals of SME in future experiments. Since antisymmetric tensor fields are likely to play s significant role in the early universe cosmology, studying their quantum aspect is a natural extension of classical analyses. In a past study \cite{aashish2018b}, it was found that issues like quantum equivalence are difficult to address in a general curved spacetime. This problem is overcome here by adopting a perturbative approach to evaluating effective action, that is also general enough to be applied to more complicated models including interaction terms.

We quantized a simple action of an antisymmetric tensor field with a nonzero vev driving potential term that introduces spontaneous Lorentz violation, using a covariant effective action approach at one-loop. The one-loop corrections were calculated in a nearly flat spacetime, at $O(\kappa\hbar)$. We revisited the issue of quantum equivalence, and found that for the non-Lorentz-violating modes (independent of vev $b_{\mu\nu}$), antisymmetric tensor field is quantum-equivalent to a vector field. However, contributions from the Lorentz violating part of the propagator leads to different terms in effective actions, and as a result, $\Delta\Gamma = \Gamma_1 - \tilde{\Gamma}_1 \neq 0$, i.e. the theories are not quantum equivalent.


\chapter{Covariant Quantum Corrections to a Scalar Field Model Inspired by Nonminimal Natural Inflation}

So far in this thesis, we have generalized the covariant quantization formalism using DeWit-Vilkovisky approach in Chapter 2, and applied it to compute one-loop effective action in a nearly flat spacetime, without quantizing gravity, in Chapter 3. Although our focus has been the rank-2 antisymmetric tensor field, the computation method and formalism developed in the previous chapters is quite general in terms of applicability to other models. In this chapter, we take the next step by including graviton loop corrections by quantizing the perturbations about Minkowski background. This time, we shift our focus to a scalar field model inspired by natural inflation. The present treatment can be applied to antisymmetric tensor fields as well, and are part of our future plans once cosmologically relevant models are developed (see Refs. \cite{almeida2019,obata2018,aashish2018c,aashish2019a} for recent developments). 

The organization of this chapter is as follows. In Sec. \ref{sec4a}, we introduce and briefly review the nonminimal natural inflation model. Sec. \ref{sec4b} covers a review of covariant effective action formalism, notations, and the methodology of our calculations. Sec. \ref{sec4c} constitutes a major part of this chapter, detailing the calculations of each contributing term mentioned in Sec. \ref{sec4b}, along with the divergent part, loop integrals, and renormalization. Some past results and their extensions have also been presented. Finally, in Sec. \ref{sec4d}, we derive the effective potential including the finite corrections from the loop integrals, and perform an order-of-magnitude estimation of quantum corrections. The contents of this chapter are part of Refs. \cite{aashish2019c,aashish2020a}

\section{\label{sec4a}Periodic nonminimal natural inflation model}
Natural inflation (NI) was first introduced by Freese \etal \cite{freese1990} as an approach where inflation arises dynamically (or \textit{naturally}) from particle physics models. We consider a recently proposed modification of the NI model, wherein a periodic nonminimal coupling term similar to NI potential is added along with a new parameter, that eventually leads to a better fit with Planck results \cite{ferreira2018}. These phenomenological implications are in no way the only motivation for considering this model in the present work. Rather, it serves as a toy model to achieve our mainly three objectives, which are as follows. First, to set up the computation using symbolic manipulation packages to evaluate one-loop covariant effective action up to quartic order terms in the background field. As a starting point, we work in the Minkowski background. Second, we aim to recover and establish past results. And third, we wish to estimate the magnitudes of quantum gravitational corrections from the finite contributions at least for the effective potential, since there are typically several thousands of terms one has to deal with.

In natural inflation models, a flat potential is effected using pseudo Nambu-Goldstone bosons arising from breaking the continuous shift symmetry of Nambu-Goldstone modes into a discrete shift symmetry. As a result, the inflation potential in a Natural inflation model takes the form,
\begin{eqnarray}
\label{eq01}
V(\phi) = \Lambda^4 \left(1 + \cos(\phi/f)\right);
\end{eqnarray} 
where the magnitude of parameter $\Lambda^4$ and periodicity scale $f$ are model dependent. Majority of natural inflation models are in tension with recent Planck 2018 results \cite{planck2018x}. However, it was shown in Ref. \cite{gerbino2017} that once neutrino properties are more consistently taken into account when analyzing the data, natural inflation does marginally agree with data. 

This work concerns a recently proposed extension of the original natural inflation model introducing a new periodic non-minimal coupling to gravity \cite{ferreira2018}. The authors in \cite{ferreira2018} showed that the new model leads to a better fit with observation data thanks to the introduction of a new parameter in the nonminimal coupling term, with $n_{s}$ and $r$ values well within $95\%$ C.L. region from combined Planck 2018+BAO+BK14 data. An important feature of this model is that $f$ becomes sub-Planckian, contrary to a super-Planckian $f$ in the original natural inflation model \cite{freese1990}, and thus addresses issues related to gravitational instanton corrections \cite{banks2003,rudelius2015a,rudelius2015b,montero2015,hebecker2017}.
 
Our objective here is to study one-loop quantum gravitational corrections to the natural inflation model with non-minimal coupling, using Vilkovisky-DeWitt's covariant effective action approach \cite{parker2009}. One of the first works considering one-loop gravitational corrections were pioneered by Elizalde and Odintsov \cite{elizalde1993,elizalde1994b,elizalde1994c,elizalde1994d}. Vilkovisky-DeWitt method was used to study effective actions in Refs. \cite{odintsov1989,odintsov1990a,odintsov1990b,odintsov1991,odintsov1993}. Unfortunately, although non-covariant effective actions can in principle be evaluated using proper time methods, calculating the covariant effective action exactly even at one-loop is highly nontrivial. Hence, we take a different route by employing a perturbative calculation of one-loop effective action, in orders of the background scalar field. This requires us to apply a couple of approximations. First, we work in the regime where potential is flat, i.e. $\phi \ll f$, which is generally true during slow-rolling inflation. Second, the background metric is set to be Minkowski. This choice is debatable, since it does not accurately represent an inflationary scenario, but has been used before \cite{saltas2017,bounakis2018} as a first step towards studying quantum corrections.

The action for the nonminimal natural inflation in the Einstein frame is given by,
\begin{eqnarray}
\label{eq02}
S = \int d^4 x \sqrt{-g}\left(-\dfrac{2 R}{\kappa^2} + \dfrac{1}{2}K(\phi)\phi {}_{;a} \phi {}^{;a} + \dfrac{V(\phi)}{(\gamma(\phi))^4}\right)
\end{eqnarray}
where, 
\begin{equation}
\label{eq03}
\gamma(\phi)^2 = 1 + \alpha\left(1+\cos\left(\dfrac{\phi}{f}\right)\right),
\end{equation}
and,
\begin{equation}
\label{eq04}
K(\phi) = \dfrac{1 + 24\gamma'(\phi)^2/\kappa^2}{\gamma(\phi)^2}. 
\end{equation}
$V(\phi)$ is as in Eq. (\ref{eq01}). Here, $\phi_{;a} \equiv \nabla_{a}\phi$ denotes the covariant derivative.
In the region where potential is flat, $\phi/f \ll 1$, and we expand all periodic functions in Eq. (\ref{eq02}) up to quartic order in $\phi$ followed by rescaling $\sqrt{k_0}\phi \to \phi$:
\begin{equation}
\label{action}
S \approx \int d^4 x \sqrt{-g} \left(- \frac{2 R}{\kappa^2} + \tfrac{1}{2} \frac{m^2}{k_0} \phi^2 + \tfrac{1}{24} \frac{\lambda}{k_0^2} \phi^4 + \tfrac{1}{2} (1 + \frac{k_1}{k_0^2} \phi^2) \phi {}_{;a} \phi {}^{;a}\right) + \mathcal{O}(\phi^5)
\end{equation}
where parameters $m, \lambda,k_0$ and $k_1$ have been defined out of $\alpha, f$ and $\Lambda^4$ in from Eq. (\ref{eq02}):
\begin{eqnarray}
\label{param}
m^2 &=& \dfrac{\Lambda^4 (2\alpha - 1)}{(1 + 2\alpha)^3 f^2};\nonumber \\
\lambda &=& \dfrac{\Lambda^4 (8\alpha^2 - 12\alpha + 1)}{(1 + 2\alpha)^4 f^4}; \nonumber \\
k_0 &=& \dfrac{1}{1 + 2\alpha}; \nonumber \\
k_1 &=& \dfrac{\alpha(\kappa^2 f^2 + 96\alpha^2 + 48\alpha)}{2\kappa^2 f^4 (1 + 2\alpha)^2}. \nonumber \\
\end{eqnarray}
We have also omitted a constant term appearing in ($\ref{action}$) because such terms are negligibly small in early universe. The action (\ref{action}) is in effect a $\phi^4$ scalar theory with derivative coupling.

\section{\label{sec4b}The one-loop effective action}
A standard procedure while calculating loop corrections in quantum field theory, is to use the well known background field method, according to which a field is split into a classical background and a quantum part that is much smaller in magnitude (and hence treated perturbatively) \cite{falkenberg1998,labus2016,ohta2016}. A by-product of this procedure is the background and gauge dependence of quantum corrections, which is why we use the gauge- and background- independent version of one-loop effective action \cite{dewitt1967b,dewitt1967c},
\begin{eqnarray}
    \label{aeaf4}
    \Gamma[\bar{\varphi}] = -\ln\int [d\zeta] [d \bar{c}^{\alpha}] [d c^{\beta}] \exp\left[-\dfrac{\zeta^{i}\zeta^{j}}{2}\Big(S_{,ij}[\bar{\varphi}] - \Gamma^{k}_{ij}S_{,k}[\bar{\varphi}]\Big) - \frac{1}{4\alpha}f_{\alpha\beta}\chi^{\alpha}\chi^{\beta} - \bar{c}^{\alpha} Q_{\alpha\beta} c^{\beta}\right].
\end{eqnarray}

The evaluation of effective action is carried out as follows. For a theory $S[\varphi]$ with fields $\varphi^{i}$, calculations are performed about a classical background $\bar{\varphi}^{i}$:  $\varphi^{i} = \bar{\varphi}^{i} + \zeta^{i}$, where $\zeta^{i}$ is the quantum part. In our case, $\varphi^{i}=\{g_{\mu\nu}(x),\phi(x)\}$; $\bar{\varphi}^{i}=\{\eta_{\mu\nu},\bar{\phi}(x)\}$ where $\eta_{\mu\nu}$ is the Minkowski metric; and, $\zeta^{i}=\{\kappa h_{\mu\nu}(x),\delta\phi(x)\}$. The fluctuations $\zeta^{i}$ are assumed to be small enough for a perturbative treatment to be valid, viz. $|\kappa h_{\mu\nu}|\ll 1; |\delta\phi|\ll |\phi|$. In this limit, the infinitesimal general coordinate transformations can be treated as gauge transformations associated with $h_{\mu\nu}$ \cite{donoghue1994,donoghue2017}. In fact, for any metric $g_{\mu\nu}(x)$, this infinitesimal transformation takes the form,
\begin{eqnarray}
    \label{aeaf0}
    \delta g_{\mu\nu} = -\delta\epsilon^\lambda g_{\mu\nu,\lambda}-\delta\epsilon^\lambda \,_{,\mu}g_{\lambda\nu}-\delta\epsilon^\lambda \,_{,\nu}g_{\lambda\mu}.
\end{eqnarray}

The computation of Eq. (\ref{aeaf4}) traditionally has involved the use of proper time method, such as employing the heat kernel technique. For Laplace type operators (coefficients of $\zeta^{i}\zeta^{j}$ in the exponential), of the form $g^{\mu\nu}\nabla_{\mu}\nabla_{\nu} + Q$ (where $Q$ does not contain derivatives), the heat kernel coefficients are known and are quite useful because they are independent of dimensionality \cite{dewitt1964}. However, these operators in general are not Laplace type, as in the present case. A class of nonminimal operators such as the one in Eq. (\ref{aeaf4}) can be transformed to minimal (Laplace) form using the generalised Schwinger-DeWitt technique \cite{barvinsky1985}, but in practice the implementation is quite complicated and specific to a given Lagrangian. Examples of such an implementation can be found in Refs. \cite{alvarez2015,steinwachs2011}. We take a different approach here, calculating the one-loop effective action perturbatively in orders of the background field. While one does not obtain exact results in a perturbative approach, unlike the heat kernel approach, it is possible to obtain accurate results up to a certain order in background fields which is of relevance for a theory in, say, the early universe. Some past examples are Refs. \cite{saltas2017,bounakis2018}. Moreover, our implementation of this method using xAct packages for Mathematica \cite{xact,xpert} is fairly general in terms of its applicability to not only scalars coupled with gravity, but also vector and tensor fields (see, for instance, Ref. \cite{aashish2019b}). A caveat at this time, is that the perturbative expansions are performed about the Minkowski background and not a general metric background. As a result, contributions from Ricci curvature corrections do not appear in the current calculations. However, a generalization to include FRW background is part our future plans. 

For convenience, we write the exponential in the first term of $\Gamma$ as,
\begin{eqnarray}
\label{aeaf5}
\exp[\cdots] &=& \exp\left\{- \left(\tilde{S}[\bar{\varphi}^{0}]+\tilde{S}[\bar{\varphi}^{1}]+\tilde{S}[\bar{\varphi}^{2}]+\tilde{S}[\bar{\varphi}^{3}]+\tilde{S}[\bar{\varphi}^{4}]\right)\right\}\nonumber \\
&\equiv & \exp\left\{- (\tilde{S}_{0}+\tilde{S}_{1}+\tilde{S}_{2}+\tilde{S}_{3}+\tilde{S}_{4})\right\}
\end{eqnarray}
$\tilde{S}_0$ yields the propagator for each of the fields $\zeta^{i}$. The rest of the terms are contributions from interaction terms, which we assume to be small. Treating $\tilde{S}_{1},...,\tilde{S}_{4}$ as perturbative, and expanding Eq. (\ref{aeaf5}), $\Gamma[\bar{\varphi}]$ can be written as,
\begin{eqnarray}
    \label{aeaf6}
    \Gamma[\bar{\varphi}] &=& -\ln\int [d\zeta] [d \bar{c}^{\alpha}] [d c^{\beta}] e^{-S_0}(1-\delta S + \dfrac{\delta S^2}{2} + \cdots) ; \nonumber \\
    &=& - \ln (1 - \langle\delta S\rangle + \dfrac{1}{2}\langle\delta S^2 \rangle + \cdots); 
\end{eqnarray}
where $\delta S = \sum_{i=1}^4 \tilde{S}_{i}$ and $\langle\cdot\rangle$ represents the expectation value in the path integral formulation. Finally, we use $\ln(1 + x)\approx x$ to find the contributions to $\Gamma$ at each order of background field. We only use the leading term in the logarithmic expansion, since all higher order terms will yield contributions from disconnected diagrams viz-a-viz $\langle\delta S\rangle^{2},$ etc which we ignore throughout our calculation. Moreover, since we are interested in terms up to quartic order in background field, we truncate the Taylor series in Eq. (\ref{aeaf6}) up to $\delta S^{4}$. With these considerations, the final contributions to $\Gamma$ at each order of $\bar{\varphi}$ is:
\begin{eqnarray}
\label{aeaf7}
\mathcal{O}(\bar{\varphi}) &:& \langle\tilde{S}_{1}\rangle;\nonumber \\
\mathcal{O}(\bar{\varphi}^2) &:& \langle\tilde{S}_{2}\rangle - \dfrac{1}{2}\langle\tilde{S}_{1}^{2}\rangle; \nonumber \\
\mathcal{O}(\bar{\varphi}^3) &:& \langle\tilde{S}_{3}\rangle - \langle\tilde{S}_{1}\tilde{S}_{2}\rangle + \dfrac{1}{6} \langle\tilde{S}_{1}^{3}\rangle;\nonumber \\
\mathcal{O}(\bar{\varphi}^4) &:& \langle\tilde{S}_{4}\rangle - \langle\tilde{S}_{1}\tilde{S}_{3}\rangle + \dfrac{1}{2}\langle\tilde{S}_{1}^{2}\tilde{S}_{2}\rangle -\dfrac{1}{2}\langle\tilde{S}_{2}^{2}\rangle - \dfrac{1}{24}\langle\tilde{S}_{1}^{4}\rangle .
\end{eqnarray}
Also, we recall that the metric fluctuations have a factor of $\kappa$. Accordingly, the terms in Eq. (\ref{aeaf7}) will also contain powers of $\kappa$. It turns out, as will be shown below, that all contributions are at most of the order $\kappa^4$. Expecting $\mathcal{O}(\kappa^4)$ terms to be significantly suppressed, we only take into account the corrections up to $\mathcal{O}(\kappa^2)$. In what follows, we will detail the evaluation of terms in Eq. (\ref{aeaf7}).


\section{\label{sec4c}Covariant one-loop corrections}
\subsection{Setup}
The first step towards writing $\Gamma[\bar{\varphi}]$ in Eq. (\ref{aeaf4}) is to identify the field space metric, given in terms of the field-space line element, 
\begin{align}
    ds^{2} &= G_{ij}d\varphi^{i} d\varphi^{j} \\ 
    \label{acqc0}
    &= \int d^{4}x d^{4}x' \left(G_{g_{\mu\nu}(x)g_{\rho\sigma}(x')}dg_{\mu\nu}(x) dg_{\rho\sigma}(x') + G_{\phi(x)\phi(x')} d\phi(x) d\phi(x')\right).
\end{align}
A prescription for identifying field space metric is to read off the components of $G_{ij}$ from the coefficients of highest derivative terms in classical action functional \cite{vilkovisky1984b}. For the scalar field $\phi(x)$, the field-space metric is chosen to be,
\begin{eqnarray}
    \label{acqc1}
    G_{\phi(x)\phi(x')}=\sqrt{g(x)}\delta(x,x');
\end{eqnarray}
For the metric $g_{\mu\nu}(x)$, a standard choice for field-space metric is \cite{parker2009,mackay2010}
\begin{align}
    \label{acqc2}
    G_{g_{\mu\nu}(x)g_{\rho\sigma}(x')}=\dfrac{\sqrt{g(x)}}{\kappa^2}\left\{ g^{\mu(\rho}(x)g^{\sigma)\nu}(x)- \frac{1}{2}g^{\mu\nu}(x)g^{\rho\sigma}(x) \right\}\delta(x,x'),
\end{align}
where the brackets around tensor indices in the first term indicate symmetrization. As a convention, we choose to include $\kappa^2$ factor in Eq. (\ref{acqc2}) to account for dimensionality of the length element in Eq. (\ref{acqc0}), although choosing otherwise is also equally valid as long as dimensionality is taken care of. The inverse metric can be derived from the identity $G_{ij}G^{jk}=\delta^{k}_{i}$:
\begin{eqnarray}
    \label{acqc3}
    G^{g_{\mu\nu}(x)g_{\rho\sigma}(x')}&=&\kappa^2\left\{ g_{\mu(\rho}(x)g_{\sigma)\nu}(x)- \frac{1}{2}g_{\mu\nu}(x)g_{\rho\sigma}(x) \right\}\delta(x,x'); \\
    \label{acqc4}
    G^{\phi(x)\phi(x')}&=&\delta(x,x').
\end{eqnarray}
Next, using Eqs. (\ref{acqc1})-(\ref{acqc4}), one can find the Vilkovisky-DeWitt connections $\Gamma^{k}_{ij}$ which has an identical definition to the Christoffel connections thanks to the Landau-DeWitt gauge choice. Out of a total of six possibilities there are three non-zero connections obtained as follows:
\begin{align}
    \label{acqc5}
    \Gamma^{g_{\lambda\tau}(x)}_{g_{\mu\nu}(x')g_{\rho\sigma}(x'')} &= \delta(x'',x')\delta(x'',x)\left[-\delta^{(\mu}_{(\lambda}g^{\nu)(\rho}(x)\delta^{\sigma)}_{\tau)} + \dfrac{1}{4}g^{\mu\nu}(x)\delta^{\rho}_{(\lambda}\delta^{\sigma}_{\tau)} + \dfrac{1}{4}g^{\rho\sigma}(x)\delta^{\mu}_{(\lambda}\delta^{\nu}_{\tau)} \right. \nonumber \\ & \left. + \dfrac{1}{4}g_{\lambda\tau}(x)g^{\mu(\rho}(x)g^{\sigma)\nu}(x) - \dfrac{1}{8} g_{\lambda\tau}(x)g^{\mu\nu}(x)g^{\rho\sigma}(x) \right] \\
    \Gamma^{g_{\lambda\tau}(x)}_{\phi(x')\phi(x'')} &= \dfrac{\kappa^2}{4}\delta(x'',x')\delta(x'',x) g_{\lambda\tau}(x) \\
    \Gamma^{\phi(x)}_{\phi(x') g_{\lambda\tau}(x'')} &= \dfrac{1}{4}\delta(x'',x')\delta(x'',x) g^{\lambda\tau}(x) = \Gamma^{\phi(x)}_{g_{\lambda\tau}(x') \phi(x'')}.
\end{align}
Note that upon substituting into Eq. (\ref{aeaf4}), all calculations here are evaluated at the background field(s) which in our case is the Minkowski metric and a scalar field $\bar{\phi}(x)$. We also recall that this rather unrestricted choice of background is allowed because of the DeWitt connections that ensure gauge and background independence. As alluded to earlier, the Landau-DeWitt gauge condition, $K_{\alpha i}[\bar{\varphi}]\zeta^{i} = 0$, is given in terms of the gauge generators $K_{\alpha i}$. Since there is only one set of transformations vi-a-viz general coordinate transformation, there exists one gauge parameter which we call $\xi^{\lambda}(x)$. In the condensed notation, this corresponds to $\delta\epsilon^{\alpha}$ where $\alpha\to (\lambda,x)$. Gauge generator on the gravity side $K^{g_{\mu\nu}}_{\lambda}(x,x')$ is read off from Eq. (\ref{aeaf0}), while $K^{\phi}_{\lambda}(x,x')$ is read off from the transformation of $\phi$:
\begin{eqnarray}
    \label{acqc6}
    \delta_{\xi}\phi = -\partial_{\mu}\phi \xi^{\lambda} .
\end{eqnarray}
Substituting in the definition of $\chi_{\alpha}[\bar{\varphi}]$ in coordinate space, we obtain
\begin{eqnarray}
    \label{acqc7}
    \chi_{\lambda}[\bar{\phi}] &=& \int d^{4}x' \left( K_{g_{\mu\nu} \lambda}(x,x') \kappa h_{\mu\nu}(x') + K_{\phi \lambda}(x,x')\delta\phi(x') \right) \nonumber \\
    &=& \dfrac{2}{\kappa}\left(\partial^{\mu}h_{\mu\lambda} - \dfrac{1}{2}\partial_{\lambda}h\right) - \omega\partial_{\lambda}\bar{\phi}\delta\phi .
\end{eqnarray}
where $\omega$ is a bookkeeping parameter, which we adopt from Ref. \cite{mackay2010}; a second such parameter $\nu$ (not to be confused with the tensor index), appears with all Vilkovisky-DeWitt connection terms to keep track of gauge (non-)invariance. That is, we write $S_{;ij}=S_{,ij} - \nu \Gamma^{k}_{ij}S_{,k}$. As shown later, playing with these parameters reproduces past non-gauge-invariant results. Here and throughout, the indices of field-space quantities like the gauge generator are raised and lowered using field-space metric defined in Eqs. (\ref{acqc1}) - (\ref{acqc4}). Lastly, we choose $f^{\alpha\beta} = \kappa^{2}\eta^{\lambda \lambda'}\delta(x,x')$ in Eq. (\ref{aeaf4}) to determine the gauge fixing term. One last piece needed before background-field-order expansions, the ghost term $Q_{\alpha\beta}$. We use the definition \cite{parker2009}, $Q_{\alpha\beta} \equiv \chi_{\alpha, i}K^{i}_{\beta}$, to obtain
\begin{eqnarray}
    \label{acqc8}
    Q_{\mu\nu} = \left(-\dfrac{2}{\kappa} \eta_{\mu\nu}\partial_{\alpha}\partial^{\alpha} + \omega\partial_{\mu}\bar{\phi}\partial_{\nu}\bar{\phi}\right)\delta(x,x').
\end{eqnarray}

\subsection{\label{loopint}Loop integrals and divergent parts}
Substituting the connections, the gauge fixing term and the ghost term along with action (\ref{action}) in Eq. (\ref{aeaf4}), and employing the notations in Eq. (\ref{aeaf5}), we obtain:
\begin{eqnarray}
    \label{s0}
    \tilde{S}_{0} &=& \intx{x} \Big[\frac{m^2 (\delta \phi)^2}{2 k_0} + \tfrac{1}{2} \delta \phi{}_{,a} \delta \phi{}^{,a} - 2 h^{ab} h^{c}{}_{c}{}_{,a}{}_{,b} -  \frac{2 \bar{c}^{a} c_{a}{}^{,b}{}_{,b}}{\kappa} + h^{ab} h^{}{}_{a}{}^{c}{}_{,b}{}_{,c} -  \frac{h^{ab} h^{}{}_{a}{}^{c}{}_{,b}{}_{,c}}{\alpha} \nonumber \\ 
&& + h^{a}{}_{a} h^{bc}{}_{,b}{}_{,c} + \frac{h^{a}{}_{a} h^{bc}{}_{,b}{}_{,c}}{\alpha}  -  \tfrac{1}{2} h^{ab} h^{}{}_{ab}{}^{,c}{}_{,c} + \tfrac{1}{2} h^{a}{}_{a} h^{b}{}_{b}{}^{,c}{}_{,c} -  \frac{h^{a}{}_{a} h^{b}{}_{b}{}^{,c}{}_{,c}}{4 \alpha} \Big]; \\
\label{s1}
\tilde{S}_{1} &=& \intx{x} \Bigl[\frac{m^2 \kappa \delta \phi h^{a}{}_{a} \bar{\phi}}{2 k_0} -  \frac{m^2 \kappa \nu \delta \phi h^{a}{}_{a} \bar{\phi}}{4 k_0} -  \tfrac{1}{2} \kappa \delta \phi h^{b}{}_{b} \bar{\phi} {}^{,a}{}_{,a} + \tfrac{1}{4} \kappa \nu \delta \phi h^{b}{}_{b} \bar{\phi} {}^{,a}{}_{,a} -  \tfrac{1}{2} \kappa \delta \phi h^{b}{}_{b}{}_{,a} \bar{\phi} {}^{,a} \nonumber \\ 
&& + \frac{\kappa \omega \delta \phi h^{b}{}_{b}{}_{,a} \bar{\phi} {}^{,a}}{2 \alpha} + \kappa \delta \phi \bar{\phi} {}^{,a} h^{}{}_{a}{}^{b}{}_{,b} -  \frac{\kappa \omega \delta \phi \bar{\phi} {}^{,a} h^{}{}_{a}{}^{b}{}_{,b}}{\alpha} + \kappa \delta \phi h^{}{}_{ab} \bar{\phi} {}^{,a}{}^{,b}\Bigr] ; \\
\label{s2}
\tilde{S}_{2} &=& \intx{x} \Bigl[ -\frac{m^2 \kappa^2 h^{}{}_{ab} h^{ab} \bar{\phi}^2}{8 k_0} + \frac{m^2 \kappa^2 h^{a}{}_{a} h^{b}{}_{b} \bar{\phi}^2}{16 k_0} + \frac{\lambda \bar{\phi}^2 (\delta \phi)^2}{4 k_0^2} -  \frac{m^2 \kappa^2 \nu \bar{\phi}^2 (\delta \phi)^2}{8 k_0} \nonumber \\ 
&& + \frac{k_1 \bar{\phi}^2 \delta \phi{}_{,a} \delta \phi{}^{,a}}{2 k_0^2} + \frac{2 k_1 \delta \phi \bar{\phi} \bar{\phi} {}_{,a} \delta \phi{}^{,a}}{k_0^2} -  \tfrac{1}{8} \kappa^2 h^{}{}_{bc} h^{bc} \bar{\phi} {}_{,a} \bar{\phi} {}^{,a} + \tfrac{1}{16} \kappa^2 \nu h^{}{}_{bc} h^{bc} \bar{\phi} {}_{,a} \bar{\phi} {}^{,a} \nonumber \\ 
&& + \tfrac{1}{16} \kappa^2 h^{b}{}_{b} h^{c}{}_{c} \bar{\phi} {}_{,a} \bar{\phi} {}^{,a} -  \tfrac{1}{32} \kappa^2 \nu h^{b}{}_{b} h^{c}{}_{c} \bar{\phi} {}_{,a} \bar{\phi} {}^{,a} + \frac{k_1 (\delta \phi)^2 \bar{\phi} {}_{,a} \bar{\phi} {}^{,a}}{2 k_0^2} -  \tfrac{1}{16} \kappa^2 \nu (\delta \phi)^2 \bar{\phi} {}_{,a} \bar{\phi} {}^{,a} \nonumber \\ 
&& + \frac{\kappa^2 \omega^2 (\delta \phi)^2 \bar{\phi} {}_{,a} \bar{\phi} {}^{,a}}{4 \alpha} + \omega c^{a} \bar{c}^{b} \bar{\phi} {}_{,a} \bar{\phi} {}_{,b} + \tfrac{1}{2} \kappa^2 h^{}{}_{a}{}^{c} h^{}{}_{bc} \bar{\phi} {}^{,a} \bar{\phi} {}^{,b} -  \tfrac{1}{4} \kappa^2 \nu h^{}{}_{a}{}^{c} h^{}{}_{bc} \bar{\phi} {}^{,a} \bar{\phi} {}^{,b} \nonumber \\ 
&& -  \tfrac{1}{4} \kappa^2 h^{}{}_{ab} h^{c}{}_{c} \bar{\phi} {}^{,a} \bar{\phi} {}^{,b} + \tfrac{1}{8} \kappa^2 \nu h^{}{}_{ab} h^{c}{}_{c} \bar{\phi} {}^{,a} \bar{\phi} {}^{,b}\Bigr]; \\
\label{s3}
    \tilde{S}_{3} &=& \intx{x} \Bigl[\frac{\kappa \lambda \delta \phi h^{a}{}_{a} \bar{\phi}^3}{12 k_0^2} -  \frac{\kappa \lambda \nu \delta \phi h^{a}{}_{a} \bar{\phi}^3}{24 k_0^2} + \frac{k_1 \kappa \nu \delta \phi h^{b}{}_{b} \bar{\phi}^2 \bar{\phi}{}^{,a}{}_{,a}}{4 k_0^2} + \frac{k_1 \kappa h^{b}{}_{b} \bar{\phi}^2 \bar{\phi}{}_{,a} \delta \phi{}^{,a}}{2 k_0^2} \nonumber \\ 
    && + \frac{k_1 \kappa \delta \phi h^{b}{}_{b} \bar{\phi} \bar{\phi}{}_{,a} \bar{\phi}{}^{,a}}{2 k_0^2} + \frac{k_1 \kappa \nu \delta \phi h^{b}{}_{b} \bar{\phi} \bar{\phi}{}_{,a} \bar{\phi}{}^{,a}}{4 k_0^2} -  \frac{k_1 \kappa h^{}{}_{ab} \bar{\phi}^2 \delta \phi{}^{,a} \bar{\phi}{}^{,b}}{k_0^2} -  \frac{k_1 \kappa \delta \phi h^{}{}_{ab} \bar{\phi} \bar{\phi}{}^{,a} \bar{\phi}{}^{,b}}{k_0^2}\Bigr] ; \\
\label{s4}
\tilde{S}_{4} &=& \intx{x} \Biggl[- \frac{\kappa^2 \lambda h^{}{}_{ab} h^{ab} \bar{\phi}^4}{96 k_0^2} + \frac{\kappa^2 \lambda h^{a}{}_{a} h^{b}{}_{b} \bar{\phi}^4}{192 k_0^2} -  \frac{\kappa^2 \lambda \nu \bar{\phi}^4 (\delta \phi)^2}{96 k_0^2} -  \frac{k_1 \kappa^2 h^{}{}_{bc} h^{bc} \bar{\phi}^2 \bar{\phi}{}_{,a} \bar{\phi}{}^{,a}}{8 k_0^2} \nonumber \\ 
&& + \frac{k_1 \kappa^2 \nu h^{}{}_{bc} h^{bc} \bar{\phi}^2 \bar{\phi}{}_{,a} \bar{\phi}{}^{,a}}{16 k_0^2} + \frac{k_1 \kappa^2 h^{b}{}_{b} h^{c}{}_{c} \bar{\phi}^2 \bar{\phi}{}_{,a} \bar{\phi}{}^{,a}}{16 k_0^2} -  \frac{k_1 \kappa^2 \nu h^{b}{}_{b} h^{c}{}_{c} \bar{\phi}^2 \bar{\phi}{}_{,a} \bar{\phi}{}^{,a}}{32 k_0^2}  \nonumber \\ 
&& -  \frac{k_1 \kappa^2 \nu \bar{\phi}^2 (\delta \phi)^2 \bar{\phi}{}_{,a} \bar{\phi}{}^{,a}}{16 k_0^2} + \frac{k_1 \kappa^2 h^{}{}_{a}{}^{c} h^{}{}_{bc} \bar{\phi}^2 \bar{\phi}{}^{,a} \bar{\phi}{}^{,b}}{2 k_0^2} -  \frac{k_1 \kappa^2 \nu h^{}{}_{a}{}^{c} h^{}{}_{bc} \bar{\phi}^2 \bar{\phi}{}^{,a} \bar{\phi}{}^{,b}}{4 k_0^2} \nonumber \\
&& -  \frac{k_1 \kappa^2 h^{}{}_{ab} h^{c}{}_{c} \bar{\phi}^2 \bar{\phi}{}^{,a} \bar{\phi}{}^{,b}}{4 k_0^2} + \frac{k_1 \kappa^2 \nu h^{}{}_{ab} h^{c}{}_{c} \bar{\phi}^2 \bar{\phi}{}^{,a} \bar{\phi}{}^{,b}}{8 k_0^2}\Biggr];
\end{eqnarray}
Here, the indices $(a,b,c,\dots)$ and $(\mu,\nu,\rho,\dots)$ are used interchangeably to denote the gauge indices. 
$\tilde{S}_0$ leads to the well known free theory propagators for gravity and massive scalar field and the ghost field respectively,
\begin{eqnarray}
    \label{prop}
    D(x,x') &=&\intp{k}e^{ik\cdot(x-x')}D(k) = \langle\delta\phi(x)\delta\phi(x')\rangle; \nonumber \\
    D_{\alpha\beta\mu\nu}(x,x') & =& \intp{k} e^{ik\cdot(x-x')}D_{\alpha\beta\mu\nu}(k) = \langle h_{\alpha\beta}(x)h_{\mu\nu}(x')\rangle ; \\
    D^{G}_{\mu\nu}(x,x') &=& \intp{k}e^{ik\cdot(x-x')}D^{G}_{\mu\nu}(k) = \langle\bar{c}_{\mu}(x)c_{\nu}(x')\rangle \nonumber ;
\end{eqnarray}
where,
\begin{eqnarray}
    \label{props}
    D(k) &=& \frac{1}{k^2+\frac{m^2}{k_0}}; \\
    \label{propg}
    D_{\alpha\beta\mu\nu}(k) &=&\frac{\delta_{\alpha\mu}\delta_{\beta\nu}+\delta_{\alpha\nu}\delta_{\beta\mu} -\delta_{\alpha\beta}\delta_{\mu\nu}}{2 k^2} + \nonumber \\ && (\alpha-1)\frac{\delta_{\alpha\mu}k_\beta k_\nu+\delta_{\alpha\nu}k_\beta k_\mu+\delta_{\beta\mu}k_\alpha k_\nu+\delta_{\beta\nu}k_\alpha k_\mu}{2k^4}; \\
    \label{propgh}
    D^{G}_{\mu\nu}(k) &=& \dfrac{1}{k^2}\delta_{\mu\nu}.
\end{eqnarray}
Looking at the structure of rest of the terms $\tilde{S}_i$, it is straightforward to conclude that all terms with odd combinations of $h_{\mu\nu}(x)$ and $\delta\phi(x)$ appearing in Eqs. (\ref{aeaf7}) will not contribute to the effective action, since $\langle h_{\alpha\beta}(x)\delta\phi(x')\rangle = 0$. Therefore, $\langle \tilde{S}_{1}\rangle = 0$ and there is no contribution at $\mathcal{O}(\bar{\phi})$ to the one-loop effective action. Similarly, $\langle \tilde{S}_{3}\rangle = \langle \tilde{S}_{1}\tilde{S}_{2}\rangle = \langle \tilde{S}_{1}^3\rangle = 0$, and hence at $\mathcal{O}(\bar{\phi}^{3})$ too, there is no contribution to the effective action. Hence, the only non-zero contributions in Eq. (\ref{aeaf7}) come at $\mathcal{O}(\bar{\phi}^{2})$ and $\mathcal{O}(\bar{\phi}^{4})$. In the latter, we ignore $\langle S_{1}^{4}\rangle$ terms since they are relevant at $\mathcal{O}(\kappa^4)$ and above while we are interested in terms up to $\kappa^2$ order. Expectation value of $\tilde{S}_{i}$ consists of local terms, and thus describes contributions from tadpole diagrams. 

The ghost term appears only in $\tilde{S}_{2}$. However, it can be shown that at $\mathcal{O}(\bar{\phi}^{2})$ it yields no nontrivial contributions, and as a result, has usually been ignored in past literature where only quadratic order corrections were considered \cite{mackay2010,saltas2017,bounakis2018}. Consider the ghost propagator (\ref{propgh}). Because there is no physical scale involved, the term containing ghost in (\ref{s2}) yields,
\begin{eqnarray}
    \label{loop00}
    \left\langle \intx{x}\omega c^{a} \bar{c}^{b} \bar{\phi} {}_{,a} \bar{\phi} {}_{,b} \right\rangle &=& \intx{x}\omega\bar{\phi} {}_{,a} \bar{\phi} {}_{,b} \langle c^{a} \bar{c}^{b} \rangle \nonumber \\
    &=& \intx{x}\omega\bar{\phi} {}_{,a} \bar{\phi} {}_{,b} \intp{k}\delta^{ab}\dfrac{1}{k^2},
\end{eqnarray}
which in four dimensions gives no physical result. The only nontrivial ghost contribution comes at quartic order in background field. 

Eventually, finding the one-loop corrections then boils down to evaluating up to $\kappa^{2}$ order, the quadratic and quartic order corrections from the following:
\begin{equation}
    \label{loop01}
    \Gamma = \langle\tilde{S}_{2}\rangle - \dfrac{1}{2}\langle\tilde{S}_{1}^{2}\rangle + \langle\tilde{S}_{4}\rangle - \langle\tilde{S}_{1}\tilde{S}_{3}\rangle + \dfrac{1}{2}\langle\tilde{S}_{1}^{2}\tilde{S}_{2}\rangle -\dfrac{1}{2}\langle\tilde{S}_{2}^{2}\rangle .
\end{equation}
In principle, solving Eq. (\ref{loop01}) broadly consists of two steps: (i) writing each term in terms of the Fourier space integral(s) of Green's functions found in Eqs. (\ref{props}) - (\ref{propgh}); and (ii) solving the resulting loop integrals. In this section, we restrict ourselves to writing just the divergent part of effective action, since there are already several thousand terms to deal with and writing their finite parts would introduce unnecessary complexity. We do consider finite part in the subsequent section, where we evaluate the effective potential after assuming all derivatives of background fields to be zero. 

\subsubsection{Calculating $\langle \tilde{S}_{i} \rangle$}
We first deal with $\langle\tilde{S}_{2}\rangle$ and $\langle\tilde{S}_{4}\rangle$. For convenience, we do not explicitly write the tensor indices of correlators, fields and their coefficients. First, the derivatives of field fluctuations are transformed to momentum space:
\begin{eqnarray}
\label{loop02}
\int d^{4}x A(x) \langle\partial^{m}\delta(x) \ \partial^{n}\delta(x')\rangle \longrightarrow \int d^{4}x \dfrac{d^{4}p}{(2\pi^{4})} A(x) (-ip)^{m} (ip)^{n} \langle \delta_{p}(x)\delta_{p}(x')\rangle,
\end{eqnarray}
where, $\delta (x)$ and $A(x)$ represent the field fluctuations and coefficients respectively. $\delta(x)$ here represents any of the fields ($\delta\phi(x),h_{\mu\nu}(x),c_{\mu}(x),\bar{c}_{\mu}(x)$), and is not to be confused with the Dirac delta function $\delta(x,x')$. $\langle \delta_{p}\delta_{p}\rangle$ represents the propagator(s) in momentum space. Then, $\langle \delta_{p}\delta_{p}\rangle$ is replaced with values of Green's function to obtain the loop integrals. For solving integrals here, we primarily use the results in Ref. \cite{bardin1999} to evaluate the divergent terms in dimensional regularization, except for some higher rank two-point integrals that appear below, which we solve by hand using well known prescriptions \cite{romao2019,george1975}. There are three types of loop integrals coming from Eqs. (\ref{s2}) and (\ref{s4}):
\begin{eqnarray}
\label{loop03}
\int d^{4}x \dfrac{d^{4}p}{(2\pi^{4})} A(x)\dfrac{1}{p^{2}} ; \quad \int d^{4}x \dfrac{d^{4}p}{(2\pi^{4})} A(x)\dfrac{p^{\mu}p^{\nu}}{p^{4}}; \quad  \int d^{4}x \dfrac{d^{4}p}{(2\pi^{4})} A(x)\dfrac{1}{p^{2}+\frac{m^2}{k_{0}}}.
\end{eqnarray}
The first two integrals are pole-less, and vanish due to the lack of a physical scale \cite{bardin1999}. The third integral is straightforward and contributes to the divergent part. 
See appendix \ref{AppendixC} for values of all integrals appearing here, including finite parts for some integrals used in the next section.

\subsubsection{Calculating $\langle \tilde{S}_{i}\tilde{S}_{j} \rangle$}
$\langle \tilde{S}_{1}\tilde{S}_{1} \rangle$ and $\langle \tilde{S}_{1}\tilde{S}_{3} \rangle$ contain terms of the form,
\begin{eqnarray}
    \label{loop04}
    & \intx{x}\intx{x'} A(x)B(x')\langle\delta\phi(x)\delta\phi(x')\rangle \langle h(x)h(x')\rangle \\
    =&\intx{x}\intp{k}\frac{d^4 k'}{2\pi^4}\frac{d^4 k''}{2\pi^4} A(x)\tilde{B}(k'') D_{\phi\phi}(k)D_{hh}(k') e^{-i(k+k')\cdot x}\delta^{(4)}(k+k'-k'') \nonumber \\
    \label{loop05}
    =& \intx{x}A(x)\intp{k}e^{-ik\cdot x}\tilde{B}(k) \intp{k'} D_{\phi\phi}(k-k')D_{hh}(k')
\end{eqnarray}
where $A(x), B(x')$ are classical coefficients, and $D_{\phi\phi}, D_{hh}$ are scalar and gravity propagators respectively; $\tilde{B}(k)$ is the Fourier transform of $B(x')$. There are also the derivatives of Eq. (\ref{loop04}) present, and are dealt with in a way similar to Eq. (\ref{loop02}), leading to factors of $k'^{\mu}$ in the loop integrals. Consequently, we encounter three types of loop integrals:
\begin{eqnarray}
    \label{loop06}
    \intp{k'} \dfrac{k'^{a}\dots k'^{b}}{(k'-k)^2 + \frac{m^2}{k_0}}; \quad \intp{k'} \dfrac{k'^{a}\dots k'^{b}}{k'^{2} ((k'-k)^2 + \frac{m^2}{k_0})}; \quad \intp{k'} \dfrac{k'^{a}\dots k'^{b}}{k'^{4} ((k'-k)^2 + \frac{m^2}{k_0})};
\end{eqnarray}
which constitute standard one-, two- and three-point $n$-rank integrals ($n=0,1,2$). 

Likewise, $\langle \tilde{S}_{2}\tilde{S}_{2} \rangle$ yields $4-$point correlators given by,
\begin{eqnarray}
    \label{loop07}
    \langle\delta\phi(x)\delta\phi(x)\delta\phi(x')\delta\phi(x')\rangle; \ \langle\delta\phi(x)\delta\phi(x)h(x')h(x')\rangle; \ \langle\delta\phi(x)\delta\phi(x)\bar{c}(x') c(x')\rangle; \nonumber \\
    \langle h(x) h(x)\bar{c}(x') c(x')\rangle; \ \langle \bar{c}(x) c(x) \bar{c}(x') c(x')\rangle; \ \langle h(x) h(x) h(x') h(x')\rangle
\end{eqnarray}
The second, third and fourth terms in (\ref{loop07}) are of the form $\langle \delta(x)\delta(x)\delta'(x')\delta'(x')\rangle$ (again, $\delta(x),\delta(x')$ denote the fields), thereby corresponding to disconnected tadpoles and hence do not give any meaningful contribution. The rest of $4-$point correlators in Eq. (\ref{loop07}) are resolved into $2-$point functions using Wick theorem \cite{peskin1995,schwartz2013}. Fortunately, the last term involving only graviton propagators can be ignored since it only contains $\mathcal{O}(\kappa^4)$ terms. Moreover, $\langle \bar{c}(x)\bar{c}(x')\rangle = \langle c(x)c(x')\rangle = 0$. Therefore, after applying Wick theorem, the final contribution in Eq. (\ref{loop07}) comes from,
\begin{eqnarray}
    \label{loop08a}
    \langle\delta\phi(x)\delta\phi(x)\delta\phi(x')\delta\phi(x')\rangle &=& \langle\delta\phi(x)\delta\phi(x')\rangle\langle\delta\phi(x)\delta\phi(x')\rangle + \langle\delta\phi(x)\delta\phi(x')\rangle\langle\delta\phi(x)\delta\phi(x')\rangle; \\
    \label{loop08b}
    \langle \bar{c}(x) c(x) \bar{c}(x') c(x')\rangle &=& \langle \bar{c}(x) c(x')\rangle \langle \bar{c}(x) c(x')\rangle.
\end{eqnarray}
Using Eq. (\ref{loop08a}) and (\ref{loop08b}) in $\langle \tilde{S}_{2}\tilde{S}_{2} \rangle$ along with Eqs. (\ref{props})-(\ref{propgh}), and Fourier transforming according to Eq. (\ref{loop02}) gives rise to up to rank-4 two-point integrals:
\begin{eqnarray}
    \label{loop09}
    \intp{k'} \dfrac{k'^{a}\dots k'^{b}}{(k'^{2}+\frac{m^2}{k_0}) ((k'-k)^2 + \frac{m^2}{k_0})};
\end{eqnarray}

\subsubsection{Calculating $\langle \tilde{S}_{i}\tilde{S}_{j}\tilde{S}_{k} \rangle$}
The last term to be evaluated is $\langle \tilde{S}_{1}\tilde{S}_{1}\tilde{S}_{2} \rangle$. It consists of six-point correlators given by,
\begin{eqnarray}
    \label{loop10}
    &\langle h(x)h(x'')\delta\phi(x)\delta\phi(x'')\bar{c}(x') c(x') \rangle ; \langle h(x)h(x'')\delta\phi(x)\delta\phi(x'')\delta\phi(x')\delta\phi(x') \rangle ; & \nonumber \\ 
    &\langle h(x)h(x'')\delta\phi(x)\delta\phi(x'') h(x')h(x') \rangle &
\end{eqnarray}
Again, the last term can be ignored since it has no terms up to $\mathcal{O}(\kappa^2)$. And the first term can be written as $\langle h(x)h(x'')\delta\phi(x)\delta\phi(x'')\rangle\langle\bar{c}(x') c(x') \rangle$, which implies disconnected diagrams and thus can also be ignored. So, ghost terms only end up in $\langle \tilde{S}_{2}\tilde{S}_{2} \rangle$. Hence, only the second term needs to be evaluated, which after applying Wick theorem similar to Eq. (\ref{loop08a}) turns out to be,
\begin{eqnarray}
    \label{loop11}
    \langle h(x)h(x'')\delta\phi(x)\delta\phi(x'')\delta\phi(x')\delta\phi(x') \rangle = & \langle h(x)h(x'')\rangle\langle\delta\phi(x)\delta\phi(x')\rangle\langle\delta\phi(x')\delta\phi(x'') \rangle \nonumber \\
    & + \langle h(x)h(x'')\rangle\langle\delta\phi(x)\delta\phi(x')\rangle\langle\delta\phi(x')\delta\phi(x'') \rangle ,
\end{eqnarray}
A typical scalar integral in $\langle \tilde{S}_{1}\tilde{S}_{1}\tilde{S}_{2} \rangle$ takes the form,
\begin{eqnarray}
    \label{loop12}
    &&\intx{x}\intx{x'}\intx{x''}\intp{k}\intp{k'}\intp{k''} A(x)B(x')C(x'')\times \nonumber \\ &&
    e^{-ik\cdot(x'-x)}e^{-ik''\cdot(x-x'')}e^{-ik'\cdot (x''-x')}D_{\phi\phi}(k)D_{\phi\phi}(k')D_{hh}(k'') \nonumber\\
   & =& \intx{x}\intp{p}\intp{k} A(x)\tilde{B}(p)e^{-ip\cdot x} \tilde{C}(k)e^{-ik\cdot x}\intp{k'} \times \nonumber \\ && D_{\phi\phi}(k'-p-k) D_{\phi\phi}(k'-k)D_{hh}(k'),
\end{eqnarray}
resulting in scalar and tensor two-, three- and four-point integrals:
\begin{eqnarray}
    \label{s112}
    \intp{k'}\dfrac{k'^{a}\dots k'^{b}}{d_{0}d_{1}d_{2}d_{3}}; \intp{k'}\dfrac{k'^{a}\dots k'^{b}}{d_{0}d_{1}d_{2}}; \intp{k'}\dfrac{k'^{a}\dots k'^{b}}{d_{0}d_{1}}.
\end{eqnarray}
where,
\begin{eqnarray}
    \label{loop13}
    d_{0} = (k'-k)^2 + \frac{m^2}{k_0}; \ d_{1} = (k'-k-p)^{2} + \frac{m^2}{k_0}; \ d_{2} = d_{3} = k'^2.
\end{eqnarray}
There are up to rank-3 four-point integrals in $\langle \tilde{S}_{1}\tilde{S}_{1}\tilde{S}_{2} \rangle$, and hence have no divergent part \cite{bardin1999}.

\subsubsection{Divergent part}
In total, there are several thousand terms that eventually add up to give the divergent part of Eq. (\ref{loop01}). After solving all the above integrals and extracting their divergent parts using dimensional regularization, we end up with Fourier transforms $\tilde{B}(k)$ (and $\tilde{C}(p)$ in case of six-point functions) with or without factors of $k^{a}$ and/or $p^{a}$. These expressions are transformed back to coordinate space as follows:
\begin{eqnarray}
    \label{divp0}
    \intx{x}\intp{p}\intp{k}A(x)\tilde{B}(p)\tilde{C}(k)e^{-ip\cdot x}e^{-ik\cdot x} k^{a}\dots k^{b} p^{\mu}\dots p^{\nu} \to \nonumber \\ \intx{x}(i\partial^{\mu})\dots(i\partial^{\nu})B(x) (i\partial^{a})\dots(i\partial^{b})C(x)
\end{eqnarray}
and likewise for other cases including $\langle \tilde{S}_{i}\tilde{S}_{j} \rangle$ and $\langle \tilde{S}_{i}\rangle$. Substituting these results for the divergent part in Eq. (\ref{loop01}), we get,
\begin{eqnarray}
    divp(\Gamma) &=& \intx{x} L \Bigg[\frac{ k_{1} m^4 \bar{\phi}^2}{2 k_{0}^4} + \frac{3 m^4 \kappa^2 \bar{\phi}^2}{4 k_{0}^2} -  \frac{ m^2 \lambda \bar{\phi}^2}{4 k_{0}^3} -  \frac{5 m^4 \kappa^2 \nu \bar{\phi}^2}{8 k_{0}^2} \nonumber \\ 
    && + \frac{3 m^4 \kappa^2 \nu^2 \bar{\phi}^2}{16 k_{0}^2} + \frac{ k_{1} m^2 \bar{\phi} \bar{\phi}{}^{,a}{}_{,a}}{2 k_{0}^3} -  \frac{3 m^2 \kappa^2 \bar{\phi} \bar{\phi}{}^{,a}{}_{,a}}{4 k_{0}} + \frac{17 m^2 \kappa^2 \nu \bar{\phi} \bar{\phi}{}^{,a}{}_{,a}}{16 k_{0}} \nonumber \\ 
    && -  \frac{3 m^2 \kappa^2 \nu^2 \bar{\phi} \bar{\phi}{}^{,a}{}_{,a}}{8 k_{0}} + \frac{ m^2 \kappa^2 \omega \bar{\phi} \bar{\phi}{}^{,a}{}_{,a}}{4 k_{0}} + \frac{ m^2 \kappa^2 \nu \omega \bar{\phi} \bar{\phi}{}^{,a}{}_{,a}}{8 k_{0}} -  \tfrac{3}{8} \kappa^2 \nu \bar{\phi} \bar{\phi}{}^{,a}{}_{,a}{}^{,b}{}_{,b} \nonumber \\ 
    && + \tfrac{3}{16} \kappa^2 \nu^2 \bar{\phi} \bar{\phi}{}^{,a}{}_{,a}{}^{,b}{}_{,b} -  \tfrac{1}{4} \kappa^2 \omega \bar{\phi} \bar{\phi}{}^{,a}{}_{,a}{}^{,b}{}_{,b} -  \tfrac{1}{8} \kappa^2 \nu \omega \bar{\phi} \bar{\phi}{}^{,a}{}_{,a}{}^{,b}{}_{,b} \nonumber \\
    && - \frac{3 k_{1}^2 m^4 \bar{\phi}^4}{256 k_{0}^6 \pi^2} -  \frac{3 k_{1} m^4 \kappa^2 \bar{\phi}^4}{32 k_{0}^4 \pi^2} + \frac{ k_{1} m^2 \lambda \bar{\phi}^4}{16 k_{0}^5 \pi^2} + \frac{ m^2 \kappa^2 \lambda \bar{\phi}^4}{32 k_{0}^3 \pi^2} \nonumber \\ 
    && -  \frac{ \lambda^2 \bar{\phi}^4}{128 k_{0}^4 \pi^2} + \frac{ k_{1} m^4 \kappa^2 \nu \bar{\phi}^4}{16 k_{0}^4 \pi^2} -  \frac{17 m^2 \kappa^2 \lambda \nu \bar{\phi}^4}{768 k_{0}^3 \pi^2} -  \frac{3 k_{1} m^4 \kappa^2 \nu^2 \bar{\phi}^4}{128 k_{0}^4 \pi^2} \nonumber 
\end{eqnarray}
\begin{eqnarray} 
    \label{divp1}
    && + \frac{ m^2 \kappa^2 \lambda \nu^2 \bar{\phi}^4}{128 k_{0}^3 \pi^2} -  \frac{19 k_{1}^2 m^2 \bar{\phi}^3 \bar{\phi} {}^{,a}{}_{,a}}{128 k_{0}^5 \pi^2} + \frac{ k_{1} \lambda \bar{\phi}^3 \bar{\phi} {}^{,a}{}_{,a}}{64 k_{0}^4 \pi^2} -  \frac{ \kappa^2 \lambda \bar{\phi}^3 \bar{\phi} {}^{,a}{}_{,a}}{64 k_{0}^2 \pi^2} \nonumber \\
    && -  \frac{ k_{1} m^2 \kappa^2 \nu \bar{\phi}^3 \bar{\phi} {}^{,a}{}_{,a}}{128 k_{0}^3 \pi^2} + \frac{3 \kappa^2 \lambda \nu \bar{\phi}^3 \bar{\phi} {}^{,a}{}_{,a}}{128 k_{0}^2 \pi^2} -  \frac{ \kappa^2 \lambda \nu^2 \bar{\phi}^3 \bar{\phi} {}^{,a}{}_{,a}}{128 k_{0}^2 \pi^2} + \frac{ k_{1} m^2 \kappa^2 \omega \bar{\phi}^3 \bar{\phi} {}^{,a}{}_{,a}}{64 k_{0}^3 \pi^2} \nonumber \\
    && -  \frac{ \kappa^2 \lambda \omega \bar{\phi}^3 \bar{\phi} {}^{,a}{}_{,a}}{192 k_{0}^2 \pi^2} -  \frac{ k_{1} m^2 \kappa^2 \nu \omega \bar{\phi}^3 \bar{\phi} {}^{,a}{}_{,a}}{128 k_{0}^3 \pi^2} + \frac{ \kappa^2 \lambda \nu \omega \bar{\phi}^3 \bar{\phi} {}^{,a}{}_{,a}}{384 k_{0}^2 \pi^2} -  \frac{11 k_{1}^2 m^2 \bar{\phi}^2 \bar{\phi} {}_{,a} \bar{\phi} {}^{,a}}{128 k_{0}^5 \pi^2} \nonumber \\
    && -  \frac{3 k_{1} m^2 \kappa^2 \bar{\phi}^2 \bar{\phi} {}_{,a} \bar{\phi} {}^{,a}}{32 k_{0}^3 \pi^2} -  \frac{3 k_{1} \lambda \bar{\phi}^2 \bar{\phi} {}_{,a} \bar{\phi} {}^{,a}}{64 k_{0}^4 \pi^2} + \frac{5 k_{1} m^2 \kappa^2 \nu \bar{\phi}^2 \bar{\phi} {}_{,a} \bar{\phi} {}^{,a}}{32 k_{0}^3 \pi^2} + \frac{ \kappa^2 \lambda \nu \bar{\phi}^2 \bar{\phi} {}_{,a} \bar{\phi} {}^{,a}}{256 k_{0}^2 \pi^2} \nonumber \\
    && -  \frac{3 k_{1} m^2 \kappa^2 \nu^2 \bar{\phi}^2 \bar{\phi} {}_{,a} \bar{\phi} {}^{,a}}{64 k_{0}^3 \pi^2} + \frac{3 k_{1} m^2 \kappa^2 \omega \bar{\phi}^2 \bar{\phi} {}_{,a} \bar{\phi} {}^{,a}}{64 k_{0}^3 \pi^2} -  \frac{ \kappa^2 \lambda \omega \bar{\phi}^2 \bar{\phi} {}_{,a} \bar{\phi} {}^{,a}}{32 k_{0}^2 \pi^2} + \frac{ k_{1} m^2 \kappa^2 \nu \omega \bar{\phi}^2 \bar{\phi} {}_{,a} \bar{\phi} {}^{,a}}{128 k_{0}^3 \pi^2} \nonumber \\ 
    && + \frac{7 k_{1}^2 \bar{\phi}^2 \bar{\phi} {}^{,b}{}_{,b}{}_{,a} \bar{\phi} {}^{,a}}{128 k_{0}^4 \pi^2} + \frac{5 k_{1} \kappa^2 \omega \bar{\phi}^2 \bar{\phi} {}^{,b}{}_{,b}{}_{,a} \bar{\phi} {}^{,a}}{384 k_{0}^2 \pi^2} -  \frac{ k_{1} \kappa^2 \nu \omega \bar{\phi}^2 \bar{\phi} {}^{,b}{}_{,b}{}_{,a} \bar{\phi} {}^{,a}}{256 k_{0}^2 \pi^2} -  \frac{ k_{1}^2 \bar{\phi}^2 \bar{\phi} {}^{,a} \bar{\phi} {}^{,b}{}_{,a}{}_{,b}}{24 k_{0}^4 \pi^2} \nonumber \\ 
    && -  \frac{3 k_{1} \kappa^2 \omega \bar{\phi}^2 \bar{\phi} {}^{,a} \bar{\phi} {}^{,b}{}_{,a}{}_{,b}}{128 k_{0}^2 \pi^2} -  \frac{ k_{1}^2 \bar{\phi}^2 \bar{\phi} {}^{,a}{}_{,a} \bar{\phi} {}^{,b}{}_{,b}}{256 k_{0}^4 \pi^2} + \frac{ k_{1} \kappa^2 \bar{\phi}^2 \bar{\phi} {}^{,a}{}_{,a} \bar{\phi} {}^{,b}{}_{,b}}{32 k_{0}^2 \pi^2} -  \frac{3 k_{1} \kappa^2 \nu \bar{\phi}^2 \bar{\phi} {}^{,a}{}_{,a} \bar{\phi} {}^{,b}{}_{,b}}{64 k_{0}^2 \pi^2} \nonumber \\ 
    && + \frac{3 k_{1} \kappa^2 \nu^2 \bar{\phi}^2 \bar{\phi} {}^{,a}{}_{,a} \bar{\phi} {}^{,b}{}_{,b}}{128 k_{0}^2 \pi^2} + \frac{ k_{1} \kappa^2 \omega \bar{\phi}^2 \bar{\phi} {}^{,a}{}_{,a} \bar{\phi} {}^{,b}{}_{,b}}{64 k_{0}^2 \pi^2} -  \frac{ k_{1} \kappa^2 \nu \omega \bar{\phi}^2 \bar{\phi} {}^{,a}{}_{,a} \bar{\phi} {}^{,b}{}_{,b}}{128 k_{0}^2 \pi^2} + \frac{ k_{1}^2 \bar{\phi} \bar{\phi} {}_{,a} \bar{\phi} {}^{,a} \bar{\phi} {}^{,b}{}_{,b}}{16 k_{0}^4 \pi^2} \nonumber \\ 
    && + \frac{ k_{1} \kappa^2 \bar{\phi} \bar{\phi} {}_{,a} \bar{\phi} {}^{,a} \bar{\phi} {}^{,b}{}_{,b}}{16 k_{0}^2 \pi^2} -  \frac{25 k_{1} \kappa^2 \nu \bar{\phi} \bar{\phi} {}_{,a} \bar{\phi} {}^{,a} \bar{\phi} {}^{,b}{}_{,b}}{256 k_{0}^2 \pi^2} + \frac{3 k_{1} \kappa^2 \nu^2 \bar{\phi} \bar{\phi} {}_{,a} \bar{\phi} {}^{,a} \bar{\phi} {}^{,b}{}_{,b}}{64 k_{0}^2 \pi^2} + \frac{3 k_{1} \kappa^2 \omega \bar{\phi} \bar{\phi} {}_{,a} \bar{\phi} {}^{,a} \bar{\phi} {}^{,b}{}_{,b}}{32 k_{0}^2 \pi^2} \nonumber \\ 
    && -  \frac{ k_{1} \kappa^2 \nu \omega \bar{\phi} \bar{\phi} {}_{,a} \bar{\phi} {}^{,a} \bar{\phi} {}^{,b}{}_{,b}}{64 k_{0}^2 \pi^2} + \frac{ k_{1}^2 \bar{\phi}^2 \bar{\phi} {}^{,a} \bar{\phi} {}_{,a}{}^{,b}{}_{,b}}{384 k_{0}^4 \pi^2} -  \frac{ k_{1} \kappa^2 \omega \bar{\phi}^2 \bar{\phi} {}^{,a} \bar{\phi} {}_{,a}{}^{,b}{}_{,b}}{192 k_{0}^2 \pi^2} -  \frac{ k_{1} \kappa^2 \nu \omega \bar{\phi}^2 \bar{\phi} {}^{,a} \bar{\phi} {}_{,a}{}^{,b}{}_{,b}}{256 k_{0}^2 \pi^2} \nonumber \\ 
    && -  \frac{ k_{1}^2 \bar{\phi}^3 \bar{\phi} {}^{,a}{}_{,a}{}^{,b}{}_{,b}}{256 k_{0}^4 \pi^2} + \frac{ k_{1} \kappa^2 \nu \bar{\phi} {}_{,a} \bar{\phi} {}^{,a} \bar{\phi} {}_{,b} \bar{\phi} {}^{,b}}{256 k_{0}^2 \pi^2} -  \frac{ k_{1} \kappa^2 \bar{\phi} \bar{\phi} {}^{,a} \bar{\phi} {}_{,a}{}_{,b} \bar{\phi} {}^{,b}}{16 k_{0}^2 \pi^2} + \frac{ k_{1} \kappa^2 \nu \bar{\phi} \bar{\phi} {}^{,a} \bar{\phi} {}_{,a}{}_{,b} \bar{\phi} {}^{,b}}{128 k_{0}^2 \pi^2} \nonumber \\ 
    && -  \frac{ k_{1} \kappa^2 \omega \bar{\phi} \bar{\phi} {}^{,a} \bar{\phi} {}_{,a}{}_{,b} \bar{\phi} {}^{,b}}{16 k_{0}^2 \pi^2} -  \frac{ k_{1}^2 \bar{\phi}^2 \bar{\phi} {}_{,a}{}_{,b} \bar{\phi} {}^{,a}{}^{,b}}{128 k_{0}^4 \pi^2} -  \frac{ k_{1} \kappa^2 \bar{\phi}^2 \bar{\phi} {}_{,a}{}_{,b} \bar{\phi} {}^{,a}{}^{,b}}{32 k_{0}^2 \pi^2} -  \frac{ k_{1} \kappa^2 \omega \bar{\phi}^2 \bar{\phi} {}_{,a}{}_{,b} \bar{\phi} {}^{,a}{}^{,b}}{32 k_{0}^2 \pi^2} \Bigg]
\end{eqnarray}
where, $L=-1/8\pi^{2}\epsilon$ ($\epsilon=n-4$) as the dimensionality $n\to 4$. 
As expected, there are no $\alpha$ dependent terms. Although not explicitly shown here, factors of $1/\alpha$ appear in individual pieces in Eq. (\ref{loop01}). However, when all contributions are added to evaluate $\Gamma$, these terms cancel so that the final result is gauge-invariant. Final result for divergent part of $\Gamma$ after removing bookkeeping parameters ($\omega\to 1$, $\nu \to 1$) in the Landau gauge ($\alpha \to 0$) leads to the covariant corrections:
\begin{eqnarray}
    \label{finalEA}
    divp(\Gamma) &=& \int d^4 x\Big[\frac{  k_{1} m^4  \bar{\phi}^2}{2  k_{0}^4} + \frac{5 m^4 \kappa^2  \bar{\phi}^2}{16  k_{0}^2} -  \frac{ m^2 \lambda  \bar{\phi}^2}{4  k_{0}^3} + \frac{  k_{1} m^2  \bar{\phi}  \bar{\phi} {}^{,a}{}_{,a}}{2  k_{0}^3} + \frac{5 m^2 \kappa^2  \bar{\phi}  \bar{\phi} {}^{,a}{}_{,a}}{16  k_{0}} \nonumber \\ 
    && -  \tfrac{9}{16} \kappa^2   \phi   \phi {}^{,a}{}_{,a}{}^{,b}{}_{,b}  - \frac{3  k_{1}^2 m^4 \bar{  \phi}^4}{32  k_{0}^6} -  \frac{7  k_{1} m^4 \kappa^2 \bar{  \phi}^4}{16  k_{0}^4} + \frac{  k_{1} m^2 \lambda \bar{  \phi}^4}{2  k_{0}^5} + \frac{13 m^2 \kappa^2 \lambda \bar{  \phi}^4}{96  k_{0}^3} \nonumber \\ 
    && -  \frac{ \lambda^2 \bar{  \phi}^4}{16  k_{0}^4} -  \frac{19  k_{1}^2 m^2 \bar{  \phi}^3 \bar{  \phi}{}^{,a}{}_{,a}}{16  k_{0}^5} + \frac{  k_{1} \lambda \bar{  \phi}^3 \bar{  \phi}{}^{,a}{}_{,a}}{8  k_{0}^4} -  \frac{ \kappa^2 \lambda \bar{  \phi}^3 \bar{  \phi}{}^{,a}{}_{,a}}{48  k_{0}^2} \nonumber \\ 
    && -  \frac{11  k_{1}^2 m^2 \bar{  \phi}^2 \bar{  \phi}{}_{,a} \bar{  \phi}{}^{,a}}{16  k_{0}^5} + \frac{9  k_{1} m^2 \kappa^2 \bar{  \phi}^2 \bar{  \phi}{}_{,a} \bar{  \phi}{}^{,a}}{16  k_{0}^3} -  \frac{3  k_{1} \lambda \bar{  \phi}^2 \bar{  \phi}{}_{,a} \bar{  \phi}{}^{,a}}{8  k_{0}^4} -  \frac{7 \kappa^2 \lambda \bar{  \phi}^2 \bar{  \phi}{}_{,a} \bar{  \phi}{}^{,a}}{32  k_{0}^2} \nonumber \\ 
    && + \frac{7  k_{1}^2 \bar{  \phi}^2 \bar{  \phi}{}^{,b}{}_{,b}{}_{,a} \bar{  \phi}{}^{,a}}{16  k_{0}^4} + \frac{7  k_{1} \kappa^2 \bar{  \phi}^2 \bar{  \phi}{}^{,b}{}_{,b}{}_{,a} \bar{  \phi}{}^{,a}}{96  k_{0}^2} -  \frac{  k_{1}^2 \bar{  \phi}^2 \bar{  \phi}{}^{,a} \bar{  \phi}{}^{,b}{}_{,a}{}_{,b}}{3  k_{0}^4} -  \frac{3  k_{1} \kappa^2 \bar{  \phi}^2 \bar{  \phi}{}^{,a} \bar{  \phi}{}^{,b}{}_{,a}{}_{,b}}{16  k_{0}^2} \nonumber \\ 
    && -  \frac{  k_{1}^2 \bar{  \phi}^2 \bar{  \phi}{}^{,a}{}_{,a} \bar{  \phi}{}^{,b}{}_{,b}}{32  k_{0}^4} + \frac{  k_{1} \kappa^2 \bar{  \phi}^2 \bar{  \phi}{}^{,a}{}_{,a} \bar{  \phi}{}^{,b}{}_{,b}}{8  k_{0}^2} + \frac{  k_{1}^2 \bar{  \phi} \bar{  \phi}{}_{,a} \bar{  \phi}{}^{,a} \bar{  \phi}{}^{,b}{}_{,b}}{2  k_{0}^4} + \frac{23  k_{1} \kappa^2 \bar{  \phi} \bar{  \phi}{}_{,a} \bar{  \phi}{}^{,a} \bar{  \phi}{}^{,b}{}_{,b}}{32  k_{0}^2} \nonumber \\ 
    && + \frac{  k_{1}^2 \bar{  \phi}^2 \bar{  \phi}{}^{,a} \bar{  \phi}{}_{,a}{}^{,b}{}_{,b}}{48  k_{0}^4} -  \frac{7  k_{1} \kappa^2 \bar{  \phi}^2 \bar{  \phi}{}^{,a} \bar{  \phi}{}_{,a}{}^{,b}{}_{,b}}{96  k_{0}^2} -  \frac{  k_{1}^2 \bar{  \phi}^3 \bar{  \phi}{}^{,a}{}_{,a}{}^{,b}{}_{,b}}{32  k_{0}^4} + \frac{  k_{1} \kappa^2 \bar{  \phi}{}_{,a} \bar{  \phi}{}^{,a} \bar{  \phi}{}_{,b} \bar{  \phi}{}^{,b}}{32  k_{0}^2} \nonumber \\ 
    && -  \frac{15  k_{1} \kappa^2 \bar{  \phi} \bar{  \phi}{}^{,a} \bar{  \phi}{}_{,a}{}_{,b} \bar{  \phi}{}^{,b}}{16  k_{0}^2} -  \frac{  k_{1}^2 \bar{  \phi}^2 \bar{  \phi}{}_{,a}{}_{,b} \bar{  \phi}{}^{,a}{}^{,b}}{16  k_{0}^4} -  \frac{  k_{1} \kappa^2 \bar{  \phi}^2 \bar{  \phi}{}_{,a}{}_{,b} \bar{  \phi}{}^{,a}{}^{,b}}{2  k_{0}^2}\Big]
\end{eqnarray} 
If instead we turn off the DV connections by setting $\nu = 0$ and choose $\alpha = 1, \omega = 1$, we recover gauge-dependent results obtained in the past by Steinwachs and Kamenshchik \cite{steinwachs2011}, where they calculated the one-loop divergences for a general scalar-tensor theory that in the single field limit (with the identifications $U=1, G=K$, and $V=V/\gamma^4$ in their notations) encompasses the model (\ref{action}). Similarly, in the case $k_1 = 0, k_0 = 1$ we recover the gauge-invariant calculations of Mackay and Toms \cite{mackay2010} (excluding cosmological constant and nonminimal coupling to gravity).

\subsection{Renormalization and Comparisons}
Not all the divergences in Eq. (\ref{finalEA}) can be absorbed by renormalizing the parameters in the classical action (\ref{action}), particularly the quartic derivatives of $\bar{\phi}(x)$, which are absent in the classical action. However, we need not worry about these UV divergences since the current framework is an effective theory approach, and we assume that such non-renormalizable terms are resolved by some high energy theory. For now, we still can construct counterterms from the classical action functional to absorb corresponding divergent parts, which will in turn induce 1-loop corrections to the parameters $\frac{m^2}{k_0}, \frac{k_1}{k_0^2}, \frac{\lambda}{k_0^2}$ of the theory (\ref{action}).

We start by re-writing Eq. (\ref{finalEA}) in the form,
\begin{eqnarray}
    \label{ren00}
    divp(\Gamma) = L \intx{x} (A \bar{\phi}\Box\bar{\phi} + B\bar{\phi}^2 + C \bar{\phi}^4 + D \bar{\phi}^2 \partial_{\mu}\bar{\phi}\partial^{\mu}\bar{\phi})
\end{eqnarray}
where we have ignored the terms not present in the classical background action. We note that the terms of the form $\bar{\phi}^{3}\Box\bar{\phi}$ in Eq. (\ref{finalEA}) are transformed to $-3\bar{\phi}^2\partial_{\mu}\bar{\phi}\partial^{\mu}\bar{\phi}$ after by-parts integration. The coefficients $A,B,C,D$ are read off from Eq. (\ref{finalEA}):
\begin{eqnarray}
    \label{ren01}
    A &=& \dfrac{5m^{2}\kappa^{2}}{16 k_{0}} + \dfrac{k_{1} m^{2}}{2 k_{0}^{3}}; \nonumber \\
    B &=& \dfrac{k_{1}m^{4}}{2 k_{0}^{4}} + \dfrac{5 m^{4}\kappa^{2}}{16 k_{0}^{2}} - \dfrac{m^{2}\lambda}{4 k_{0}^{3}}; \nonumber \\
    C &=& - \dfrac{3 k_{1}^{2}m^{4}}{32 k_{0}^{6}} - \dfrac{7 k_{1}m^{4}\kappa^{2}}{16 k_{0}^{4}} + \dfrac{k_{1}m^{2}\lambda^2}{k_{0}^{5}} + \dfrac{13 m^2 \kappa^2 \lambda}{96 k_0^3} - \dfrac{\lambda^2}{16 k_0^4}; \nonumber \\
    D &=& \dfrac{23 k_1^2 m^2}{8 k_0^5} + \dfrac{9 k_1 m^2 \kappa^2}{16 k_0^3} - \dfrac{3 k_1 \lambda}{4 k_0^4} - \dfrac{5\kappa^2 \lambda}{32 k_0^2}.
\end{eqnarray}
Taking into account the field Renormalization $\bar{\phi}\to Z^{1/2}\bar{\phi}$, the classical background Lagrangian reads,
\begin{eqnarray}
    \label{ren02}
    \mathcal{L}_{Z} = -\dfrac{1}{2} Z \bar{\phi}\Box\bar{\phi} + \frac{1}{2}\dfrac{m^2}{k_0} Z \bar{\phi}^2 + \frac{\lambda}{24 k_0^2} Z^2 \bar{\phi}^4 + \frac{1}{2}\frac{k_1}{k_0^2} Z^2 \bar{\phi}^2 \partial_{\mu}\bar{\phi}\partial^{\mu}\bar{\phi}
\end{eqnarray}
Suppose, the renormalized Lagrangian is given in terms of renormalized parameters as follows,
\begin{eqnarray}
    \label{ren03}
    \mathcal{L}_{r} = -\dfrac{1}{2} \bar{\phi}\Box\bar{\phi} + \frac{1}{2}\left(\dfrac{m^2}{k_0}\right)_{r} \bar{\phi}^2 + \frac{1}{24}\left(\frac{\lambda}{k_0^2}\right)_{r} \bar{\phi}^4 + \frac{1}{2}\left(\frac{k_1}{k_0^2}\right)_{r} \bar{\phi}^2 \partial_{\mu}\bar{\phi}\partial^{\mu}\bar{\phi}
\end{eqnarray}
where $(\cdot)_{r}$ represents the renormalized parameter. The counterterm Lagrangian is then defined as $\delta\mathcal{L} = \mathcal{L}_{r} - \mathcal{L}_{Z}$. Accordingly, the counterterms for field and other parameters are as follows:
\begin{eqnarray}
    \label{ren04}
    \delta_{Z} = Z-1; & \quad \delta\left(\dfrac{m^2}{k_0}\right) = \dfrac{m^2}{k_0}Z - \left(\dfrac{m^2}{k_0}\right)_{r} ; \nonumber \\
    \delta\left(\dfrac{\lambda}{k_0^2}\right) = \dfrac{\lambda}{k_0^2} Z^2 - \left(\dfrac{\lambda}{k_0^2}\right)_{r}; & \quad \delta\left(\dfrac{k_1}{k_0^2}\right) = \dfrac{k_1}{k_0^2} Z^2 - \left(\dfrac{k_1}{k_0^2}\right)_{r}.
\end{eqnarray}
These counterterms are fixed by demanding that $divp(\Gamma) = -\intx{x}\delta\mathcal{L}$. With some algebraic manipulations, the counterterms read,
\begin{eqnarray}
    \label{ren05}
    \delta_{Z} = -\dfrac{A}{4\pi^2 \epsilon} ; & \quad \delta\left(\dfrac{m^2}{k_0}\right) = \dfrac{B}{4\pi^2 \epsilon} ; \nonumber \\
    \delta\left(\dfrac{\lambda}{k_0^2}\right) = \dfrac{3 C}{\pi^2 \epsilon}; & \quad \delta\left(\dfrac{k_1}{k_0^2}\right) = \dfrac{D}{4\pi^2\epsilon}. 
\end{eqnarray}
Using Eq. (\ref{ren05}) in (\ref{ren04}), we find the one-loop corrections to coupling parameters in terms of the coefficients $A,B,C,D$, 
\begin{eqnarray}
    \label{ren06}
    \Delta \left(\dfrac{m^2}{k_0}\right) = \dfrac{m^2 A}{4\pi^2 k_0 \epsilon} + \dfrac{B}{4\pi^2 \epsilon}; \nonumber \\
    \Delta\left(\dfrac{\lambda}{k_0^2}\right) = \dfrac{3 C}{\pi^2 \epsilon} + \dfrac{\lambda A}{2\pi^2 k_0^2 \epsilon}; \\
    \Delta \left(\dfrac{k_1}{k_0^2}\right) = \dfrac{k_1 A}{2\pi^2 k_0^2\epsilon} + \dfrac{D}{4\pi^2\epsilon}. \nonumber 
\end{eqnarray}
For the sake of comparisons, and also as a crosscheck, we point out that upon choosing $\nu=0,\alpha=1,\omega=0$ in the case $k_1 = 0, k_0 = 1$, the gauge-dependent one-loop quantum gravitational correction to $\phi^4$ theory first calculated by Rodigast and Schuster \cite{rodigast2010a} is recovered: $\Delta\lambda = \frac{\kappa^2}{4\pi^2\epsilon}(m^2\lambda - 3\lambda^2/4\kappa^2)$. Note that, all gravitational corrections in Eq. (\ref{ren06}) appear with a factor of $\kappa$, while the ones without it are nongravitational corrections that could in principle be obtained from flat space quantum field theory. Also, in the gauge covariant version of the same case (viz. $\nu=1,\alpha=0,\omega=1$ with $k_1 = 0, k_0 = 1$), our results match that of Pietrykowski \cite{artur2013}. 

In a similar spirit, we would like to shed some light on the extensions of the work of Ref. \cite{mackay2010}. There, a self-interacting scalar field with nonminimal coupling to gravity (of the form $\xi R \phi^2/2$) was considered and the corresponding field and mass renormalizations were studied. The action in Ref. \cite{mackay2010} matches ours if we put $k_1 = 0, k_0 = 1$ and add $\xi R \phi^2/2$. However, corrections to quartic coupling including contributions from the nonminimal coupling have not been calculated so far. Without going into the details (see Appendix \ref{AppendixD}) we present here the covariant one-loop corrections to quartic coupling $\lambda$ so as to complete the analysis of Ref. \cite{mackay2010}, 
\begin{eqnarray}
    \label{ren07}
    \Delta \lambda = \dfrac{3\lambda^2}{16\pi^2\epsilon} + \dfrac{\kappa^2}{\pi^2\epsilon}\left(\dfrac{9}{16}m^2\lambda + \dfrac{21}{8}m^2\lambda\xi^2 - \dfrac{3}{2}m^2\lambda\xi\right). 
\end{eqnarray}

\section{\label{sec4d}Effective potential}
It is evident from the analysis so far that extracting any more information, in the form of finite corrections for example, is a cumbersome task. A resolution to this problem lies in making a reasonable compromise, wherein the derivatives of background fields are ignored basis the assumption that either the background field is constant due to a symmetry or it is slowly varying. The resulting effective action is known as effective potential. One of the first instances of this workaround is the well known Coleman Weinberg potential \cite{coleman1973,weinberg1973}. This approximation holds up especially during inflation, where the slow-rolling condition requires fields to be slowly varying. In this section, we evaluate the effective potential of the theory (\ref{action}) including finite terms and infer cosmological implications. 

We begin by substituting $\partial^{\mu}\bar{\phi}=0$ in Eqs. (\ref{s1})-(\ref{s4}), resulting in,
\begin{eqnarray}
    \label{es1}
    \tilde{S}_{1} &=& \intx{x} \Bigl[\frac{m^2 \kappa \delta \phi h^{a}{}_{a} \bar{\phi}}{2 k_0} -  \frac{m^2 \kappa \nu \delta \phi h^{a}{}_{a} \bar{\phi}}{4 k_0} \Bigr]; \\
\label{es2}
\tilde{S}_{2} &=& \intx{x} \Bigl[ -\frac{m^2 \kappa^2 h^{}{}_{ab} h^{ab} \bar{\phi}^2}{8 k_0} + \frac{m^2 \kappa^2 h^{a}{}_{a} h^{b}{}_{b} \bar{\phi}^2}{16 k_0} + \frac{\lambda \bar{\phi}^2 (\delta \phi)^2}{4 k_0^2} -  \frac{m^2 \kappa^2 \nu \bar{\phi}^2 (\delta \phi)^2}{8 k_0} \nonumber \\ 
&& + \frac{k_1 \bar{\phi}^2 \delta \phi{}_{,a} \delta \phi{}^{,a}}{2 k_0^2} \Bigr]; \\
\label{es3}
    \tilde{S}_{3} &=& \intx{x} \Bigl[\frac{\kappa \lambda \delta \phi h^{a}{}_{a} \bar{\phi}^3}{12 k_0^2} -  \frac{\kappa \lambda \nu \delta \phi h^{a}{}_{a} \bar{\phi}^3}{24 k_0^2} \Bigr]; \\
\label{es4}
\tilde{S}_{4} &=& \intx{x} \Bigl[- \frac{\kappa^2 \lambda h^{}{}_{ab} h^{ab} \bar{\phi}^4}{96 k_0^2} + \frac{\kappa^2 \lambda h^{a}{}_{a} h^{b}{}_{b} \bar{\phi}^4}{192 k_0^2} -  \frac{\kappa^2 \lambda \nu \bar{\phi}^4 (\delta \phi)^2}{96 k_0^2} \Bigr].
\end{eqnarray}
Using the above expressions in Eq. (\ref{loop01}) and following the steps outlined in the Sec. \ref{loopint}, we obtain the covariant effective potential,
\begin{eqnarray}
    \label{efpot0}
\Gamma_{eff}[\bar{\phi}] &=& \dfrac{1}{8\pi^2}\intx{x} [A_{1}\frac{1}{\epsilon}\bar{\phi}^2 + A_{2}\bar{\phi}^2 + B_{1}\frac{1}{\epsilon}\bar{\phi}^4 + B_{2}\bar{\phi}^4]
\end{eqnarray} 
where, $A_1$ and $B_1$ are the same as $B$ and $C$ from Eq. (\ref{ren01}) respectively, and,  
\begin{eqnarray}
    \label{efpot1}
    A_{2} &=& (\gamma + \log(\pi)) (- \frac{ k_1 m^4}{4  k_0^4} -  \frac{5 m^4 \kappa^2}{32  k_0^2} + \frac{ m^2 \lambda}{8  k_0^3})  + \frac{3 k_1 m^4}{8  k_0^4} + \frac{ m^4 \kappa^2}{4  k_0^2} -  \frac{ m^2 \lambda}{8  k_0^3} \nonumber \\ && + (- \frac{ k_1 m^4}{4  k_0^4} -  \frac{5 m^4 \kappa^2}{32  k_0^2} + \frac{ m^2 \lambda}{8  k_0^3}) \log(\frac{m^2}{ k_0 \mu^2}) \nonumber \\ && - \frac{1}{\bar{\phi}} \intp{k} e^{-ik\cdot x} \tilde{\bar{\phi}} \left( \frac{3 m^4 \kappa^2 \log\bigl(1 + \frac{ k_0  k^2}{m^2}\bigr)}{32  k_0^2} -  \frac{3 m^6 \kappa^2 \log\bigl(1 + \frac{ k_0  k^2}{m^2}\bigr)}{32  k_0^3  k^2}\right); \\
    \label{b2}
    B_{2} &=& - \frac{9  k_1^2 m^4}{128  k_0^6} -  \frac{5  k_1 m^4 \kappa^2}{16  k_0^4} + \frac{  k_1 m^2 \lambda}{4  k_0^5} + \frac{25 m^2 \kappa^2 \lambda}{192  k_0^3} -  \frac{ \lambda^2}{16  k_0^4} \nonumber \\ && + (\gamma + \log(\pi)) (\frac{3  k_1^2 m^4}{64  k_0^6} + \frac{7  k_1 m^4 \kappa^2}{32  k_0^4} -  \frac{  k_1 m^2 \lambda}{4  k_0^5} -  \frac{13 m^2 \kappa^2 \lambda}{192  k_0^3} + \frac{ \lambda^2}{32  k_0^4}) \nonumber \\ && + (\frac{3  k_1^2 m^4}{64  k_0^6} + \frac{5  k_1 m^4 \kappa^2}{32  k_0^4} -  \frac{  k_1 m^2 \lambda}{8  k_0^5} -  \frac{13 m^2 \kappa^2 \lambda}{192  k_0^3} + \frac{ \lambda^2}{32  k_0^4}) \log(\frac{m^2}{ k_0 \mu^2}) \nonumber \\ && 
    - \frac{1}{\bar{\phi}^3}\intp{k} e^{-ik\cdot x} \tilde{\bar{\phi}^3} \Bigg( \frac{ m^2 \kappa^2 \lambda \log\bigl(1 + \frac{ k_0 k^2}{m^2}\bigr)}{32  k_0^3}\Bigg) \nonumber \\ && + \frac{1}{\bar{\phi}^2}\intp{k} e^{-ik\cdot x} \tilde{\bar{\phi}^2} \Bigg( -\frac{ m^2 \kappa^2 \lambda \log\Bigl(\frac{1 + \bigl(1 + \frac{4 m^2}{ k_0 k^2}\bigr)^{1/2}}{-1 + \bigl(1 + \frac{4 m^2}{ k_0 k^2}\bigr)^{1/2}}\Bigr) \bigl(1 + \frac{4 m^2}{ k_0 k^2}\bigr)^{1/2}}{32  k_0^3} + \nonumber \\ &&  \frac{ \lambda^2 \log\Bigl(\frac{1 + \bigl(1 + \frac{4 m^2}{ k_0 k^2}\bigr)^{1/2}}{-1 + \bigl(1 + \frac{4 m^2}{ k_0 k^2}\bigr)^{1/2}}\Bigr) \bigl(1 + \frac{4 m^2}{ k_0 k^2}\bigr)^{1/2}}{32  k_0^4} - \frac{ m^4 \kappa^2 \lambda \log\bigl(1 + \frac{ k_0 k^2}{m^2}\bigr)}{32  k_0^4 k^2} \nonumber \\ && + \frac{  k_1^2 \arctan\Bigl(\frac{ k_0^{1/2} k}{\bigl(4 m^2 -   k_0 k^2\bigr)^{1/2}}\Bigr) k^{3} \bigl(4 m^2 -   k_0 k^2\bigr)^{1/2}}{64  k_0^{9/2}} \Bigg) \nonumber \\ && + \frac{1}{\bar{\phi}}\intp{p} e^{-ip\cdot x} \tilde{\bar{\phi}} \frac{3  k_1 m^4 \kappa^2 \log\Bigl(\frac{1 + \bigl(1 + \frac{4 m^2}{ k_0 p^2}\bigr)^{1/2}}{-1 + \bigl(1 + \frac{4 m^2}{ k_0 p^2}\bigr)^{1/2}}\Bigr) \bigl(1 + \frac{4 m^2}{ k_0 p^2}\bigr)^{1/2}}{32  k_0^4} .
\end{eqnarray}
The logarithmic terms appearing in expressions above are dealt with as follows. In the context of the present problem and the effective theory treatment, we restrict ourselves to the condition $k\ll 10^{-6}M_p$ so that $\frac{k_0 k^2}{m^2}\ll 1$ (more on this later) using the order-of-magnitude estimates of parameters in Eq. (\ref{param}) from the results of \cite{ferreira2018}. Hence, logs involving this fraction can be expanded in a Taylor series. On the other hand, $\sqrt{1 + \frac{m^2}{k_0 k^2}}\approx \sqrt{\frac{m^2}{k_0 k^2}}$. For the $\arctan(\cdots)$ term, we use $\arctan(x)\approx x$ for small $x$. After these expansions, all terms with factors of $k$ will vanish since we assume the derivatives of $\bar{\phi}$ to be zero. Hence, all the integrands of momenta integrals in Eqs. (\ref{efpot1},\ref{b2}) reduce to c-numbers times Fourier transforms of $\bar{\phi}^{n}$. Using these simplifications, the coefficients $A_{2}$ and $B_{2}$ are obtained as,
\begin{eqnarray}
    \label{efpot2}
    A_{2} &=& (\gamma + \log(\pi)) (- \frac{ k_1 m^4}{4  k_0^4} -  \frac{5 m^4 \kappa^2}{32  k_0^2} + \frac{ m^2 \lambda}{8  k_0^3})  + \frac{3 k_1 m^4}{8  k_0^4} + \frac{ m^4 \kappa^2}{4  k_0^2} -  \frac{ m^2 \lambda}{8  k_0^3} \nonumber \\ && + \dfrac{3 m^4 \kappa^2}{32 k_0^2} + (- \frac{ k_1 m^4}{4  k_0^4} -  \frac{5 m^4 \kappa^2}{32  k_0^2} + \frac{ m^2 \lambda}{8  k_0^3}) \log(\frac{m^2}{ k_0 \mu^2}); \nonumber \\
    B_{2} &=& - \frac{9  k_1^2 m^4}{128  k_0^6} -  \frac{5  k_1 m^4 \kappa^2}{16  k_0^4} + \frac{  k_1 m^2 \lambda}{4  k_0^5} + \frac{25 m^2 \kappa^2 \lambda}{192  k_0^3} -  \frac{ \lambda^2}{16  k_0^4} \nonumber \\ &&  -\dfrac{m^2\kappa^2\lambda}{32 k_0^3} + \dfrac{\lambda^2}{32 k_0^4} - \dfrac{m^2 \kappa^2 \lambda }{32 k_0^3} + \dfrac{3 k_1 m^4 \kappa^2}{32 k_0^4} \nonumber \\ && + (\gamma + \log(\pi)) (\frac{3  k_1^2 m^4}{64  k_0^6} + \frac{7  k_1 m^4 \kappa^2}{32  k_0^4} -  \frac{  k_1 m^2 \lambda}{4  k_0^5} -  \frac{13 m^2 \kappa^2 \lambda}{192  k_0^3} + \frac{ \lambda^2}{32  k_0^4}) \nonumber \\ && + (\frac{3  k_1^2 m^4}{64  k_0^6} + \frac{5  k_1 m^4 \kappa^2}{32  k_0^4} -  \frac{  k_1 m^2 \lambda}{8  k_0^5} -  \frac{13 m^2 \kappa^2 \lambda}{192  k_0^3} + \frac{ \lambda^2}{32  k_0^4}) \log(\frac{m^2}{ k_0 \mu^2}) 
\end{eqnarray}
The counterterms for quadratic and quartic terms have a similar form to Eq. (\ref{ren05}), so that the effective potential can be written in terms of renormalized parameters which can be calculated from Eq. (\ref{ren06}) with $A=0$. The effective action takes the form,
\begin{eqnarray}
    \label{efpot3}
    V_{eff} = \dfrac{1}{2}\dfrac{m^2}{k_0} \bar{\phi}^2 + \dfrac{1}{4!}\dfrac{\lambda}{k_0^2}\bar{\phi}^4 + A_{2}\bar{\phi}^2 + B_{2}\bar{\phi}^4.
\end{eqnarray}

\subsection{Estimating the magnitude of corrections}
Making a definitive statement about cosmological implications of quantum corrected potential requires an analysis in the FRW background, which unfortunately is out of scope of the present work. However, we can get an order-of-magnitude estimate of the quantum corrections to the effective potential using the values of parameters $k_0, k_1, m^2, \lambda$ from the results of Ref. \cite{ferreira2018}. 

From the action (\ref{action}), the Einstein equations are given by,
\begin{eqnarray}
    \label{om0}
    3 H^2 &=& \dfrac{\kappa^2}{8}\left(-3\dot{\bar{\phi}}^{2} - 3 \dfrac{k_1}{k_0^2}\bar{\phi}^2\dot{\bar{\phi}}^2 + \dfrac{m^2}{k_0}\bar{\phi}^2 + \dfrac{\lambda}{12 k_0^2}\bar{\phi}^4\right); \nonumber \\
    2\dot{H} + 3 H^2 &=& \dfrac{\kappa^2}{8}\left(-\dot{\bar{\phi}}^{2} - \dfrac{k_1}{k_0^2}\bar{\phi}^2\dot{\bar{\phi}}^2 + \dfrac{m^2}{k_0}\bar{\phi}^2 + \dfrac{\lambda}{12 k_0^2}\bar{\phi}^4\right),
\end{eqnarray}
from which we obtain in the de-Sitter limit ($\dot{H}\sim\dot{\phi}\sim 0$),
\begin{eqnarray}
    \label{om1}
    3H^2 = \dfrac{\kappa^2}{8}\left( \dfrac{m^2}{k_0}\bar{\phi}^2 + \dfrac{\lambda}{12 k_0^2}\bar{\phi}^4 \right). 
\end{eqnarray}
The field equation for $\bar{\phi}$ reads,
\begin{eqnarray}
    \label{om2}
    (a + \frac{k_1}{k_0^2}\bar{\phi}^2)\ddot{\bar{\phi}} + \dfrac{k_1}{k_0^2}\bar{\phi}\dot{\bar{\phi}}^2 + (3aH + \frac{2k_1}{k_0^2}H\bar{\phi}^2)\dot{\bar{\phi}} - \dfrac{m^2}{k_0}\bar{\phi} - \dfrac{\lambda}{6k_0^2}\bar{\phi}^3 = 0.
\end{eqnarray}
Applying the de-Sitter conditions, Eq. (\ref{om2}) yields the de-Sitter value of $\bar{\phi}$, 
\begin{eqnarray}
    \label{om3}
    \bar{\phi}_{0}^2 = - \dfrac{6 k_0 m^2}{\lambda}.
\end{eqnarray}
Using Eq. (\ref{om3}) in (\ref{om1}), we find the de-Sitter value of Hubble parameter $H_{0}$:
\begin{eqnarray}
    \label{om4}
    H_0^2 = - \dfrac{\kappa^2 m^4}{8\lambda}.
\end{eqnarray}
Clearly, the condition for existence of de-Sitter solutions is $\lambda < 0$. Demanding this condition in Eq. (\ref{param}), along with $m^2>0$ and $\bar{\phi}<f$, leads to a constraint on the parameter $\alpha$ of the original theory (\ref{eq02}): $0.5 < \alpha \lesssim 1$. Following the results of \cite{ferreira2018}, we choose $0.5\lesssim\alpha\lesssim 0.6 \sim \mathcal{O}(1)$. Near this value of $\alpha$, $f\sim M_{p}=1/\kappa$ and $\Lambda \sim 10^{16} GeV$. Substituting these in Eqs. (\ref{param}), we find $m^2 \sim \Lambda^{4}/f^{2} \sim 10^{-12}M_{p}^{2}$; $\lambda \sim \Lambda^{4}/f^{4} \sim 10^{-12}$. Similarly, $k_0 \sim 1$ while $k_{1}\sim M_{p}^{-2}$. This also implies that in the low energy limit where momenta $k\ll 10^{13}$ GeV $\ll M_p$, $k_{1}k^2/k_{0}^2 \ll \lambda/k_{0}^2$, i.e. the derivative coupling term is suppressed.

From the above, we can estimate the order of magnitude contributions of terms in $A_{2}$ and $B_{2}$ at $\mathcal{O}(\bar{\phi}^2)$ and $\mathcal{O}(\bar{\phi}^4)$ respectively. We estimate the magnitude of each type of term present at both orders. At quadratic order in background field, we find,
\begin{eqnarray}
    \label{om5}
    \dfrac{\kappa^2 m^4}{k_{0}^2} \sim \dfrac{\lambda m^2}{k_0^3} \sim \dfrac{k_{1}m^4}{k_0^4} \sim 10^{14} GeV^2 .
\end{eqnarray}
Similarly, at quartic order in background field,
\begin{eqnarray}
    \label{om6}
    \dfrac{\kappa^2 m^4 k_1}{k_0^4} \sim \dfrac{\kappa^2 m^2 \lambda}{k_0^3} \sim \dfrac{\lambda^2}{k_0^4} \sim \dfrac{m^4 k_1^2}{k_0^6} \sim \dfrac{k_1 m^2 \lambda}{k_0^5} \sim 10^{-24} .
\end{eqnarray}
Quite an interesting observation here is that the magnitudes of gravitational (terms with a factor of $\kappa^2$) and non-gravitational (terms without $\kappa$) corrections turn out to be exactly the same for both quadratic and quartic order contributions. However, the corresponding quantum corrections are expectedly smaller by an order of $10^{-12}$ compared to $m^2$ and $\lambda$, as can also be checked using the loop counting parameter for de Sitter inflation $H_{0}^{2}/M_{Pl}^{2}$ with $H_0 \sim 10^{13}GeV$ and $M_{Pl}\sim 10^{19}GeV$.

\section{\label{end}Summary}
The nonminimal natural inflation model in consideration here is approximately described by a massive scalar field model with quartic self interaction and a derivative coupling in the region where $\phi/f<1$. We study one-loop corrections to this theory, about a Minkowski background, using a covariant effective action approach developed by DeWitt-Vilkovisky. The one-loop divergences and corresponding counterterms have been obtained. Along the way, we also recover several past results, both gauge-invariant non-gauge-invariant, for similar theories. In one such exercise, we obtain the $\phi^4$ coupling correction in a theory with nonminimal coupling of scalar field to gravity, originally considered in Ref. \cite{mackay2010} and thereby extend their result. 

Finite corrections have been taken into account for the calculation of effective potential, where we assume that the background field changes sufficiently slowly so that all derivatives of background field(s) can be ignored. Although cosmologically relevant inferences are not feasible as long as the metric background is Minkowski and not FRW, we can still estimate approximately the magnitudes of quantum corrections. Using the range of parameters applicable to our model, we find that the gravitational and non-gravitational corrections are of same order of magnitudes, while still being expectedly small compared to $m^2$ and $\lambda$. 

This is quite an interesting observation, since one would naively assume that gravitational corrections are $\kappa^2$ suppressed and thus would necessarily be small. There is thus enough motivation to go a step further, and calculate gravitational corrections in the FRW background so that cosmologically relevant inferences can be derived.

\chapter{Conclusions and Outlook}
\label{conclusion}

In this thesis, we used the DeWitt-Vilkovisky's (DV) covariant effective action formalism that is gauge invariant and background field invariant, to study formal and quantum gravitational aspects of quantum fields in curved spacetime. 

We systematically developed perturbative computation of one-loop effective action in a series of steps. First, we generalized the formalism to include quantization of rank-2 antisymmetric tensor and similar theories where gauge parameters have additional symmetries. We presented a geometric interpretation of quantization procedure for theories where gauge parameters are linearly dependent, which is simpler and more evolved than the traditional procedure followed in previous works viz. \cite{buchbinder1992} and \cite{shapiro2016}. Specifically, we generalized the calculation of ghost determinant using the geometric picture and solved the problem of degeneracy in the ghost determinant. This led to a general formula of covariant effective action that can be applied to any theory with or without linearly dependent gauge parameters (for instance, rank-2 and higher antisymmetric tensor field which has this problem). 

Second, we developed one-loop computation including classical metric perturbations through an application to antisymmetric tensor with spontaneous Lorentz violation. We explicitly showed using a perturbative application of DeWitt-Vilkovisky method that quantum equivalence of massive antisymmetric theory with a massive vector theory breaks down in presence of spontaneous Lorentz violating potential, validating inferences from a past result \cite{seifert2010a}. The use of conventional approaches like the proper-time (or heat kernel) method is not straightforward in this case because of nonminimal operators in functional determinants, so we calculated the one-loop effective action perturbatively in a nearly flat spacetime. 

In the third and final part of the thesis, we focus on the cosmological applications of DV method by including quantum gravitational corrections. We calculated one-loop effective action perturbatively for a scalar field model inspired by a recently proposed nonminimal natural inflation model \cite{ferreira2018} in Minkowski background up to quartic order in background fields using xAct package for Mathematica. 
The advantage of perturbative approach is the ease in implementation as a code and its scope for generalizing to other models while still producing correct results at a given order. For instance, quartic order calculations involve dealing with thousands of terms and performing loop integrals on each of them. In doing so, we obtained both gauge-dependent and gauge-invariant versions of corrections to $\phi^{4}$ coupling's $\beta-$function and confirmed several past results obtained using different methods. We also found that the gravitational and non-gravitational corrections to the effective potential are of the same orders of magnitude. This result is quite interesting, since one would naively assume gravitational (one-loop) corrections to be Planck-mass-squared suppressed. Of course, due to the Minkowski background, our calculations do not include the contributions from Ricci correction terms that in principle appear in the one-loop effective action.

The method utilized in this work is quite general in its applicability, and these calculations can be easily extended to calculate one-loop corrections at higher order in background fields including antisymmetric tensors in future. There is thus scope in near future for continued development of the existing one-loop code  to include FRW background so that cosmologically relevant computations can finally be made. Applications will include building upon the cosmological aspects of antisymmetric tensor fields, and several comparative studies involving different formulations of gravity. Moreover, several comparative theoretical studies can be performed to establish and compare existing alternative approaches like diagrammatic calculations and a recently developed gravitational quantum field theory formalism \cite{wu2015,wu2016}.

A few short term projects are also part of our future plans. For example, there is still considerable debate about gravitational corrections to beta function. Gauge dependent corrections to $\phi^4$ beta functions were first obtained in \cite{rodigast2010a}. A gauge invariant result was obtained in \cite{artur2013}. Later, in \cite{gonzalez2017} it was claimed that gravitational corrections have no physical significance because field redefinitions can cause the beta functions to vanish. However, in the DV approach one should be able to obtain beta functions that are also invariant under field redefinitions. Hence, an investigation using DV approach will likely settle the debate on gravitational corrections to $\phi^4$ coupling. Similarly, a second problem of interest is that of quantum corrections to Newtonian potential. Donoghue and collaborators in a long series of papers calculated corrections to Newtonian potential in Minkowski background from Feynman diagrams involving graviton loops (final results in \cite{donoghue2003}). However, gauge (in)dependence of these corrections have not yet been explicitly checked. First step towards this was undertaken in \cite{codello2010,codello2016} where (gauge-dependent) effective action was calculated partially. We plan to complete this analysis in a series of steps: (i) calculating gauge-dependent one-loop effective action while quantizing matter field and reproduce diagrammatic results; (ii) Finding corrections to Newtonian potential from each term in the effective action; and (ii) calculating gauge invariant effective action using the heat kernel method followed by gauge invariant corrections to Newtonian potential. This analysis will unambiguously establish the results for corrections to Newtonian potential.  

A third research direction in the long term, building upon the progress through this thesis, is to work towards identifying possible quantum gravity candidate signals, including but not limited to Lorentz violation, using observations as well as laboratory experiments. Detection of Lorentz violation is vital for developing effective theories such as SME. Since these corrections are expected to be Planck suppressed, it is interesting to study Lorentz violation in cosmological context so that future high-precision cosmological probes can be used for its detection\footnote{In fact, there is a lot to be done in the gravity sector as can be inferred from the complete database of known results documented in \cite{kostelecky2010}}.

\singlespacing
\addcontentsline{toc}{chapter}{References}

\bibliographystyle{styles/hunsrt}
\renewcommand{\bibname}{References}
\bibliography{endmatter/references,endmatter/ref}

\appendix
\appendixpage


\chapter{Projection operators for $B_{\mu\nu}$}
\label{AppendixA}

The basic projection operators for an antisymmetric tensor are defined as \cite{colatto2004},
\begin{equation} \begin{array}{rcl} \hspace{-0.8pc}P^{(1)}_{\mu\nu,\alpha\beta}&=&\displaystyle\frac{1}{2}(\theta_{\mu\alpha}\theta_{\nu\beta}-\theta_{\mu\beta}\theta_{\nu\alpha}),\\ \hspace{-0.8pc}P^{(2)}_{\mu\nu,\alpha\beta}&=&\displaystyle\frac{1}{4}(\theta_{\mu\alpha}\omega_{\nu\beta}-\theta_{\nu\alpha}\omega_{\mu\beta}-\theta_{\mu\beta}\omega_{\nu\alpha}+\theta_{\nu\beta}\omega_{\mu\alpha}), \end{array} \end{equation}
where,
\begin{equation} \theta_{\mu\nu}=\eta_{\mu\nu}-\omega_{\mu\nu},\quad \omega_{\mu\nu}=\frac{\partial_{\mu}\partial_{\nu}}{\Box}, \end{equation}
are the longitudinal and transverse projection operators along the momentum. To account for the Lorentz violation induced by nonzero vev, four new operators need to be introduced as follows \cite{maluf2019}:
\begin{eqnarray} 
P^{(3)}_{\mu\nu,\alpha\beta}&=& P^{\perp}_{\mu\nu, \alpha\beta},\\
P^{(4)}_{\mu\nu,\alpha\beta}&=& \frac{1}{2}\left( \omega_{\mu\lambda} \,P^{\parallel}_{\nu\lambda, \alpha\beta} - \omega_{\nu\lambda}\, P^{\parallel}_{\mu\lambda, \alpha\beta} \right),\\
P^{(5)}_{\mu\nu,\alpha\beta}&=& \frac{1}{2}\left( \omega_{\alpha\lambda} \,P^{\parallel}_{\mu\nu, \beta\lambda} - \omega_{\beta\lambda}\, P^{\parallel}_{\mu\nu, \alpha\lambda} \right),\\
P^{(6)}_{\mu\nu,\alpha\beta}&=& \frac{1}{4}\left( \omega_{\mu\alpha} \,P^{\parallel}_{\nu\rho, \beta\sigma}\,\omega^{\rho\sigma} - \omega_{\nu\alpha}\, P^{\parallel}_{\mu\rho, \beta\sigma}\,\omega^{\rho\sigma}\right. \nonumber\\ && {}-\,\left.\omega_{\mu\beta}\, P^{\parallel}_{\nu\rho, \alpha\sigma}\,\omega^{\rho\sigma} + \omega_{\nu\beta}\, P^{\parallel}_{\mu\rho, \alpha\sigma}\,\omega^{\rho\sigma} \right).
 \end{eqnarray}
The operators $P^{(1)}_{\mu\nu,\alpha\beta}, \cdots , P^{(6)}_{\mu\nu,\alpha\beta}$ obey a closed algebra \cite{maluf2019}. 

The identity element is given by,
\begin{equation} 
\mathcal{I}_{\mu\nu, \alpha\beta} = \frac{1}{2}(\eta_{\mu\alpha}\eta_{\nu\beta}-\eta_{\mu\beta}\eta_{\nu\alpha})=\left[P^{(1)}+P^{(2)}\right]_{\mu\nu,\alpha\beta}. 
\end{equation}


\chapter{Effective Action for Antisymmetric Tensor with Spontaneous Lorentz violation in operator form}
\label{AppendixB}

Following the method developed in Ref. \cite{aashish2018a}, the calculation of ghost determinants proceeds as follows. We rewrite $\chi_{\xi_{\nu}}$ as,
\begin{eqnarray}
\label{beaa4}
\chi_{\xi_{\nu}}[B^{\mu\nu}_{\xi_{\nu}}, C^{\mu}_{\xi_{\nu}}] = \chi_{\xi_{\nu}}[B^{\mu\nu},C^{\mu},{\xi_{\nu}},\Lambda,\check{\chi}_{\psi}],
\end{eqnarray}
which yields,
\begin{eqnarray}
\label{beaa5}
\chi_{\xi_{\nu}} = n_{\mu\nu}n_{\rho\sigma}\nabla^{\mu}B^{\rho\sigma} + \alpha C_{\nu}
+ 2n_{\mu\nu}n_{\rho\sigma}\nabla^{\mu}\nabla^{\rho}\xi^{\sigma} + \nabla_{\nu}\nabla_{\mu}\xi^{\mu} - \alpha^{2}\xi_{\nu} - \nabla_{\nu}\check{\chi}_{\psi}.
\end{eqnarray}
Then, using the definition of $Q'^{\xi_{\mu}}_{\xi_{\nu}}$, we get
\begin{eqnarray}
\label{beaa6}
Q'^{\xi_{\nu}}_{\xi_{\alpha}} = \left(\dfrac{\delta\chi_{\xi_{\nu}}}{\delta\xi_{\alpha}}\right)_{\xi_{\mu} = 0} &=& 2n_{\mu\nu}n_{\rho\alpha}\nabla^{\mu}\nabla^{\rho} + \nabla_{\nu}\nabla_{\alpha} - \alpha^{2}\delta_{\nu\alpha}.
\end{eqnarray}
A straightforward calculation leads to other non-zero components of ghost determinant,
\begin{eqnarray}
\label{beaa7}
Q'^{\Lambda}_{\Lambda} = \dfrac{\delta\chi_{\Lambda}}{\delta\Lambda} = \Box_{x} - \alpha^{2} \\
\label{beaa8}
\check{Q}^{\psi}_{\psi} \equiv \dfrac{\delta\check{\chi}_{\psi}}{\delta\psi} = \Box_{x} - \alpha^{2}
\end{eqnarray}
Using the definition of effective action obtained in \cite{aashish2018a}, 
\begin{eqnarray}
\label{beaa9}
\exp(i\Gamma[\bar{B},\bar{C}]) = \int\prod_{\mu}dC_{\mu}\prod_{\rho\sigma}dB_{\rho\sigma}\prod_{x}d\Phi
\det(Q'^{\Lambda}_{\Lambda})\det(Q'^{\xi_{\nu}}_{\xi_{\alpha}})(\det\check{Q}^{\psi}_{\psi})^{-1}\times \nonumber \\  
 \exp\left\{i \Bigg(\int d v_{x} \mathcal{L}_{2}^{GF}\Bigg)  + (\bar{B}_{\mu\nu}-B_{\mu\nu})\dfrac{\delta}{\delta \bar{B}_{\mu\nu}}\Gamma[\bar{B},\bar{C}] \right. \nonumber \\ \left. + (\bar{C}_{\mu}-C_{\mu})\dfrac{\delta}{\delta \bar{C}_{\mu}}\Gamma[\bar{B},\bar{C}] \right\},
\end{eqnarray}
The 1-loop effective action is obtained as,
\begin{eqnarray}
\label{ceaa0}
\Gamma_{2}^{(1)} = \frac{i\hbar}{2}\Big[\ln\det(D_{2} - \alpha^{2}n^{\mu\nu}n_{\rho\sigma}) - \ln\det(D_{1}-\alpha^{2}) + \ln\det(\Box_{x} - \alpha^{2})\Big]
\end{eqnarray}
where,
\begin{eqnarray}
\label{cea1}
D_{2}{}^{\mu\nu}_{\ \ \rho\sigma}B^{\rho\sigma} &\equiv & \nabla_{\alpha}\nabla^{\alpha}B^{\mu\nu} + \nabla_{\alpha}\nabla^{\mu}B^{\nu\alpha} + \nabla_{\alpha}\nabla^{\nu}B^{\alpha\mu} + 2 n^{\mu\nu}n_{\rho\sigma}n^{\alpha\sigma}n_{\beta\gamma}\nabla^{\rho}\nabla_{\alpha}B^{\beta\gamma}, \nonumber \\
D_{1}{}^{\mu}_{\ \nu}C^{\nu} &\equiv & 2 n^{\nu\mu}n_{\rho\sigma}\nabla_{\nu}\nabla^{\rho}C^{\sigma} + \nabla^{\mu}\nabla_{\nu}C^{\nu}.
\end{eqnarray}
It is to be noted that the coefficient of $\alpha^{2}$ in the first term in Eq. (\ref{ceaa0}) ensures that massive modes correspond to field components along vacuum expectation tensor $n_{\mu\nu}$ and massless modes correspond to transverse components. An interesting observation here is the last term, which is unaffected by $n_{\mu\nu}$. In case of no SLV, the last term causes the quantum discontinuity when going from massive to massless case \cite{shapiro2016}. 

To compare Eq. (\ref{ceaa0}) with the effective action of classically equivalent Lagrangian, the Lagrangian in (\ref{eq3}) is treated with the St{\"u}ckelberg procedure to obtain,
\begin{eqnarray}
\label{ceaa1}
\tilde{\mathcal{L}}_{1} &=& \dfrac{1}{4}\Big(\tilde{n}_{\mu\nu}F^{\mu\nu}\Big)^{2} - \dfrac{1}{2}\alpha^{2}(C_{\mu} + \dfrac{1}{\alpha}\nabla_{\mu}\Phi)^{2}
\end{eqnarray}
The above Lagrangian is invariant under transformations,
\begin{eqnarray}
\label{ceaa2}
C_{\mu}\longrightarrow C_{\mu} + \nabla_{\mu}\Lambda, \quad \Phi\longrightarrow \Phi - \alpha\Lambda.
\end{eqnarray}
With the gauge condition Eq. (\ref{aeaa9}), the gauge fixed Lagrangian reads,
\begin{eqnarray}
\label{ceaa3}
\tilde{\mathcal{L}}_{1}^{GF} = \dfrac{1}{2}C_{\mu}D_{1}C^{\mu} - \dfrac{1}{2}\alpha^{2}C_{\mu}C^{\mu} + \dfrac{1}{2}\Phi (\Box_{x} - \alpha^{2})\Phi.
\end{eqnarray}
where, 
\begin{eqnarray}
\label{deaa0}
D_{1}C_{\mu} = -2\tilde{n}_{\nu\mu}\tilde{n}_{\rho\sigma}\nabla^{\nu}\nabla^{\rho}C^{\sigma} + \nabla_{\mu}\nabla_{\nu}C^{\nu}.
\end{eqnarray}
It is straightforward to check that the 1-loop effective action is,
\begin{eqnarray}
\label{ceaa4}
\Gamma_{1}^{(1)} = \frac{i\hbar}{2}\Big[\ln\det(D_{1}-\alpha^{2}) - \ln\det(\Box_{x} - \alpha^{2})\Big].
\end{eqnarray}
Similar to Eq. (\ref{ceaa0}), the scalar term is unaffected by $n_{\mu\nu}$ and the operator $D_{1}$ possesses a non-trivial structure. The expression for $D_{1}$ has a striking resemblance to that of $D_{1}$, which has opposite sign in the first term and $n_{\mu\nu}$ instead of $\tilde{n}_{\mu\nu}$.

\section{\label{appendixB1}The Problem with Quantum Equivalence}
To compare Eqs. (\ref{ceaa0}) and (\ref{ceaa4}), we define the difference in 1-loop effective actions  given by,
\begin{eqnarray}
\label{aqeq0}
\Delta\Gamma &=& \Gamma_{2}^{(1)} - \Gamma_{1}^{(1)} \nonumber \\
&=& \frac{i\hbar}{2}\Big[\ln\det(D_{2} - \alpha^{2}n^{\mu\nu}n_{\rho\sigma}) - \ln\det(D_{1}-\alpha^{2}) - \ln\det(D_{1}-\alpha^{2})\nonumber \\ 
&& + 2\ln\det(\Box_{x} - \alpha^{2})\Big].
\end{eqnarray}
In contrast, the corresponding difference in 1-loop effective action in the case of massive antisymmetric and vector fields, with mass $m$, with no spontaneous Lorentz violation is given by \cite{buchbinder2008},
\begin{eqnarray}
\label{aqeq1}
\Delta\Gamma' = \frac{i\hbar}{2}\Big[\ln\det(\Box_{2} - m^{2}) - 2\ln\det(\Box_{1}-m^{2}) + 2\ln\det(\Box_{x} - m^{2})\Big],
\end{eqnarray}
where,
\begin{eqnarray}
\label{aqeq2}
\Box_{2}B_{\mu\nu} &=& \Box_{x}B_{\mu\nu} - [\nabla^{\rho},\nabla_{\nu}]B_{\mu\rho} - [\nabla^{\rho},\nabla_{\mu}]B_{\rho\nu}, \nonumber \\
\Box_{1}C^{\mu} &=& \Box_{x}C^{\nu} - [\nabla^{\nu},\nabla_{\mu}]C^{\mu}.
\end{eqnarray}
This comparison between cases with and without SLV is quite insightful, because it helps in understanding how the functional operators change due to the presence of Lorentz violating terms. In the later case, the operator for St{\"u}ckelberg vector field and that for vector field of equivalent Lagrangian are equal, while in the former case they are not, as was noted earlier. Moreover, operators in Eq. (\ref{aqeq0}) do not contain the commutator terms due to presence of $n_{\mu\nu}$, and hence do not simplify in flat spacetime unlike their counterparts in Eq. (\ref{aqeq1}).

In flat spacetime, it can be explicitly checked that Eq. (\ref{aqeq1}) vanishes, taking into account the number of independent components of respective fields (eight, four and one for antisymmetric, vector and scalar fields respectively), because the commutators in Eq. (\ref{aqeq2}) vanish and hence the operators $\Box_{2}$, $\Box_{1}$, and $\Box_{x}$ are identical. Inferring quantum equivalence is thus trivial. However, this is clearly not the case in Eq. (\ref{aqeq0}) due to the non-trivial structure of operators $D_{2}$ and $D_{1}$. This can be demonstrated in a rather simple example when a special choice of tensor $n_{\mu\nu}$ is considered. It can be shown that in Minkowski spacetime, $n_{\mu\nu}$ can be chosen to have a special form
\begin{eqnarray}
\label{afse0}
n_{\mu\nu} = 
\left(\begin{matrix}
0 & -a & 0 & 0\\
a & 0 & 0 & 0\\
0 & 0 & 0 & b\\
0 & 0 & -b & 0
\end{matrix}\right),
\end{eqnarray}
where $a$ and $b$ are real numbers, provided at least one of the quantities $x_{1}\equiv -2(a^{2}-b^{2})$ and $x_{2}\equiv 4ab$ are non-zero \cite{altschul2010}. For simplicity, and dictated by the requirements for non-trivial monopole solutions \cite{seifert2010b}, we may choose $b=0$. Further, the constraint $n_{\mu\nu}n^{\mu\nu}=1$ implies that $a=1/\sqrt{2}$. Therefore, the only non-zero components of $n_{\mu\nu}$ are $n_{10}=1/\sqrt{2}$ and $n_{01}=-1/\sqrt{2}$. For the dual tensor $\tilde{n}_{\mu\nu}$, the non-zero components are $\tilde{n}_{32}=-1/\sqrt{2}$ and $\tilde{n}_{23}=1/\sqrt{2}$. Substituting in Eqs. (\ref{cea1}) and (\ref{deaa0}), one obtains, for the non-zero components of $n_{\mu\nu}$ and $\tilde{n}_{\mu\nu}$,
\begin{eqnarray}
\label{afse1}
D_{1}C^{2} &=& \partial_{2}^{2}C^{2} - \partial_{3}^{2}C^{2} + 2\partial_{3}\partial^{2}C^{3} + \partial^{2}\partial_{i}C^{i}, \nonumber \\
D_{1}C^{3} &=& - \partial_{2}^{2}C^{3} + \partial_{3}^{2}C^{3} + 2\partial^{3}\partial_{2}C^{2} + \partial^{3}\partial_{i}C^{i}, \nonumber \\
D_{1}C^{0} &=&  \partial_{0}^{2}C^{0} + \partial_{1}^{2}C^{0} + \partial_{0}\partial_{j}C^{j}, \\
D_{1}C^{1} &=& \partial_{0}^{2}C^{1} + \partial_{1}^{2}C^{1} + \partial_{1}\partial_{j}C^{j} , \nonumber \\
D_{2}B^{10} &=& \Box_{x}B^{10} + \partial_{j}\left(\partial_{1}B^{0j} + \partial_{0}B^{j1}\right) = - D_{2}B^{01}, \nonumber
\end{eqnarray}
where, $j=2,3$ and $i=0,1$. The remaining components of operators $D_{2}$, $D_{1}$ and $D_{1}$ are given by,
\begin{eqnarray}
\label{afse2}
D_{2}B^{jk} &=& \Box_{x}B^{jk} + \partial_{\mu}\partial^{j}B^{k\mu} + \partial_{\mu}\partial^{k}B^{\mu j}, \quad B^{jk}\neq B^{10} \nonumber \\
D_{1}C^{l} &=& \partial^{l}\partial_{\nu}C^{\nu}, \quad l=2,3 \\
D_{1}C^{k} &=& \partial^{k}\partial_{\nu}C^{\nu}, \quad k=0,1. \nonumber
\end{eqnarray}
An interesting feature here, compared to the case of Eq. (\ref{aqeq1}), is that Eqs. (\ref{afse1}) and (\ref{afse2}) substituted in Eq. (\ref{aqeq0}) show explicitly that $\Delta\Gamma$ does not vanish. However,  functional determinants in Eq. (\ref{aqeq0}) do not have field dependence and can only contribute as infinite (regularization-dependent) constants \cite{simon1977,dunne2008}. Hence, each determinant in Eq. (\ref{aqeq0}) can be normalized to identity and will thus be equal to each other. And once again, taking into account the degrees of freedom of corresponding tensor, vector and scalar fields, similar to Eq. (\ref{aqeq1}), they will cancel for all physical processes. This proves the quantum equivalence of theories (\ref{ceaa0}) and (\ref{ceaa4}) in flat spacetime. 


\chapter{Loop Integrals}
\label{AppendixC}

Most of the loop integrals are calculated using the well known PV reduction method \cite{bardin1999}. Some integrals, namely (\ref{int4},\ref{int5}) are calculated the general method outlined in Ref. \cite{romao2019}. Finite parts have been calculated for integrals needed for evaluating the effective potential.

Integrals in $\langle\tilde{S}_{2}\rangle$,$\langle\tilde{S}_{4}\rangle$:
\begin{eqnarray}
    \intp{k}\dfrac{k_{\mu}k_{\nu}}{k^2+\frac{m^2}{k_0}} &=& \dfrac{g_{\mu\nu}}{16\pi^2}\Big(\frac{m^4}{8 k_0^2} -  \frac{m^4 \bigl(-1 -  \frac{2}{\epsilon} + \gamma + \log(\pi) + \log(\frac{m^2}{k_0 \mu^2})\bigr)}{4 k_0^2}\Big) \\
    \intp{k}\dfrac{k_{\mu}}{k^2+\frac{m^2}{k_0}} &=& 0 \\
    \intp{k}\dfrac{1}{k^2+\frac{m^2}{k_0}} &=& \dfrac{1}{16\pi^2}\frac{m^2 \bigl(-1 -  \frac{2}{\epsilon} + \gamma + \log(\pi) + \log(\frac{m^2}{k_0 \mu^2})\bigr)}{k_0}
\end{eqnarray}
Integrals in $\langle\tilde{S}_{1}\tilde{S}_{1}\rangle$, $\langle\tilde{S}_{1}\tilde{S}_{3}\rangle$:
\begin{eqnarray}
    \intp{k'}\dfrac{k'^{\mu} k'^{\nu}}{k'^4 ((k'-k)^2+\frac{m^2}{k_0})} &=& \dfrac{1}{16\pi^2} \biggl(\tfrac{1}{4} g^{\mu\nu} \bigl(\frac{2}{\epsilon} -  \gamma -  \log(\pi)\bigr) \nonumber \\ && + \frac{k^{\mu} k^{\nu} \Bigl(\tfrac{1}{2} \bigl(\frac{2}{\epsilon} -  \gamma -  \log(\pi)\bigr) + \tfrac{1}{2} \bigl(- \frac{2}{\epsilon} + \gamma + \log(\pi)\bigr)\Bigr)}{2 k^2}\biggr) \\
    \intp{k'}\dfrac{k'^{\mu} k'^{\nu}}{k'^2 ((k'-k)^2+\frac{m^2}{k_0})} &=& \dfrac{1}{16\pi^2} \tfrac{1}{3} k^{\mu} k^{\nu} \bigl(\frac{2}{\epsilon} -  \gamma -  \log(\pi)\bigr) \nonumber \\ && -  \tfrac{1}{4} g^{\mu\nu} \bigl(\frac{2}{\epsilon} -  \gamma -  \log(\pi)\bigr) \bigl(\frac{m^2}{k_0} + \tfrac{1}{3} k^2\bigr)
\end{eqnarray}

\begin{eqnarray}
    \intp{k'}\dfrac{k'^{\mu}}{k'^2 ((k'-k)^2+\frac{m^2}{k_0})} &=& \dfrac{1}{16\pi^2} \dfrac{k^{\mu}}{2}(\frac{2}{\epsilon}-\gamma - \log(\pi)) \\
    \intp{k'}\dfrac{k'^{2}}{((k'-k)^2+\frac{m^2}{k_0})} &=& \dfrac{1}{16\pi^2} \Bigg[4 \Bigl(\frac{m^4}{8 k_0^2} - \frac{m^4 \bigl(-1 -  \frac{2}{\epsilon} + \gamma + \log(\pi) + \log(\frac{m^2}{k_0 \mu^2})\bigr)}{4 k_0^2}\Bigr) \nonumber \\ && + \frac{m^2 \bigl(-1 -  \frac{2}{\epsilon} + \gamma + \log(\pi) + \log(\frac{m^2}{k_0 \mu^2})\bigr) (k^2)}{k_0}\Bigg] \\
    \intp{k'}\dfrac{1}{k'^2 ((k'-k)^2+\frac{m^2}{k_0})} &=& \dfrac{1}{16\pi^2} \Big(2 + \frac{2}{\epsilon} -  \gamma -  \log(\pi) -  \log(\frac{m^2}{k_0 \mu^2}) \nonumber \\ && -  \log\bigl(1 + \frac{k_0 (k^2)}{m^2}\bigr) \bigl(1 + \frac{m^2}{k_0 (k^2)}\bigr)\Big)\\
    \intp{k'}\dfrac{1}{((k'-k)^2+\frac{m^2}{k_0})} &=& \dfrac{1}{16\pi^2} \frac{m^2 \bigl(-1 -  \frac{2}{\epsilon} + \gamma + \log(\pi) + \log(\frac{m^2}{k_0 \mu^2})\bigr)}{k_0}
\end{eqnarray}
Integrals in $\langle\tilde{S}_{2}\tilde{S}_{2}\rangle$:
\begin{eqnarray}
    \label{int4}
    \intp{k'}\dfrac{k'^{4}}{(k'^2 + \frac{m^2}{k_0}) ((k'-k)^2+\frac{m^2}{k_0})} = \dfrac{1}{16\pi^2} \Bigg(\frac{9 m^4}{16 k_0^2} + \frac{3 m^4}{4 \epsilon k_0^2} -  \frac{3 m^4 \gamma}{8 k_0^2} -  \frac{3 m^4 \log(\pi)}{8 k_0^2} \nonumber \\  -  \frac{3 m^4 \log(\frac{m^2}{k_0 \mu^2})}{8 k_0^2} + \frac{7 m^2 (k^2)}{8 k_0} + \frac{7 m^2 (k^2)}{4 \epsilon k_0} \nonumber \\  - \frac{7 m^2 \gamma (k^2)}{8 k_0} - \frac{7 m^2 \log(\pi) (k^2)}{8 k_0} -  \frac{7 m^2 \log(\frac{m^2}{k_0 \mu^2}) (k^2)}{8 k_0} + \tfrac{1}{8} (k^2)^2 \nonumber \\  + \frac{(k^2)^2}{8 \epsilon} -  \tfrac{1}{16} \gamma (k^2)^2 -  \tfrac{1}{16} \log(\pi) (k^2)^2 -  \tfrac{1}{16} \log(\frac{m^2}{k_0 \mu^2}) (k^2)^2 \nonumber \\ -  \dfrac{\arctan\Bigl(\frac{k_0^{1/2} (k^2)^{1/2}}{\bigl(4 m^2 -  k_0 (k^2)\bigr)^{1/2}}\Bigr) (k^2)^{3/2} \bigl(4 m^2 -  k_0 (k^2)\bigr)^{1/2}}{8 k_0^{1/2}}\Bigg)\\
    \label{int5}
     \intp{k'}\dfrac{k'^{2}k'^{\mu}}{(k'^2 + \frac{m^2}{k_0}) ((k'-k)^2+\frac{m^2}{k_0})} = -\dfrac{1}{16\pi^2} (\frac{2}{\epsilon}-\gamma - \log(\pi))\dfrac{3m^2}{2 k_0}k^{\mu}
\end{eqnarray}

\begin{eqnarray}
\label{int1}
    \intp{k'}\dfrac{k'^{\mu}k'^{\nu}}{(k'^2 + \frac{m^2}{k_0}) ((k'-k)^2+\frac{m^2}{k_0})} = \dfrac{1}{16\pi^2} \nonumber \\ g^{\mu\nu} \biggl(\frac{m^2 \bigl(-1 -  \frac{2}{\epsilon} + \gamma + \log(\pi) + \log(\frac{m^2}{k_0 \mu^2})\bigr)}{6 k_0} + \tfrac{1}{18} \bigl(- \frac{6 m^2}{k_0} -  (k^2)\bigr) \nonumber \\ -  \frac{\bigl(\frac{2}{\epsilon} -  \gamma -  \log(\pi)\bigr) \Bigl(\frac{2 m^4}{k_0^2} + (k^2)^2 - 2 \bigl(\frac{m^4}{k_0^2} -  \frac{2 m^2 (k^2)}{k_0}\bigr)\Bigr)}{12 (k^2)}\biggr) \nonumber \\ + k^{x1} k^{x2} \biggl(\frac{m^2 \bigl(-1 -  \frac{2}{\epsilon} + \gamma + \log(\pi) + \log(\frac{m^2}{k_0 \mu^2})\bigr)}{3 k_0 (k^2)} + \frac{\frac{6 m^2}{k_0} + (k^2)}{18 (k^2)} \nonumber \\ + \frac{\bigl(\frac{2}{\epsilon} -  \gamma -  \log(\pi)\bigr) \Bigl(\frac{2 m^4}{k_0^2} -  \frac{3 m^2 (k^2)}{k_0} + (k^2)^2 - 2 \bigl(\frac{m^4}{k_0^2} -  \frac{2 m^2 (k^2)}{k_0}\bigr)\Bigr)}{3 (k^2)^2}\biggr)\\
    \label{int2}
    \intp{k'}\dfrac{k'^{\mu}}{(k'^2 + \frac{m^2}{k_0}) ((k'-k)^2+\frac{m^2}{k_0})} = -\dfrac{1}{16\pi^2} k^{\mu} \frac{1}{2}\Big(-\frac{2}{\epsilon} + \gamma + \log(\pi)\Big)\\
	\label{int3}     
     \intp{k'}\dfrac{1}{(k'^2 + \frac{m^2}{k_0}) ((k'-k)^2+\frac{m^2}{k_0})} = \dfrac{1}{16\pi^2} \Big(2 + \frac{2}{\epsilon} -  \gamma -  \log(\pi) -  \log(\frac{m^2}{k_0 \mu^2}) \nonumber \\ -  \log\Bigl(\frac{1 + \bigl(1 + \frac{4 m^2}{k_0 (k^2)}\bigr)^{1/2}}{-1 + \bigl(1 + \frac{4 m^2}{k_0 (k^2)}\bigr)^{1/2}}\Bigr) \bigl(1 + \frac{4 m^2}{k_0 (k^2)}\bigr)^{1/2}\Big)
\end{eqnarray}
Integrals of type (\ref{int1},\ref{int2},\ref{int3}) are also present in $\langle\tilde{S}_{1}\tilde{S}_{1}\tilde{S}_{2}\rangle$. The rest of the integrals are, 
\begin{eqnarray}
\intp{k'}\dfrac{k'^{\mu}k'^{\nu}k'^{\rho}}{d_0 d_1 d_2} &=& \dfrac{1}{16 \pi^2} \dfrac{1}{12} \Big(\frac{2}{\epsilon} - \gamma - \log(\pi)\Big) \nonumber \\ && \times(g^{\nu \rho} (2 k^{\mu} + p^{\mu}) + g^{\rho\mu } (2 k^{\nu } + p^{\nu }) + g^{\mu \nu } (2 k^{\rho} + p^{\rho}) \\
\intp{k'}\dfrac{k'^{\mu}k'^{\nu}}{d_0 d_1 d_2} &=& \dfrac{1}{16 \pi^2} g^{\mu\nu} \Big(\frac{2}{\epsilon} - \gamma - \log(\pi)\Big)
\end{eqnarray}


\chapter{One-loop corrections to scalar field model with nonminimal coupling to gravity}
\label{AppendixD}

In this appendix, we present the essential steps while calculating the divergent part of a scalar theory with non-minimal coupling to gravity. The contributions to $divp(\Gamma)$ at quartic order in background scalar field, including nonminimal coupling ($\xi$) terms, have not been reported before. The final result Eq. (\ref{toms}) serves as the extension of results of Ref. \cite{mackay2010}. 

The perturbative expansion of action $S[\phi]$ in terms of the order of background field, $\bar{\phi}$, is obtained as\footnote{$\nu$ and $\omega$ are the bookkeeping parameters as usual.},
\begin{eqnarray}
    S_{0} &=& \tfrac{1}{2} m^2 (\delta \phi)^2 + \tfrac{1}{2} \delta \phi{}_{,a} \delta \phi{}^{,a} -  \frac{2 \bar{c}^{a} c_{a}{}^{,b}{}_{,b}}{\kappa} + h^{ab} h^{}{}_{a}{}^{c}{}_{,b}{}_{,c} \nonumber \\ 
    && -  \frac{h^{ab} h^{}{}_{a}{}^{c}{}_{,b}{}_{,c}}{\alpha} -  h^{a}{}_{a} h^{bc}{}_{,b}{}_{,c} + \frac{h^{a}{}_{a} h^{bc}{}_{,b}{}_{,c}}{\alpha} -  \tfrac{1}{2} h^{ab} h^{}{}_{ab}{}^{,c}{}_{,c} \nonumber \\ 
    && + \tfrac{1}{2} h^{a}{}_{a} h^{b}{}_{b}{}^{,c}{}_{,c} -  \frac{h^{a}{}_{a} h^{b}{}_{b}{}^{,c}{}_{,c}}{4 \alpha} \\
    S_{1} &=& \tfrac{1}{2} m^2 \kappa \delta \phi h^{a}{}_{a} \bar{\phi} -  \tfrac{1}{4} m^2 \kappa \nu \delta \phi h^{a}{}_{a} \bar{\phi} -  \tfrac{1}{2} \kappa \delta \phi h^{b}{}_{b} \bar{\phi}{}^{,a}{}_{,a} + \tfrac{1}{4} \kappa \nu \delta \phi h^{b}{}_{b} \bar{\phi}{}^{,a}{}_{,a} \nonumber \\ &&-  \tfrac{1}{2} \kappa \delta \phi h^{b}{}_{b}{}_{,a} \bar{\phi}{}^{,a} + \frac{\kappa \omega \delta \phi h^{b}{}_{b}{}_{,a} \bar{\phi}{}^{,a}}{2 \alpha} + \kappa \delta \phi h^{}{}_{ab} \bar{\phi}{}^{,a}{}^{,b} \nonumber \\ 
    && + \kappa \delta \phi \bar{\phi}{}^{,a} h^{}{}_{a}{}^{b}{}_{,b} -  \frac{\kappa \omega \delta \phi \bar{\phi}{}^{,a} h^{}{}_{a}{}^{b}{}_{,b}}{\alpha} + \kappa \xi \delta \phi \bar{\phi} h^{ab}{}_{,a}{}_{,b} -  \kappa \xi \delta \phi \bar{\phi} h^{a}{}_{a}{}^{,b}{}_{,b} ;
\end{eqnarray}

\begin{eqnarray}
    S_{2} &=& - \tfrac{1}{8} m^2 \kappa^2 h^{}{}_{ab} h^{ab} \bar{\phi}^2 + \tfrac{1}{16} m^2 \kappa^2 h^{a}{}_{a} h^{b}{}_{b} \bar{\phi}^2 + \tfrac{1}{4} \lambda \bar{\phi}^2 (\delta \phi)^2 -  \tfrac{1}{8} m^2 \kappa^2 \nu \bar{\phi}^2 (\delta \phi)^2 \nonumber \\ 
    && -  \tfrac{1}{8} \kappa^2 \nu \xi h^{}{}_{bc} h^{bc} \bar{\phi} \bar{\phi}{}^{,a}{}_{,a} + \tfrac{1}{16} \kappa^2 \nu \xi h^{b}{}_{b} h^{c}{}_{c} \bar{\phi} \bar{\phi}{}^{,a}{}_{,a} + \tfrac{3}{8} \kappa^2 \nu \xi \bar{\phi} (\delta \phi)^2 \bar{\phi}{}^{,a}{}_{,a} -  \tfrac{1}{8} \kappa^2 h^{}{}_{bc} h^{bc} \bar{\phi}{}_{,a} \bar{\phi}{}^{,a} \nonumber \\ 
    && + \tfrac{1}{16} \kappa^2 \nu h^{}{}_{bc} h^{bc} \bar{\phi}{}_{,a} \bar{\phi}{}^{,a} -  \tfrac{1}{8} \kappa^2 \nu \xi h^{}{}_{bc} h^{bc} \bar{\phi}{}_{,a} \bar{\phi}{}^{,a} + \tfrac{1}{16} \kappa^2 h^{b}{}_{b} h^{c}{}_{c} \bar{\phi}{}_{,a} \bar{\phi}{}^{,a} -  \tfrac{1}{32} \kappa^2 \nu h^{b}{}_{b} h^{c}{}_{c} \bar{\phi}{}_{,a} \bar{\phi}{}^{,a} \nonumber \\ 
    && + \tfrac{1}{16} \kappa^2 \nu \xi h^{b}{}_{b} h^{c}{}_{c} \bar{\phi}{}_{,a} \bar{\phi}{}^{,a} -  \tfrac{1}{16} \kappa^2 \nu (\delta \phi)^2 \bar{\phi}{}_{,a} \bar{\phi}{}^{,a} + \tfrac{3}{8} \kappa^2 \nu \xi (\delta \phi)^2 \bar{\phi}{}_{,a} \bar{\phi}{}^{,a} + \frac{\kappa^2 \omega^2 (\delta \phi)^2 \bar{\phi}{}_{,a} \bar{\phi}{}^{,a}}{4 \alpha} \nonumber \\ 
    && + \omega c^{a} \bar{c}^{b} \bar{\phi}{}_{,a} \bar{\phi}{}_{,b} + \tfrac{1}{2} \kappa^2 \xi h^{ab} \bar{\phi}^2 h^{c}{}_{c}{}_{,a}{}_{,b} -  \tfrac{1}{8} \kappa^2 \xi \bar{\phi}^2 h^{c}{}_{c}{}_{,b} h^{a}{}_{a}{}^{,b} + \tfrac{1}{2} \kappa^2 h^{}{}_{a}{}^{c} h^{}{}_{bc} \bar{\phi}{}^{,a} \bar{\phi}{}^{,b} \nonumber \\ 
    && -  \tfrac{1}{4} \kappa^2 \nu h^{}{}_{a}{}^{c} h^{}{}_{bc} \bar{\phi}{}^{,a} \bar{\phi}{}^{,b} + \tfrac{1}{2} \kappa^2 \nu \xi h^{}{}_{a}{}^{c} h^{}{}_{bc} \bar{\phi}{}^{,a} \bar{\phi}{}^{,b} -  \tfrac{1}{4} \kappa^2 h^{}{}_{ab} h^{c}{}_{c} \bar{\phi}{}^{,a} \bar{\phi}{}^{,b} + \tfrac{1}{8} \kappa^2 \nu h^{}{}_{ab} h^{c}{}_{c} \bar{\phi}{}^{,a} \bar{\phi}{}^{,b} \nonumber \\ 
    && -  \tfrac{1}{4} \kappa^2 \nu \xi h^{}{}_{ab} h^{c}{}_{c} \bar{\phi}{}^{,a} \bar{\phi}{}^{,b} -  \tfrac{1}{2} \kappa^2 \xi \bar{\phi}^2 h^{ab}{}_{,a} h^{}{}_{b}{}^{c}{}_{,c} + \tfrac{1}{2} \kappa^2 \xi \bar{\phi}^2 h^{a}{}_{a}{}^{,b} h^{}{}_{b}{}^{c}{}_{,c} -  \kappa^2 \xi h^{ab} \bar{\phi}^2 h^{}{}_{a}{}^{c}{}_{,b}{}_{,c} \nonumber \\ 
    && + \tfrac{1}{4} \kappa^2 \xi h^{a}{}_{a} \bar{\phi}^2 h^{bc}{}_{,b}{}_{,c} + \tfrac{1}{2} \kappa^2 \nu \xi h^{}{}_{a}{}^{c} h^{ab} \bar{\phi} \bar{\phi}{}_{,b}{}_{,c} -  \tfrac{1}{4} \kappa^2 \nu \xi h^{a}{}_{a} h^{bc} \bar{\phi} \bar{\phi}{}_{,b}{}_{,c} + \tfrac{1}{2} \kappa^2 \xi h^{ab} \bar{\phi}^2 h^{}{}_{ab}{}^{,c}{}_{,c} \nonumber \\ 
    && -  \tfrac{1}{4} \kappa^2 \xi h^{a}{}_{a} \bar{\phi}^2 h^{b}{}_{b}{}^{,c}{}_{,c} -  \tfrac{1}{4} \kappa^2 \xi \bar{\phi}^2 h^{}{}_{ac}{}_{,b} h^{ab}{}^{,c} + \tfrac{3}{8} \kappa^2 \xi \bar{\phi}^2 h^{}{}_{ab}{}_{,c} h^{ab}{}^{,c}; \\
    S_{3} &=& \tfrac{1}{12} \kappa \lambda \delta \phi h^{a}{}_{a} \bar{\phi}^3 -  \tfrac{1}{24} \kappa \lambda \nu \delta \phi h^{a}{}_{a} \bar{\phi}^3 ; \\
    S_{4} &=& - \tfrac{1}{96} \kappa^2 \lambda h^{}{}_{ab} h^{ab} \bar{\phi}^4 + \tfrac{1}{192} \kappa^2 \lambda h^{a}{}_{a} h^{b}{}_{b} \bar{\phi}^4 -  \tfrac{1}{96} \kappa^2 \lambda \nu \bar{\phi}^4 (\delta \phi)^2 .
\end{eqnarray}

The propagators derived from $S_{0}$ have similar form as Eqs. (\ref{acqc3}) and (\ref{acqc4}), with appropriate parameters. With manipulations similar to the one carried out in Chap. 4, we finally find the divergent part of one-loop effective potential, which contributes to the corrections to $\bar{\phi}^4$ coupling: 
\begin{eqnarray}
    \label{toms}
    divp(\Gamma) &=& \tfrac{5}{16} m^4 \kappa^2 \bar{\phi}^2 -  \tfrac{1}{4} m^2 \lambda \bar{\phi}^2 -  \tfrac{3}{4} m^4 \kappa^2 \xi \bar{\phi}^2 + \tfrac{3}{4} m^4 \kappa^2 \xi^2 \bar{\phi}^2 \nonumber \\ && + \tfrac{13}{96} m^2 \kappa^2 \lambda \overline{\phi}^4 -  \tfrac{1}{16} \lambda^2 \overline{\phi}^4 -  \tfrac{1}{2} m^2\kappa^2 \lambda \xi \overline{\phi}^4 + \tfrac{3}{4} m^2 \kappa^2\lambda \xi^2 \overline{\phi}^4
\end{eqnarray}
Reading off the coefficients $A$ and $C$ from Eq. (\ref{toms}), similar to Eq. (\ref{ren01}) and using Eq. (\ref{ren06}), we obtain $\Delta\lambda$, given by Eq. (\ref{ren07}).



\end{document}